\documentclass[]{pasj01}
\usepackage{url}
\usepackage{color}
\draft

\begin{document} 
%\Received{}%{yyyy/mm/dd}
%\Accepted{}%{yyyy/mm/dd}
%\Published{yyyy/mm/dd}

%\title{Sub-Jupiter Mass Planets to Five G- and K-type Giant Stars}
\title{Regular Radial Velocity Variations in Nine G- and K-type Giant Stars: Eight Planets and One Planet Candidate}

%%% begin:list of authors
% Do NOT capitalize all letters in "textsc".
\author{
Huan-Yu \textsc{Teng}\altaffilmark{1,*},
Bun'ei \textsc{Sato}\altaffilmark{1},
Takuya \textsc{Takarada}\altaffilmark{2},
Masashi \textsc{Omiya}\altaffilmark{2,3},
Hiroki \textsc{Harakawa}\altaffilmark{4},
Hideyuki \textsc{Izumiura}\altaffilmark{5},
Eiji \textsc{Kambe}\altaffilmark{4},
Yoichi \textsc{Takeda}\altaffilmark{6},
Michitoshi \textsc{Yoshida}\altaffilmark{4},
Yoichi \textsc{Itoh}\altaffilmark{7},
Hiroyasu \textsc{Ando}\altaffilmark{3}, and
Eiichiro \textsc{Kokubo}\altaffilmark{8}
}

\email{teng.h.aa@m.titech.ac.jp}
%\thanks{Example: Present Address is xxxxxxxxxx}}
%%% end:list of authors

\altaffiltext{1}{Department of Earth and Planetary Sciences, School of Science, Tokyo Institute of Technology, 2-12-1 Ookayama, Meguro-ku, Tokyo 152-8551, Japan}
\altaffiltext{2}{Astrobiology Center, National Institutes of Natural Sciences, 2-21-1 Osawa, Mitaka, Tokyo 181-8588, Japan}
\altaffiltext{3}{National Astronomical Observatory of Japan, National Institutes of Natural Sciences, 2-21-1 Osawa, Mitaka, Tokyo 181-8588, Japan}
\altaffiltext{4}{Subaru Telescope, National Astronomical Observatory of Japan, National Institutes of Natural Sciences, 650 North A’ohoku Pl., Hilo, HI, 96720, USA}
\altaffiltext{5}{Okayama Branch Office, Subaru Telescope, National Astronomical Observatory of Japan, National Institutes of Natural Sciences, Kamogata, Asakuchi, Okayama 719-0232, Japan}
\altaffiltext{6}{11-2 Enomachi, Naka-ku, Hiroshima-shi, 730-0851}
\altaffiltext{7}{Nishi-Harima Astronomical Observatory, Center for Astronomy, University of Hyogo, 407-2, Nishigaichi, Sayo, Hyogo 679-5313, Japan}
\altaffiltext{8}{The Graduate University for Advanced Studies
(SOKENDAI), 2-21-1 Osawa, Mitaka, Tokyo 181-8588, Japan}
%\altaffiltext{9}{Earth-Life Science Institute, Tokyo Institute of Technology, 2-12-1 Ookayama, Meguro-ku, Tokyo 152-8551, Japan}

%% `\KeyWords{}' always has to be placed before `\maketitle'.
 %Do NOT move this preamble from here!

\maketitle

\begin{abstract}
We report the detection of radial velocity variations in nine evolved G- and K-type giant stars. The observations were conducted at Okayama Astrophysical Observatory. Planets or planet candidates can best explain these regular variations. However, a coincidence of near 280-day variability among five of them prevents us from fully ruling out stellar origins for some of the variations, since all nine stars behave similarly in stellar properties. In the planet hypotheses to the RV variations, the planets (including one candidate) may survive close to the boundary of the so-called ``planet desert'' around evolved stars, having orbital periods between 255 and 555 days. Besides, they are the least-massive giant planets detected around G- and K-type giant stars, with minimum masses between 0.45$M_{\rm{J}}$ and 1.34$M_{\rm{J}}$. We further investigated other hypotheses for our detection, yet none of them can better explain regular RV variation. With our detection, it is convinced that year-long regular variation with amplitude down to 15 $\rm{m\ s^{-1}}$ for G- and K-type giant stars is detectable. Moreover, we performed simulations to further confirm the detectability of planets around these stars. Finally, we explored giant planets around intermediate-mass stars, and likewise found a 4 Jupiter mass gap (e.g. \cite{Santos2017}), which is probably a boundary of the giant planet population.
\end{abstract}

\section{Introduction}\label{sec:intro}
Since the first extrasolar planet (exoplanet) 51 Peg b was detected, more than 4000 exoplanets in 3000 planetary systems have been discovered around different types of stars. These exoplanets were detected by different methods, among which radial velocity (RV) method (Doppler method) contributed near 800 of them. 
For these planets, about 60\% are detected around solar-mass ($ 0.7-1.5 M_{\odot}$) stars, while only about 1\% are detected around intermediate-mass ($1.5-5.0 M_{\odot}$) stars because of the difficulty in detecting planets around them. 
However, since the properties of the circumstellar disks such as mass and lifetime around intermediate-mass stars are different from those around solar-mass and low-mass stars \citep{Haisch2001a, Haisch2001b, Andrews2013, Ribas2015}, intermediate-mass stars are valuable observational objects in terms of constraining the mechanism and timescale of planet formation.
Currently, there are two plausible scenarios to solve the puzzle, the core accretion scenario (e.g. \cite{Safronov1969, Goldreich1973, Hayashi1985, Pollack1996}) 
and the disk instability scenario (e.g. \cite{Kuiper1951, Cameron1978}).
Therefore, more observational evidence becomes urgent in understanding when different procedures apply in giant planet formation.

To observe intermediate-mass stars, it is usually hard to search for planets with the RV method around the main sequence (MS) stars. This is because intermediate-mass MS stars exhibit few absorption lines due to their high surface temperature and fast self-rotation \citep{Lagrange2009}. While evolved intermediate-mass stars, which are the evolved counterparts of late-F to early-B dwarfs and spectroscopically G- or K-type subgiants or giants, have lower temperature and slower rotation for which it is more appropriate for RV measurements. 
So far, several groups have been surveying these evolved intermediate-mass stars with the RV method: the Okayama Planet Search Program \citep{Sato2005} and its collaborative survey “EAPS-Net” \citep{Izumiura2005}, the BOAO K-giant survey \citep{Han2010}, the Lick G- and K-giant survey \citep{Frink2001,Hekker2006}, the ESO planet search program \citep{Setiawan2003}, the Tautenburg Observatory Planet Search \citep{Hatzes2005,Dollinger2007}, the Penn State Toru\'n Planet Search \citep{Niedzielski2007} and its RV follow-up program Tracking Advanced Planetary Systems \citep{Niedzielski2015}, the survey “Retired A Stars and Their Companions” \citep{Johnson2006}, the Pan-Pacific Planet Search \citep{Wittenmyer2011}, and the EXoPlanet aRound Evolved StarS project \citep{Jones2011}. Thanks to the large-scale surveys, around 150 planets have been discovered around evolved stars.

An interesting problem of ``Planet Desert” happened to these evolved intermediate-mass stars; namely, close-in ($a < 0.6\ \rm{au}$), low-mass ($M_{\rm{p}} < 0.6 M_{\rm{J}}$), planets were seldom found \citep{Johnson2007,Lillo-Box2016,Medina2018}. 
Thanks to precise RV measurements and transiting observations (e.g. \textit{Kepler} space telescope \cite{Borucki2012}), few planets with semi-major axis and minimum mass respectively lower than 0.1 au and 0.1 $M_{\rm{J}}$ mass have been discovered around intermediate-mass evolved stars \citep{Huber2013, Johnson2010}. Especially for evolved giant stars, most detected planets reside at more than 0.5 au from their hosts \citep{Jones2014}.
According to a simulation given by \citet{Alibert2011}, the desert could be attributed to a scaling of the proto-planetary disk mass with the mass of the central star. Nevertheless, when the mass of a central star is getting lower, a few warm planets less massive than 1 $M_{\rm J}$ could survive around or closer than 1 au to their central stars. Therefore observing intermediate-mass stars could be a key point in constraining mass relation between proto-planetary disk and central star.
In addition, among all detected giant exoplanets, they are more likely to reside around metal-rich stars. However, specifically for these evolved stars, many studies give the same planet-metallicity correlation that they all prefer high metallicities (e.g. \cite{Reffert2015}; \cite{Jones2016}; \cite{Wittenmyer2017a}), while some others (e.g. \cite{Jofre2015}) give no clear correlation. In an up-to-date research given by \citet{Santos2017}, the distribution of giant planet companions is likely to present more than one population with a change in regime around 4 $M_{\rm{J}}$, above which host stars tend to be more metal-poor and more massive. On the other hand, stars hosting planets below this limit show the well-known metallicity-giant planet frequency correlation. In other words, the giant planet-metallicity correlation around evolved stars still needs improvements.

In this work, we study nine evolved G- and K-type giant stars that emerged from the Okayama Planet Search Program \citep{Sato2005}. These nine stars exhibit RV amplitude of only a few tens meters per second, and five of these stars coincidentally show regular RV variations at nearly 280 days. According to our investigation, these variations are best explained by planet nature, yet we could not fully rule out the stellar origin of the 280-day variability.
The rest of this paper is organized as follows. The observations are described in section \ref{sec:obs} and the stellar properties are described in section \ref{sec:prop}. In section \ref{sec:ana}, we detailedly present our analysis to RV measurements, line profile and chromospheric activity, period search, and Keplerian orbital fit. Then in section \ref{sec:res}, we show our Keplerian fitting results by stars. Finally, in section \ref{sec:discuss} we give discussions about alternative explanations and then we summarize the paper. In addition, detailed analyses on IP variability, instrumental stability, and individual stars are respectively given in Appendix \ref{sec:bisip}, \ref{sec:inststable}, and \ref{sec:stars}.

\section{Observations}\label{sec:obs}
In this work, all spectra of the nine stars (HD 360, $\epsilon$ Psc, HD 10975, HD 79181, HD 99283, $\upsilon$ Leo, HD 161178, HD 219139, and $\gamma$ Psc) were obtained by 1.88-m reflector with HIgh Dispersion Echelle Spectrograph (HIDES: \cite{Izumiura1999}) at Okayama Astrophysical Observatory.
The first spectrum was obtained in 2001 under the Okayama Planet Search Program \citep{Sato2005}, which focuses on radial velocity measurements to late-G (or early-K) giant stars, and aims at opening the world of planets around intermediate-mass stars. As for the instrument, an iodine cell was placed in the HIDES optical path, which provided numerous iodine absorption lines in the range of 5000-5800 \AA ~as a reference for precise radial velocity measurements. To cover these iodine absorption lines, the wavelength region of HIDES was firstly set to cover 5000-6100 \AA ~with one $2 \mathrm{K} \times 4 \mathrm{K}$ CCD. In December 2007, the CCD of HIDES was upgraded from the single one to a mosaic of three, which widened the wavelength region to 3700-7500 \AA ~(3700-5000 \AA, ~5000-5800 \AA ~and ~5800-7500 \AA ~for each respectively). The upgrade enabled us to simultaneously measure the level of stellar activities (e.g. Ca \emissiontype{II} HK lines) and line profiles as well as radial velocities.
In 2010, a new high-efficiency fiber-link system with its own iodine cell was installed to the HIDES which greatly enhanced the overall throughput \citep{Kambe2013}. In this research, the spectra were obtained by both conventional slit mode with CCD pre- and post-upgrade (hereafter HIDES-S), and fiber-link mode (hereafter HIDES-F). 

In the case of HIDES-S observations, the slit width was set to 200 $\mu$m ($\timeform{0.76''}$) corresponding to the resolution $\mathit{R} = \lambda / \Delta \lambda \sim 67000$ by about 3.3-pixel sampling. In the case of HIDES-F observations, the width of the sliced image was $\timeform{1.05''}$ corresponding to the resolution $\mathit{R} \sim 55000$ by about 3.8-pixel sampling. There was an offset in radial velocity measurements between two modes due to their respective iodine in their optical path, hence we treated two modes as two different instruments in our research. 

We adopted the data which gained signal-to-noise ratio (S/N) over approximately 100 per pixel at $\sim$5500 \AA ~within 1800 seconds. For the faintest star (HD 10975, $V=5.93$) in this work, the typical exposure time and S/N is 1200 seconds and over 150, 500 seconds over 200, in the case of HIDES-S and HIDES-F observations, respectively.

The reduction of these echelle spectra (i.e. bias subtraction, flat fielding, scattered light subtraction, and spectrum extraction) was performed with IRAF\footnote{IRAF is distributed by the National Optical Astronomy Observatories, which is operated by the Association of Universities for Research in Astronomy, Inc. under a cooperative agreement with the National Science Foundation, USA} packages in a standard way.

Especially, for HIDES-F spectra, there were severe aperture overlaps among 3700--4000 \AA ~owing to the use of an image slicer, therefore the scattered light for these apertures could not be subtracted with IRAF task $\mathtt{apscatter}$. Consequently, we discard these overlapped apertures and did not analyze the Ca \emissiontype{II} H lines for HIDES-F spectra.

\section{Stellar properties}\label{sec:prop}
\begin{table*}[p]
\begin{center}	
\rotatebox{90}{\begin{minipage}{1.0\vsize}
\begin{center}
\tbl{Stellar parameters.\footnotemark[$*$]}{
\begin{tabular}{lrrrrrrrrrr}
%%%%%%%%%%%%%%%%%%%%%%%%%%%%%%%%%%%%%%%%%%%%%%%%			
\hline\hline
& HD 360 & $\epsilon\ \rm{Psc}$ & HD 10975 & HD 79181 & HD 99283 & $\upsilon\ \rm{Leo}$ & HD 161178 & HD 219139 & $\gamma\ \rm{Psc}$ & source\\ 
\hline 
$\pi\ (\rm{mas})$& $9.0115^{\rm{a}}$ & $17.8074^{\rm{a}}$ & $8.6049^{\rm{a}}$ & $9.6870^{\rm{a}}$ & $9.1832^{\rm{a}}$ & $18.3461^{\rm{a}}$ & $9.3052^{\rm{a}}$ & $9.4433^{\rm{a}}$ & $23.64^{\rm{b}}$  & a, b \\ 
$V$& $5.98$ & $4.28$ & $5.93$ & $5.72$ & $5.70$ & $4.30$ & $5.87$ & $5.85$ & $3.70$  & \textit{Hippacors}   \\ 
$B-V$& $1.03$ & $0.96$ & $0.99$ & $0.97$ & $1.00$ & $1.01$ & $1.02$ & $1.00$ & $0.92$  & \textit{Hippacors} \\ 
%$J$& $4.49$ & $2.60$ & $4.36$ & $4.22$ & $4.25$ & $2.59$ & $4.30$ & $4.05$ & $2.02$  & 2MASS   \\ 
%$H$& $3.78$ & $2.23$ & $3.87$ & $3.57$ & $3.72$ & $2.10$ & $3.77$ & $3.50$ & $1.49$  & 2MASS  \\ 
%$K$& $3.65$ & $2.00$ & $3.88$ & $3.54$ & $3.57$ & $2.01$ & $3.66$ & $3.39$ & $1.44$  & 2MASS \\ 
Spec. type & G8 III& K0 III& K0 III& G8 III& K0 III & G9 III& G9 III& G5 III& G7 III & \textit{Hippacors} \\ 
$T_{\mathrm{eff,sp}}\ ({\mathrm{K}})$ & $4850$ & $4829$ & $4866$ & $4842$ & $4883$ & $4835$ & $4766$ & $4860$ & $4802$  & Takeda08  \\ 
$\mathrm{[Fe/H]}_{\mathrm{sp}} \mathrm{(dex)}$ & $-0.08$ & $-0.31$ & $-0.17$ & $-0.29$ & $-0.17$ & $-0.19$ & $-0.20$ & $-0.19$ & $-0.62$  & Takeda08  \\ 
$\log g_{\star,sp}\ ({\mathrm{cgs}})$ & $2.62$ & $2.30$ & $2.47$ & $2.47$ & $2.59$ & $2.47$ & $2.33$ & $2.50$ & $2.25$  & Takeda08  \\ 
$v\sin i (\mathrm{km}\ \mathrm{s}^{-1})$ & $1.86$ & $1.79$ & $1.69$ & $1.84$ & $1.87$ & $1.78$ & $1.78$ & $2.04$ & $1.67$  & Takeda08   \\ 
$T_{\mathrm{eff}}\ ({\mathrm{K}})$ & $4770$ & $4824$ & $4840$ & $4862$ & $4886$ & $4836$ & $4786$ & $4831$ & $4742$  & This work  \\ 
${\mathrm{[Fe/H]\ (dex)}}$ & $-0.06_{-0.10}^{+0.17}$ & $-0.31_{-0.10}^{+0.10}$ & $-0.21_{-0.08}^{+0.10}$ & $-0.28_{-0.09}^{+0.08}$ & $-0.16_{-0.10}^{+0.06}$ & $-0.20_{-0.08}^{+0.09}$ & $-0.21_{-0.07}^{+0.09}$ & $-0.20_{-0.07}^{+0.09}$ & $-0.62_{-0.11}^{+0.10}$  & This work  \\ 
$\log g_{\star}\ ({\mathrm{cgs}})$ & $2.59_{-0.17}^{+0.05}$ & $2.34_{-0.13}^{+0.07}$ & $2.48_{-0.09}^{+0.09}$ & $2.45_{-0.07}^{+0.09}$ & $2.57_{-0.08}^{+0.07}$  & $2.46_{-0.09}^{+0.09}$ & $2.38_{-0.03}^{+0.07}$ & $2.48_{-0.07}^{+0.08}$ & $2.33_{-0.07}^{+0.08}$  & This work  \\ 
$L_{\star}\ (L_{\odot})$ & $55.16_{-2.81}^{+2.14}$ & $75.52_{-21.93}^{+130.72}$ & $61.64_{-5.83}^{+6.80}$ & $61.61_{-6.17}^{+6.68}$ & $64.51_{-5.36}^{+9.46}$ & $63.10_{-7.48}^{+41.77}$ & $56.36_{-5.34}^{+5.94}$ & $62.32_{-5.06}^{+7.33}$ & $58.41_{-3.38}^{+8.23}$  & This work \\ 
$M_{\star}\ (M_{\odot})$ & $1.69_{-0.53}^{+0.15}$ & $1.41_{-0.50}^{+1.45}$ & $1.41_{-0.35}^{+0.34}$ & $1.28_{-0.28}^{+0.32}$ & $1.76_{-0.30}^{+0.31}$  & $1.48_{-0.38}^{+0.90}$ & $1.06_{-0.17}^{+0.25}$ & $1.46_{-0.29}^{+0.31}$ & $0.99_{-0.13}^{+0.17}$  & This work  \\ 
$R_{\star}\ (R_{\odot})$ & $10.86_{-0.28}^{+0.22}$ & $12.49_{-2.18}^{+7.97}$ & $11.16_{-0.38}^{+0.45}$ & $11.06_{-0.43}^{+0.44}$ & $11.21_{-0.39}^{+0.66}$ & $11.22_{-0.45}^{+3.15}$ & $10.95_{-0.54}^{+0.42}$ & $11.22_{-0.27}^{+0.55}$ & $11.20_{-0.24}^{+0.22}$  & This work \\
%Age (age) & $1.53_{-0.25}^{+3.41}$ & $2.00_{-1.68}^{+6.07}$ & $2.44_{-1.12}^{+3.80}$ & $3.13_{-1.54}^{+4.30}$ & $1.31_{-0.34}^{+0.91}$  & $2.00_{-1.41}^{+3.12}$ & $6.84_{-3.59}^{+5.92}$ & $2.24_{-0.95}^{+2.41}$ & $6.88_{-2.87}^{+4.21}$  & This work  \\ 
$\sigma_{\mathrm{HIP}}$ & 0.011 & 0.006 & 0.008 & 0.006 & 0.007 & 0.006 & 0.005 & 0.006 & 0.005 & \textit{Hippacors}\\
\hline\hline
%%%%%%%%%%%%%%%%%%%%%%%%%%%%%%%%%%%%%%%%%%%%%%%%			
\end{tabular}
}
\begin{tabnote}
\hangindent6pt\noindent
\hbox to6pt{\footnotemark[$*$]\hss}\unskip% 
As for parallax measurements, mark ``a'' refers to \textit{Gaia EDR3}, and mark ``b'' refers to \textit{Hippacors}. 
Takeda08 refers to \citet{Takeda2008}, and the footnote ``sp'' refers to spectral determination.
\end{tabnote}
\label{tab1}
\end{center}
\end{minipage}}
\end{center}	
\end{table*}
%\fi

The nine stars in this research include HD 360, $\epsilon$ Psc, HD 10975, HD 79181, HD 99283, $\upsilon$ Leo, HD 161178, HD 219139, and $\gamma$ Psc. We collected spectral type and $V$-band magnitudes from the \textit{Hipparcos} catalog \citep{ESA1997}, and we collect parallax from either \textit{Hipparcos} (ESA: \cite{vanLeeuwen2007}) or \textit{Gaia EDR3} \citep{Gaia2016,Gaia2021,Lindegren2021} according to precision. 
%The apparent $J$-, $H$-, and $K$-band magnitude are gathered from \textit{2MASS} catalog \citep{Cutri2003}.
The atmospheric parameters (effective temperature $T_{\rm{eff,sp}}$, surface gravity $\log g_{\rm{sp}}$, and Fe abundance [Fe/H]$_{\rm{sp}}$) were determined by \citet{Takeda2008} by measuring equivalent width of Fe \emissiontype{I} and Fe \emissiontype{II} lines of iodine-free stellar spectra. \citet{Takeda2008} also determined projected rotational velocity, $v\sin i$ with the automatic spectrum-fitting technique \citep{Takeda1995}.\\
\begin{figure}
 \begin{center}
  \includegraphics[width=8.0cm]{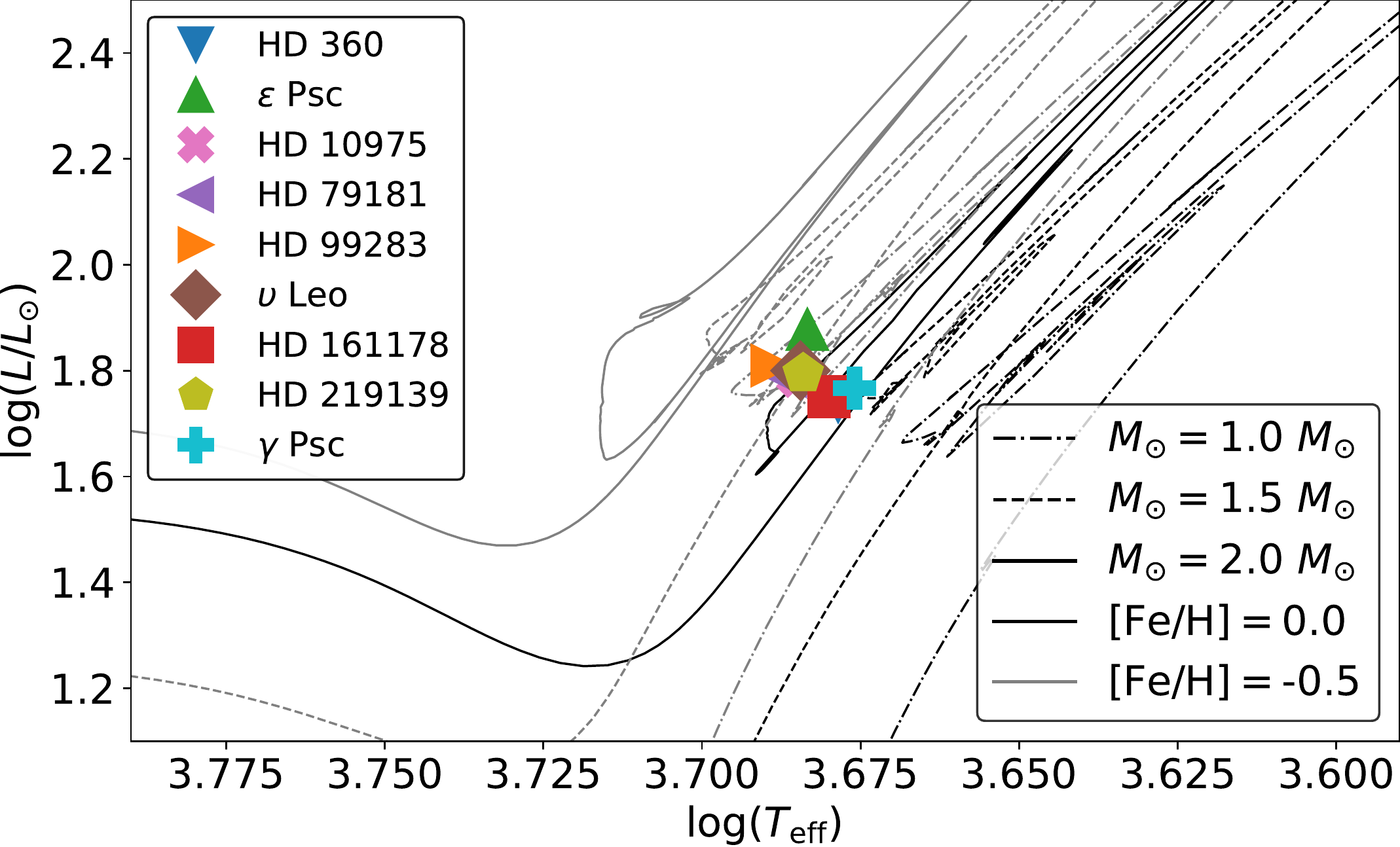} 
 \end{center}
\caption{HR diagram with HD 360, $\epsilon$ Psc, HD 10975, HD 79181, HD 99283, $\upsilon$ Leo, HD 161178, HD 219139, and $\gamma$ Psc. Some evolutionary tracks for masses of 1.0$M_{\odot}$, 1.5$M_{\odot}$ and 2.0$M_{\odot}$ are also shown. The black lines and grey lines correspond to [Fe/H] = 0.0 and [Fe/H] = -0.5, respectively.}\label{fig:HR}
\end{figure}

\begin{figure}
 \begin{center}
  \includegraphics[width=8.0cm]{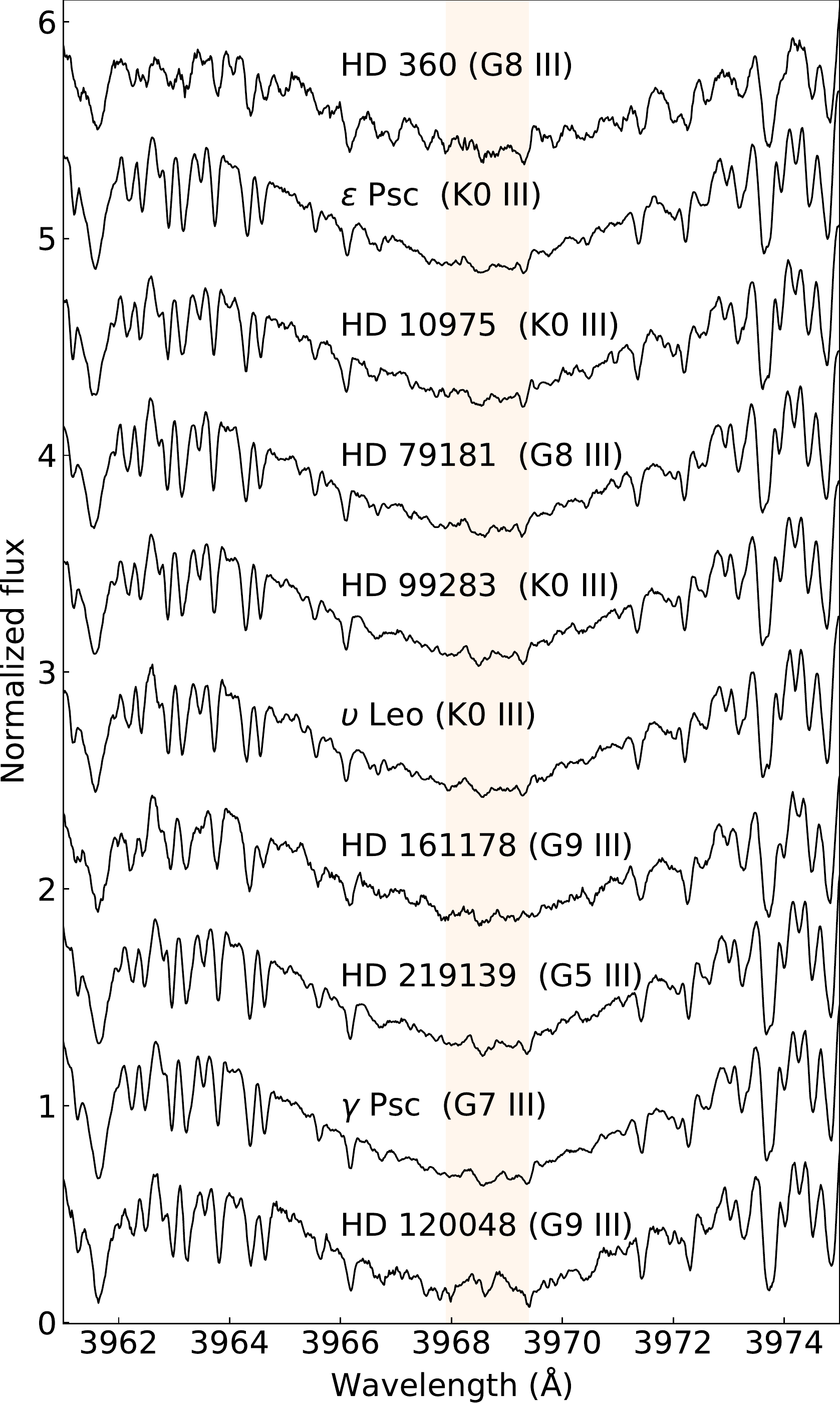} 
 \end{center}
\caption{Spectra in the region of Ca \emissiontype{II} H lines of 
HD 360, $\epsilon$ Psc, HD 10975, HD 79181, HD 99283, $\upsilon$ Leo, HD 161178, HD 219139, and $\gamma$ Psc with a comparison of the active star HD 120048. 
Vertical offsets are added to each normalized spectrum for clarity.}\label{fig:CaIIH}
\end{figure}

However, according to a work given by \citet{Takeda2015}, stellar masses in \citet{Takeda2008} tend to be overestimated by up to a factor of $\lesssim 2$, especially for giant stars located near the red clump regions in the Hertzsprung-Russell (HR) diagram. For this reason, we re-estimated stellar parameters by adopting the Bayesian estimation method with theoretical isochrones with \texttt{isoclassify} \citep{Huber2017, Berger2020}.  
Using the \texttt{isoclassify} ``direct'' mode, $R_{\star}$ and $L_{\star}$ were calculated from posterior probability distributions by applying the Stefan-Boltzmann law with spectroscopic $T_{\mathrm{eff,sp}}$ and $V$-band as photometry as inputs. In this step, distances were sampled following the parallax a posterior from either \textit{Hipparcos} or \textit{Gaia EDR3}. For each distance sample, the extinction $A_{V}$ was calculated with map given by \citet{Green2019} as implemented in \texttt{mwdust} package by \citet{Bovy2016}. Each sample was combined with independent random normal samples for apparent magnitude and effective temperature $T_{\mathrm{eff}}$. The bolometric corrections were derived by linearly interpolating $T_{\mathrm{eff}}$, $\mathrm{[Fe/H]}$, $\log g_{\star}$, and $A_{V}$ in the \texttt{MIST}/C3K grid \citep{ConroyInPrep}. The derived distances, extinctions, and bolometric corrections were iterated until convergence. Using the \texttt{isoclassify} ``grid'' mode, parameters include $T_{\mathrm{eff}}$, $\mathrm{[Fe/H]}$, $\log g_{\star}$, $L_{\star}$, $R_{\star}$, and $M_{\star}$ were estimated with posterior distribution by integrating over \texttt{MIST} isochrone \citep{Paxton2011, Paxton2013, Paxton2015, Choi2016, Dotter2016, Paxton2018}. Consequently, with masses estimated, some stars were re-classified as solar mass stars, or namely, low-mass giant stars. The properties of the nine stars in this paper are listed in table \ref{tab1} and plotted on the HR diagram in Figure \ref{fig:HR}.
These nine stars are known to be stable in photometry with a level of $\sigma_{\rm{HIP}} = 0.006$--$0.011\ \rm{mag}$ (ESA: \cite{vanLeeuwen2007}), and they are chromospherically inactive with no significant emission in the core of the Ca \emissiontype{II} H lines as it is shown in Figure \ref{fig:CaIIH}. Another chromospherically active G-type giant star (HD 120048) serves as a comparison to the nine stars. It is together shown in Figure \ref{fig:CaIIH}, where we can see clear emission lines in the core of the Ca \emissiontype{II} H lines.

\section{Analysis}\label{sec:ana}
\subsection{Radial velocity}
For precise RV measurements, we used spectra covering 5000 to 5800 \AA ~in which I$_{\rm{2}}$ absorption lines are superimposed by an I$_{\rm{2}}$ cell. We computed RV variations following the method described in \citet{Sato2002} and \citet{Sato2012} which was based on a method by \citet{Butler1996}. We modeled the spectra (I$_{\rm{2}}$ superposed stellar spectra) by using the stellar template and high-resolution I$_{\rm{2}}$ spectra which were convolved with the instrumental profile (IP) of the spectrograph. 
IP was described as one central Gaussian envelope and 10 satellite Gaussian envelopes, where we fixed the widths of 0.9 pixels and positions of 0.9 pixels and set the height as free parameters. Further analysis of IP variability is given in Appendix \ref{sec:bisip}.
The stellar template spectra used for HIDES-S were obtained by deconvolving pure stellar spectra with the IP estimated from I$_{\rm{2}}$ superposed B type star or flat spectra. The stellar template spectra used for HIDES-F were obtained by HIDES-F High-resolution mode ($\mathit{R} \sim 100000$) observations without I$_{\rm{2}}$ in its optical path. RV offsets of HIDES-S and HIDES-F were set to the mean level of the full RV time series as two free parameters (see Section \ref{sec:kepfit}). Hundreds of segments at a typical width of 150 pixels were set in the observed spectra in the I$_{\rm{2}}$ absorption region, and the final RV values and their measurement errors were taken from the average of measurement in each segment. The analysis of RV stability of the instruments is given in Appendix \ref{sec:inststable}. 

\subsection{Line profile and chromospheric activity}
Fake planetary signal is possibly masqueraded by wavelength shift due to spectral line profile deformation. To evaluate the possibility, we perform line profile analyses following the same procedure in \citet{Takarada2018}.
The iodine-free spectra within the wavelength range of 4000-5000 are adopted for this analysis. 
First, we calculate the weighted cross-correlation function (CCF: \cite{Baranne1996,Pepe2002}) and a numerical mask.
CCF is constructed by shifting the mask as a function of Doppler velocity, and its weight is determined by the depth of the spectral line. The numerical mask is generated from \texttt{SPECTRUM} \citep{Gray1994} for a G-type giant star with approximately 800 lines.
Next, we calculate the bisector inverse span (BIS: \cite{Dall2006}) of the CCF. The BIS is defined as the offset of averaged velocity between the upper region (5\%--15\% from the continuum of the CCF) and the lower region (85\%--95\% from the continuum of the CCF). Thus, we adopt the BIS as a measurement of line-profile asymmetry. Moreover, the full width at half maximum (FWHM: e.g. \cite{Suarez2020}) of CCF for bisector by each instrument is used as another line profile indicator. In order to remove the instrumental difference between instrument, we define $\rm{BIS}^{\prime}=\rm{BIS}-\overline{BIS}$, and $\rm{FWHM}^{\prime}=\rm{FWHM}-\overline{FWHM}$ for each instrument, where overline means the mean value. 

Stellar chromospheric activity could be one of the explanations for RV variations. For G- and K-type stars, Ca \emissiontype{II} HK lines, whose flux of line cores trace efficiently chromospheric activity \citep{Duncan1991}, have been widely used as activity indicators. For HIDES, the signal-to-noise ratio of Ca \emissiontype{II} K lines are not high enough for the use as the indicator. Hence following \citet{Sato2013}, we defined Ca \emissiontype{II} H index $S_H$ as
\begin{eqnarray}
S_H &=& \frac{F_H}{F_B+F_R}
\end{eqnarray}
where $F_H$ is a total flux in a wavelength bin
0.66 \AA ~wide centered on the H line, $F_B$ and $F_R$ are those in bins 1.1 \AA ~wide centered on minus and plus 1.2 \AA ~from the center of the H line, respectively. The errors were estimated based on photon noise. If RV variation contained a dominant component of stellar activity, there would be a significant correlation between RV and $S_H$, e.g. HD 120048 \citep{Sato2013}.

\subsection{Period search}\label{sec:period}
To search periodicity in the time series, we performed Generalized Lomb-Scargle (hereafter GLS) periodogram \citep{Zechmeister2009} with $\mathtt{Python}$ package $\mathtt{astropy}$. To assess the significance of the periodicity, we applied False Alarm Probability (FAP) with approximation method developed by \citet{Baluev2008} in the same package. 
For some stars, to further investigate the true period of the candidate, Stacked Bayesian Generalized Lomb-Scargle periodogram (SBGLS: \cite{Mortier2017}) was performed so that we could check the evolution of the periods with increasing data amount. 
The main idea of this method is stacking Bayesian GLS periodograms (BGLS\footnote{BGLS describes the probability that a full sine function with a specific frequency is present in the data, where weights derived from errors on the data and a constant offset is included.}: \cite{Mortier2015}) from the first $n$ points of data and normalizing periodograms with their respective minimum values. In this way, the power of a signal can be compared over time, and the signal of stable and strong(er/est) power is closer to the true one. The SBGLS has been adopted in similar studies of planet searching around giant stars, e.g. \citep{Pinto2020}.

\subsection{Keplerian orbital fit}\label{sec:kepfit}
\begin{table*}
\tbl{Priors for MCMC sampling.}{%
\begin{tabular}{lcccc}
\hline\hline
Prameter & Prior & Minimum & Maximum & knee value \\ 
\hline
$P\ (\rm{d})$ & Jeffrey's & 1 & 10000 & --\\
$K\ (\rm{m\>s^{-1}})$ & Modified Jeffrey's & 1.01 & 100 & 1 \\
$\sqrt{e} \cos \omega$ & Uniform & 0 & 1 & -- \\
$\sqrt{e} \sin \omega$ & Uniform & 0 & 1 & -- \\
$\gamma_{\rm{inst}}$ & Uniform & -200 & 200 & -- \\
$s_{\rm{inst}}$ & Modified Jeffrey's & 1.01 & 100 & 1 \\
\hline
\end{tabular}}
\begin{tabnote}
\hangindent6pt\noindent
\hbox to6pt{\footnotemark[$*$]\hss}\unskip% 
The subscript ``inst'' refers to a certain instrument.
\end{tabnote}
\label{tabprior}
\end{table*}

Orbital motion of planet(s) can be account for stable periodic RV variation of a star. Therefore, Keplerian orbital fit was performed on each star in order to investigate the validity of planetary nature. 
The best-fit Keplerian orbit for the data was derived by maximum likelihood maximization using $\mathtt{Python}$ package $\mathtt{scipy}$ with a truncated Newton \textsf{TNC} method and a \textsf{Nelder-Mead} method \citep{Nelder1965}. 
The Keplerian model was generated by $\mathtt{RadVel}$ \citep{Fulton2017, Fulton2018}.
The free parameters included Keplerian orbital elements (orbital period $P$, RV semi-amplitude $K$, the combination of eccentricity $e$ and argument of periastron $\omega$, $\sqrt{e} \cos \omega$ and $\sqrt{e} \sin \omega$ and time of inferior conjunction $T_{\rm{c}}$\footnote{The pariastron passage $T_{\rm{p}}$ is converted from inferior conjunction $T_{\rm{c}}$ after parameter fitting.}), RV offset of different instruments $\gamma_{\rm{inst}}$ and extra Gaussian noise $s_{\rm{inst}}$. The uncertainties for these parameters were derived by Markov Chain Monte Carlo (MCMC) sampling using $\mathtt{emcee}$ \citep{Foreman-Mackey2013}. 
Gelman-Rubin statistic (G-R hereafter; \cite{Gelman2013}) was used to check convergence of the chain during and after burn-in. After the chain firstly reached the G-R level lower than 1.01, we considered the chain marginally mixed well and stopped burn-in. After the chain passed five checkpoints of G-R $<1.01$ and independent samples $T_z$ \citep{Ford2006} greater than 1000, we stopped chain and finished sampling. For the priors, we selected Jeffrey's prior for $P$, uniform distribution for $\sqrt{e} \cos \omega$, $\sqrt{e} \sin \omega$ and $\gamma_{\rm{inst}}$, and modified Jeffrey's prior for $K$ and $s_{\rm{inst}}$. The detailed values for each prior are summarized in Table \ref{tabprior}. After fitting, we converted the best-fit and uncertainties of $\sqrt{e} \cos \omega$ and $\sqrt{e} \sin \omega$ into $e$ and $\omega$ as the final results. To have an evaluation of a long-term trend (liner and quadratic plus linear), which might introduce a possible outer planet, we calculated the reduced Chi-square $\chi^{2}_{\rm{red}}$ ($\chi^{2}_{\mathrm{red}} = \chi^{2} / N_{\mathrm{DoF}}$, where $N_{\mathrm{DoF}}$ is the degree of freedom) and BIC value (\cite{Schwarz1978}; $\mathrm{BIC} = -2\ln{(\hat{L})} + n_{\mathrm{pars}} \ln{(n_{\mathrm{data}})} $, where $\hat{L}$ is the maximized value of the likelihood function, $n_{\mathrm{pars}}$ is the number of parameters, and $n_{\rm{data}}$ is the number of data)
for model with trend and without trend. $\chi^{2}_{\rm{red}}$ shows the goodness of fits, and BIC value indicates the difference between models. We ruled out those fittings with $\chi^{2}_{\rm{red}}$ far from 1. 
To determine if we should use a model with a long-term trend, we compared the BIC values of different models. Only if the BIC value of the model with a higher-order trend showed a lower value with a difference greater than 10 \citep{Kass1996}, we would take the higher-order trend into further discussion. Finally, we make a comparison against the null hypothesis, which only includes RV jitters and possible linear trends. Here, we also adopt BIC as a criterion. Similarly, only if the Keplerian model has a lower BIC value with a difference greater than 10, we believe the Keplerian model is significantly better than the null hypothesis. 

We also tried to check the consistency in orbital parameters derived from HIDES-S data and HIDES-F ones, respectively. However, for each star, our observations were sparsely distributed over 15 years, especially for those data obtained by HIDES-S. In such a case, we could not perform Keplerian orbital fits by using only HIDES-S data. Thus, we could not clearly know if the best-fit RV semi-amplitude and orbital period apparently vary in time, which may suggest the possibility of non-planet origin for the observed RV variations.
For simplicity instead, we tested the variability of scattering of original RV data. We divided the data of each instrument into three and calculated the root mean square (rms) of the former, middle, and the latter third of them for RVs and JDs. For HD360, HD 10975, HD 99283, and $\gamma$ Psc observed by HIDES-S, we only calculated the rms of all the data due to their small number of observations (35, 28, 32, and 34 for HD 360, HD 10975, HD 99283, and $\gamma$ Psc respectively). The rms of JDs were used for illustrating the scatter of observations, in other words, the sparseness of observations. A more constant rms value of RVs could make the Keplerian model more convincing. Concerning solar-like oscillations, poor data sampling, and instrumental drift within a short period of time, the rms values of RVs varying within a few $\rm{m\>s^{-1}}$ should be accepted.

\section{Results}\label{sec:res}
In this section, we present the solutions of the Keplerian orbits of our nine stars in Okayama Planet Search Program: HD 360, $\epsilon$ Psc, HD 10975, HD 79181, HD 99283, $\upsilon$ Leo, HD 161178, HD 219139, and $\gamma$ Psc.
According to Bayesian analysis, all nine stars favor the one-Keplerian model rather than the null hypothesis.
The fitted orbital periods range from 255 to 555 days, and fitted minimum masses range from 0.45$M_{\rm{J}}$ to 1.34$M_{\rm{J}}$ with RV semi-amplitude between 11.7 $\rm{m\ s^{-1}}$ and 34.0 $\rm{m\ s^{-1}}$.
With further analyses in RV time series, line profile, and activity indicator, $\epsilon$ Psc can be only labeled as a planet-harboring candidate rather than a confirmed planet-harboring star. This is because we found BIS moderately correlated RV, and meanwhile, we found BIS varies at similar periods to the RV periodicity. 
In addition, the possibility that the periodicity varying by time could not be ruled out, thus $\epsilon$ Psc was not confirmed as a planet host.
As for the other eight stars, they are confirmed to be host planets, and the star HD 360, which showed a strong RV linear trend, is likely
to host another outer companion.
The orbital parameters of eight planets and one planet candidate and are summarized in Table \ref{dtab},
and the detailed analysis for each individual star is presented in Appendix \ref{sec:stars} .

\begin{table*}[p]
\begin{center}	
\rotatebox{90}{\begin{minipage}{1.0\vsize}
\begin{center}
\tbl{Orbital parameters.\footnotemark[$*$]}{
\begin{tabular}{lcccccccccc}\hline\hline
Parameters  & HD 360 b  & $\epsilon\ \rm{Psc}$ b  & HD 10975 b  & HD 79181 b  & HD 99283 b  & $\upsilon\ \rm{Leo}$ b  & HD 161178 b  & HD 219139 b  & $\gamma\ \rm{Psc}$ b  \\ 
\hline 
$P\ (\rm{d})$  & $273.1_{-0.8}^{+1.6}$ & $255.3_{-1.4}^{+2.1}$ & $283.8_{-0.1}^{+12.7}$ & $273.1_{-0.4}^{+1.3}$ & $310.4_{-1.7}^{+5.2}$ & $385.2_{-1.3}^{+2.8}$ & $279.3_{-0.8}^{+1.0}$ & $275.5_{-1.0}^{+2.3}$ & $555.1_{-2.5}^{+6.0}$  \\ 
$K\ (\rm{m\ s^{-1}})$ & $16.8_{-2.6}^{+1.8}$ & $20.6_{-2.9}^{+2.1}$ & $12.3_{-3.5}^{+0.2}$ & $17.7_{-2.2}^{+1.6}$ & $20.4_{-3.5}^{+1.9}$ & $11.7_{-1.6}^{+1.7}$ & $17.0_{-2.3}^{+1.6}$ & $19.2_{-2.6}^{+1.8}$ & $34.0_{-5.0}^{+3.9}$  \\ 
$e$ & $0.139_{-0.104}^{+0.108}$ & $0.278_{-0.186}^{+0.122}$ & $0.442_{-0.389}^{+0.040}$ & $0.259_{-0.206}^{+0.067}$ & $0.200_{-0.146}^{+0.102}$ & $0.320_{-0.218}^{+0.134}$ & $0.044_{-0.019}^{+0.132}$ & $0.110_{-0.081}^{+0.087}$ & $0.204_{-0.141}^{+0.114}$  \\ 
$\omega\ (^{\circ})$ & $133_{-63}^{+119}$ & $237_{-36}^{+26}$ & $135_{-35}^{+153}$ & $318_{-32}^{+48}$ & $291_{-41}^{+62}$ & $131_{-31}^{+36}$ & $300_{-70}^{+180}$ & $318_{-67}^{+113}$ & $44_{-56}^{+34}$  \\ 
$T_{\rm{p}}$(JD$-2450000$) & $5184.5_{-92.8}^{+44.9}$ & $5420.0_{-85.8}^{+6.9}$ & $5450.7_{-182.9}^{+29.1}$ & $5123.0_{-11.6}^{+103.8}$ & $5291.1_{-25.6}^{+256.1}$ & $5279.6_{-34.2}^{+27.9}$ & $5296.6_{-12.0}^{+218.4}$ & $5240.1_{-12.2}^{+198.9}$ & $5394.6_{-63.5}^{+60.6}$  \\ 
$M_{\rm{p}}\sin i\ (M_{\rm{J}})$ & $0.75_{-0.15}^{+0.12}$ & $0.77_{-0.10}^{+0.16}$ & $0.45_{-0.15}^{+0.06}$ & $0.64_{-0.16}^{+0.06}$ & $0.97_{-0.25}^{+0.06}$ & $0.51_{-0.26}^{+0.06}$ & $0.57_{-0.16}^{+0.02}$ & $0.78_{-0.20}^{+0.05}$ & $1.34_{-0.31}^{+0.02}$  \\ 
$a\ (\rm{AU})$ & $0.98_{-0.03}^{+0.11}$ & $0.88_{-0.10}^{+0.11}$ & $0.95_{-0.08}^{+0.06}$ & $0.90_{-0.08}^{+0.07}$ & $1.08_{-0.07}^{+0.05}$ & $1.18_{-0.32}^{+0.11}$ & $0.85_{-0.07}^{+0.05}$ & $0.94_{-0.07}^{+0.06}$ & $1.32_{-0.08}^{+0.05}$  \\ 
\hline
$s_{\rm{s}}\ (\rm{m\ s^{-1}})$ & $16.7_{-0.7}^{+4.3}$ & $22.7_{-1.6}^{+3.3}$ & $10.2_{-0.5}^{+3.0}$ & $14.1_{-0.5}^{+2.3}$ & $17.5_{-1.7}^{+4.0}$ & $12.5_{-0.7}^{+2.6}$ & $15.9_{-0.8}^{+2.1}$ & $16.6_{-1.0}^{+3.3}$ & $27.7_{-2.1}^{+6.5}$  \\ 
$s_{\rm{f}}\ (\rm{m\ s^{-1}})$ & $12.3_{-0.7}^{+1.9}$ & $14.5_{-0.5}^{+2.1}$ & $9.6_{-0.5}^{+1.6}$ & $14.7_{-1.2}^{+1.7}$ & $11.3_{-0.4}^{+2.5}$ & $8.4_{-0.4}^{+1.5}$ & $14.8_{-0.8}^{+1.9}$ & $12.5_{-0.6}^{+2.0}$ & $21.0_{-1.1}^{+4.2}$  \\ 
$\gamma_{\rm{s}}\ (\rm{m\ s^{-1}})$ & $11.6_{-3.7}^{+3.0}$ & $-2.2_{-2.8}^{+3.6}$ & $-0.5_{-2.4}^{+2.2}$ & $0.1_{-1.7}^{+1.7}$ & $-0.3_{-4.1}^{+3.5}$ & $1.2_{-2.5}^{+1.8}$ & $2.7_{-2.1}^{+2.0}$ & $1.1_{-3.4}^{+2.5}$ & $0.5_{-5.2}^{+4.0}$  \\ 
$\gamma_{\rm{f}}\ (\rm{m\ s^{-1}})$ & $-27.9_{-3.4}^{+3.8}$ & $0.7_{-1.7}^{+1.8}$ & $1.1_{-1.7}^{+1.0}$ & $0.4_{-1.7}^{+2.0}$ & $-0.1_{-1.7}^{+1.9}$ & $-1.2_{-1.6}^{+1.3}$ & $-0.4_{-1.8}^{+1.8}$ & $-4.9_{-1.5}^{+2.0}$ & $-2.5_{-3.2}^{+3.6}$  \\ 
$\dot{\gamma}\ (\rm{m\ s^{-1}\ d^{-1}} )$  & $0.018_{-0.002}^{+0.002}$ & - & - & - & - & - & - & - & -  \\ 
$\ddot{\gamma}\ (\rm{m\ s^{-1}}\ d^{-2})$  & - & - & - & - & - & - & - & - & -  \\ 
\hline 
$\rm{rms}\ (\rm{m\ s^{-1}})$ & $14.8$ & $18.9$ & $10.7$ & $15.5$ & $14.4$ & $11.2$ & $16.0$ & $15.1$ & $24.6$  \\ 
$\chi^{2}_{\rm{red}}$ & $1.142$ & $1.083$ & $1.119$ & $1.075$ & $1.143$ & $1.115$ & $1.072$ & $1.107$ & $1.140$  \\ 
$\rm{BIC}$ & $885.5$ & $1295.4$ & $793.5$ & $1420.6$ & $712.4$ & $848.1$ & $1346.1$ & $967.1$ & $868.5$  \\
$\Delta\rm{BIC}$ & $-25.0$ & $-32.7$ & $-10.0$ & $-46.5$ & $-19.2$ & $-22.3$ & $-33.5$ & $-30.2$ & $-26.2$ \\
\hline
$N_{\rm{obs,s}}$ & $35$ & $60$ & $31$ & $92$ & $28$ & $46$ & $77$ & $42$ & $35$  \\ 
$N_{\rm{obs,f}}$ & $67$ & $84$ & $67$ & $72$ & $54$ & $59$ & $77$ & $69$ & $54$  \\ 
$t_{\rm{start}}$ (Year/Month) & 2004/01 & 2002/02 & 2004/01 & 2001/03 & 2004/01 & 2001/12 & 2001/03 & 2004/01 & 2002/02 \\
$t_{\rm{end}}$ (Year/Month)   & 2017/12 & 2017/11 & 2017/12 & 2017/12 & 2017/11 & 2017/11 & 2017/12 & 2017/12 & 2017/12 \\
\hline
Detection & Planet & Candidate & Planet & Planet & Planet & Planet & Planet & Planet & Planet \\
\hline\hline
\end{tabular}
}
\begin{tabnote}
\hangindent6pt\noindent
\hbox to6pt{\footnotemark[$*$]\hss}\unskip% 
The subscript ``s'' and ``f'' refer to results related to HIDES-S data and HIDES-F data respectively. $\Delta\rm{BIC}$ is defined as $\Delta\rm{BIC}=\rm{BIC}-\rm{BIC}_{0}$, where $\rm{BIC}_{0}$ is the BIC value of null hypothesis.
\end{tabnote}
\label{dtab}
\end{center}
\end{minipage}}
\end{center}	
\end{table*}

\section{Discussion}\label{sec:discuss}
According to Keplerian fitting, planetary orbital motion can be an interpretation to periodic variation in the RV time series of eight G- and K-type giant stars from Okayama Planet Search Program: HD 360, $\epsilon$ Psc, HD 10975, HD 79181, HD 99283, $\upsilon$ Leo, HD 161178, HD 219139, and $\gamma$ Psc. While with further analysis in RV time series, line profiles, and stellar activity indicator, it is indicated that the RV variation of $\epsilon$ Psc might intrinsically from the star itself. Therefore we only label $\epsilon$ Psc as a planet-harboring candidate.
In the planetary explanation, nine planetary companions including the candidate orbit their host stars with semimajor axes around 1 au, and seven of nine have the lowest minimum masses ever discovered around solar-mass to intermediate-mass giant stars. Interestingly, RV variations of five stars (HD 360, HD 10975, HD 79181, HD 161178, HD 219139) have similar RV variability around 280 days.
Is it a coincidence? 
In the subsequent part of this section, we first put forward another two alternative scenarios to explain the regular RV variation. Then, we respectively discuss eccentricities and detectability of planets around G- and K-type giant stars. Finally, we discuss the metallicity of planet-harboring intermediate-mass stars. 

\subsection{Stellar rotation of G- and K-type giant stars}
\begin{figure}
 \begin{center}
  \includegraphics[width=8.0cm]{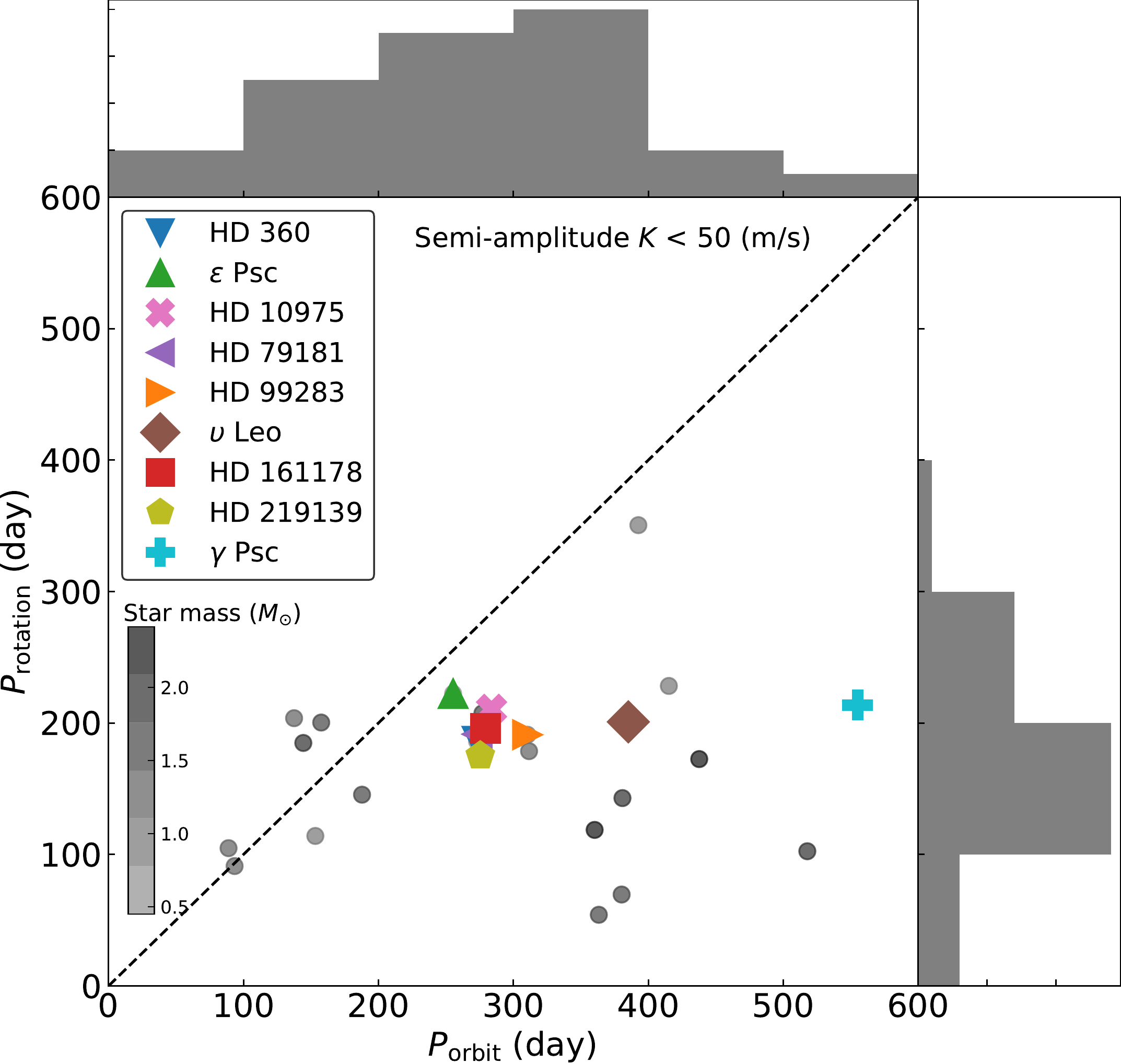} 
 \end{center}
\caption{Stellar rotational against planetary orbital period. Filled dots are G- and K-type planet-hosting giant stars with their planets showing RV semi-amplitude weaker than 50 $ \rm{m\>s^{-1}}$  (Data are collected on 2019 May 30)}\label{fig:Prot_Porb}
\end{figure}

From the Lick survey, \citet{Saar1998} proposed the relationship between rotational velocity with stellar jitter among dwarf stars. However, we know less about G- and K-type giant stars for their rotations, since these stars are usually very slow rotators with rotating periods of a few hundred days. Strong activity such as starspots accompanying rotation could probably cause RV variation similar to the Keplerian RV variation. For the nine stars in this work, yet we did not find significant periodicity from the Hipparcos photometry and Ca \emissiontype{II} H index $S_H$, where we could not recognize their stellar rotation periods. 
Therefore, at least for the nine stars in this work, there is little evidence that activity accompanied with stellar rotation is the source of regular RV variations, yet the lack of decade-long chromatic RV measurements and photometry could not fully discard activity-related variability.
Furthermore, we collect known planet-harboring G- and K-type giant stars showing RV semi-amplitude less than $50\ \rm{m\>s^{-1}}$, and compare the expected rotational periods with their planet orbiting periods. Since the stellar rotations are given in projected velocities $v\sin i$, we therefore take the mean value\footnote{The average of $\sin i$ is $\overline{\sin i} = (1/\pi) \int_{0}^{\pi}\sin i \mathrm{d}i \simeq 0.637$} of $\sin i \simeq 0.637$ and converted velocities into days.
As seen in Figure \ref{fig:Prot_Porb}, most of the reported planets including the nine ones in this research (under planetary nature) have orbital periods longer than the expected stellar rotation periods. We also see that the expected rotation periods concentrate on 200$\sim$300 days, while there's no concentration in the histogram of the orbital period. We calculated Pearson’s correlation coefficient for these data, and it returns $r=0.08$ suggesting a weak correlation. 
Statistically, it implies that the similar RV variations in G- and K-type giant stars are not related to stellar rotation, while we doubt a small number of individuals might refer to it.

\subsection{Long Secondary Periods in Giant Stars}
Long secondary periods (LSP) is a mysterious phenomenon that happens to some very high luminosity (usually $L > 300 L_{\odot}$ \cite{Saio2015}) giant stars. Typically, this LSP ranges about 200 days to 1500 days and co-exists with a primary period which is 8 times shorter. Observations showed RV variation of a few $\rm{km\>s^{-1}}$ \citep{Wood2004, Nicholls2009} and photometry variation of up to 1 mag \citep{Saio2015}, and periods in both patterns consist well with each other. 
\citet{Hatzes2018} showed a new form of stellar variability in a K-type giant star $\gamma$ Dra (K5 III, $B-V=1.53$), and \citet{Reichert2019} showed similar variation in another K-type giant star Aldebaran (K5 III). The true mechanism of LSP is not yet fully understood. \citet{Saio2015} proposed that LSP could relate to oscillatory convective modes in red giants. The variability in these two stars is also possibly related to oscillatory convective modes and behaves similar to planetary signals in RV, showing an RV semi-amplitude over $100\ \rm{m\>s^{-1}}$. 
\citet{Dollinger2021} also raised the feasibility that oscillations and other phenomena could mimic exoplanets especially for stars larger than 21$R_{\odot}$.
Compared to K-type giant stars, whose RV variations are usually up to several hundred $\rm{m\>s^{-1}}$, our late-G and early-K giant stars with $0.8 < B-V < 1.0$ and $R \sim 10 R_{\odot}$ in this research have better intrinsic stability in RV at a level of $\sigma \sim 20\ \rm{m\>s^{-1}}$ \citep{Henry2000, Sato2005}. \citet{Soszynski2021} proposed the binarity origin of LSP by characterization of the near-infrared light curves, which illustrates that the nature for the LSP modulation is obscured by a dusty cloud orbiting the red giant star with a substellar or stellar companion. In this scenario, the companion originates from a planet and accrete a huge amount of mass from the envelope of its host star. If so, we could assume that our targets are very precursors of substellar or stellar companions, which does not negate planet explanation of RV variations of our stars in this study. 

Nonetheless, in an up-to-date research, \citet{Heeren2021} showed additional regular variations with a period of 291 days in the spectroscopic binary $\epsilon$ Cyg. We notice this period is close to our detection of near 300-day periods. Importantly, the main star $\epsilon$ Cyg A (K0 III, $1.1 M_{\odot}$) shows similar stellar properties \citep{Heeren2021} to our targets. In their work, they ruled out planetary hypothesis by detection of apparently time-varying RV amplitude and orbital period, and failure in orbital stability analysis. Although it was proposed that the 291-day regular variation of $\epsilon$ Cyg A seemed to be a possibly extreme example of a heartbeat system, they did not fully exclude the mechanism of LSP. 
Combining stars in \citet{Heeren2021} and this work, they all share similar stellar properties and periodic RV variation. So, here we may leave a question: Are they originate from the same intrinsic variation of stars?
We should take notice of one essential difference that eight of our nine stars are single stars. Although $\epsilon$ Psc is a suspected double star with separation around 1000 au (0.03 arcsec; \cite{Mason2001}), the separation is more than 60 times wider than Cyg A and B, for which we do not consider the heartbeat.
Thus, the most likely heartbeat scenario for the $\epsilon$ Cyg system can be easily rejected to our stars. Since \citet{Heeren2021} did not fully eliminate the possibility of LSP, it is reasonable to suspect they are all of LSP origin.  

In a word, even if the stellar properties of our stars do not comfort well with the currently known LSP star, due to the probably incomplete knowledge of LSP, we could not fully rule out the LSP explanation. 
Furthermore, our nine stars in this work share similar stellar properties. As we can see in the HR diagram (Figure \ref{fig:HR}), all nine stars concentrate around $\log (T_{\mathrm{eff}}) \simeq 3.68$ and $\log (L/L_{\odot}) \simeq 1.8$. We infer it is most likely to be the result of target screening of the Okayama Planet Search Program \citep{Sato2005}. Yet due to similarities in stellar properties, we still reserve the explanation that the nearly 280-day periodic variation might relate to another new form of intrinsic stellar variability of G- and K-type giant stars. 

\subsection{Eccentricity of Planets around G- and K-type Giant Stars}
\begin{figure}
 \begin{center}
  \includegraphics[width=8.0cm]{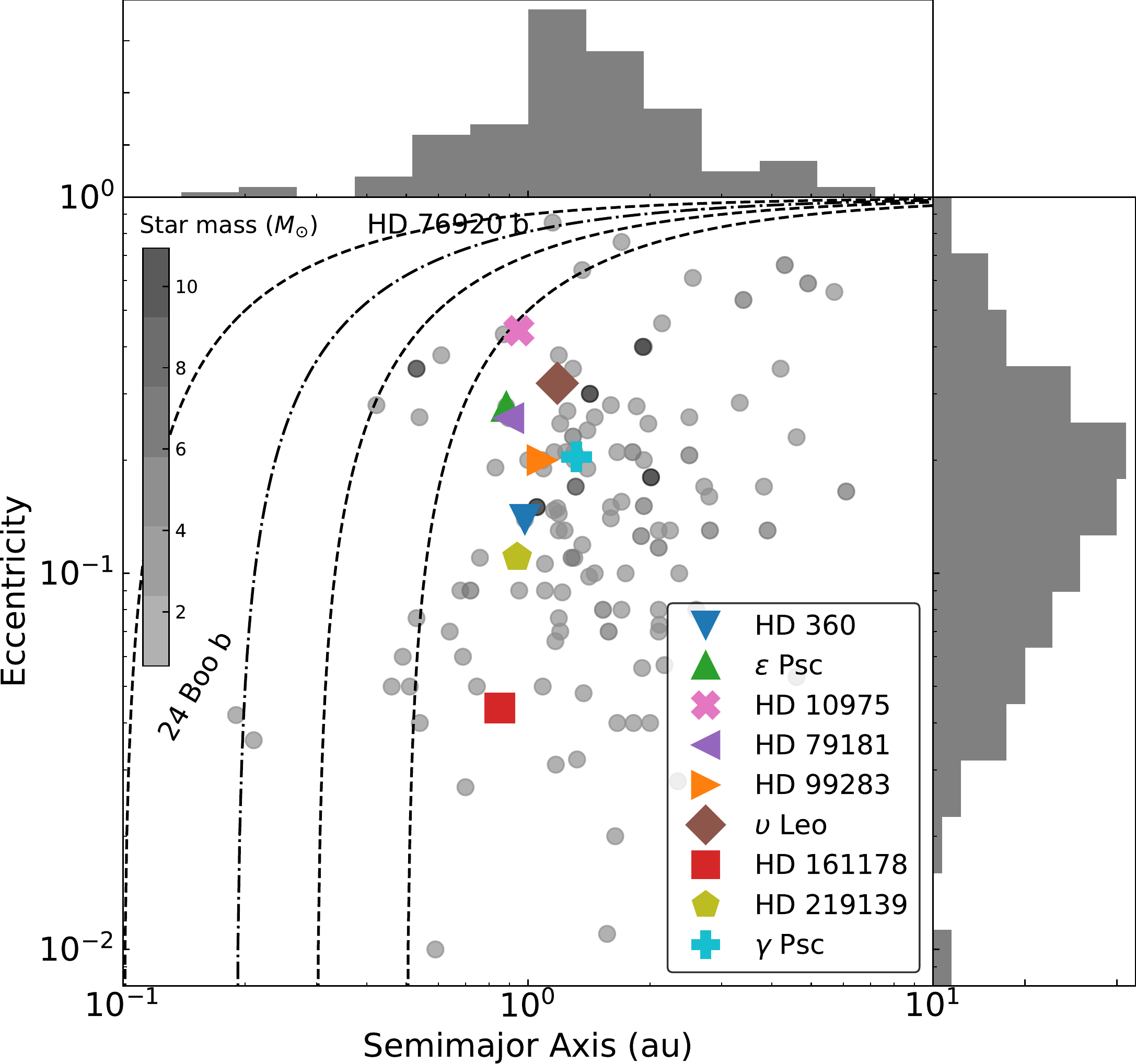} 
 \end{center}
\caption{Eccentricity as a function of semimajor axis. Filled circles are planets around G- and K-type giant stars with their darkness representing the mass of their host stars. The dash-dotted curve represents periastron distance $q=0.2\ \rm{au}$, and dashed curves represents periastron distance $q$ equal to 0.1, 0.3, 0.5 $\rm{au}$ from left to right respectively. The periastron distance is $q=a(1-e)$, where $a$ is the semimajor axis and $e$ is eccentricity. All axes and histograms are in logarithm. (Data are collected on 2019 May 30)}\label{fig:sma_ecc}
\end{figure}

According to Keplerian fitting results HD 161178 b ($e=0.044$) orbit their hosts in near circular ($e<0.1$) orbits while HD 360 b ($e=0.139$), the candidate $\epsilon$ Psc b ($e=0.278$), HD 10975 b ($e=0.442$), HD 79181 b ($e=0.264$), HD 99283 b ($e=0.198$), $\upsilon$ Leo b ($e=0.320$), HD 219139 b ($e=0.110$), and $\gamma$ Psc b ($e=0.204$) orbit their hosts in possible eccentric orbits. 
HD 10975 b refers to a rare case among current observations, having best-fit eccentricity exceeds 0.4. Besides, this planet is the least massive giant planet ever found around a giant star and it has the shortest orbital period among highly eccentric planets around giant stars.
Figure \ref{fig:sma_ecc} shows eccentricity as a function of the semimajor axis. Among planets found orbiting intermediate-mass and solar-mass evolved stars, 38\% were found with eccentricity lower than 0.1, and 85\% were found with eccentricity lower than 0.3. Eccentricity in logarithm shows that orbits of giant planets around G- and K-type giant stars are seldom extremely circular, and 72\% locates between 0.045 and 0.25. Additionally, we adopted a K-S test (\cite{Kolmogorov1933}; \cite{Smirnov1948}) on eccentricities of planets around solar-mass and intermediate-mass evolved stars against the main sequence stars. The test showed a probability of 0 which convinced that they are apparently different distributions. Besides, we found that only three planets: 24 Boo b \citep{Takarada2018}, HD 76920 b \citep{Wittenmyer2017b}, and $\it{Kepler}$-91 (\cite{Lillo-Box2014}; \cite{Sato2015}) were found residing within the periastron distance of 0.2 au, where it includes the most eccentric planet orbiting an evolved star. 

Several scenarios are proposed to explain the origin of the high eccentricities of the exoplanets. One of them is the Kozai-Lidov mechanism (\cite{Kozai1962}; \cite{Lidov1962}), in which a planet can be driven from nearly circular orbit to a highly eccentric orbit due to an inclined, external perturber such as a stellar companion and an outer planetary companion. This seems to be the scenario of two well-known extremely eccentric planets: HD 20782 b (\cite{Jones2006}; \cite{Udry2019}) and HD 80606 b (\cite{Wittenmyer2007}; \cite{Stassun2017}) whose systems contain stellar companions. Since no stellar companion was found around HD 10975 b, yet no RV trend suggesting the existence of outer bodies was found in the residuals, the Kozai-Lidov mechanism should not be the true scenario for this star.

Planet-planet scattering is another possible explanation. 
We assume that HD 10975 b has been scattered into the current highly-eccentric orbit by another unobserved planet in the system. If so, there should be another signal in the star's RV time series.
Yet, except for the signal of HD 10975 b, we did not find any long-term trend or any periodicity in the RV residuals, which means this hypothetical planet does not appear nearby HD 10975 system ``at this moment". In this case, we reserve the scenario that an outer body was ejected out of the system ``a long time ago". 
Then, we assume that the unobserved planet has been ejected away from the system, or it has been scattered into an inner orbit and soon engulfed by its host. 
BD+48 740, a Lithium rich star, is supposed to be the case that the existence of a highly eccentric planet was caused by planet-planet scattering, and consequently, one planetary companion was engulfed by its host star \citep{Adamow2012,Adamow2018}. However, by measuring the Lithium abundance of HD 10975 \citep{Liu2014}, we found an upper limit of $A(\rm{Li}) \leq 0.35$, where no Lithium overabundance happened against other giant stars. This suggests that probably there is not a planet scattered into the host star due to the interaction with HD 10975 b. In brief, although we did not find any evidence, planet-planet scattering is a satisfactory explanation.

\subsection{Detectability of Less-Massive Planets around G- and K-type Giant Stars}
\begin{figure}
 \begin{center}
  \includegraphics[width=8.0cm]{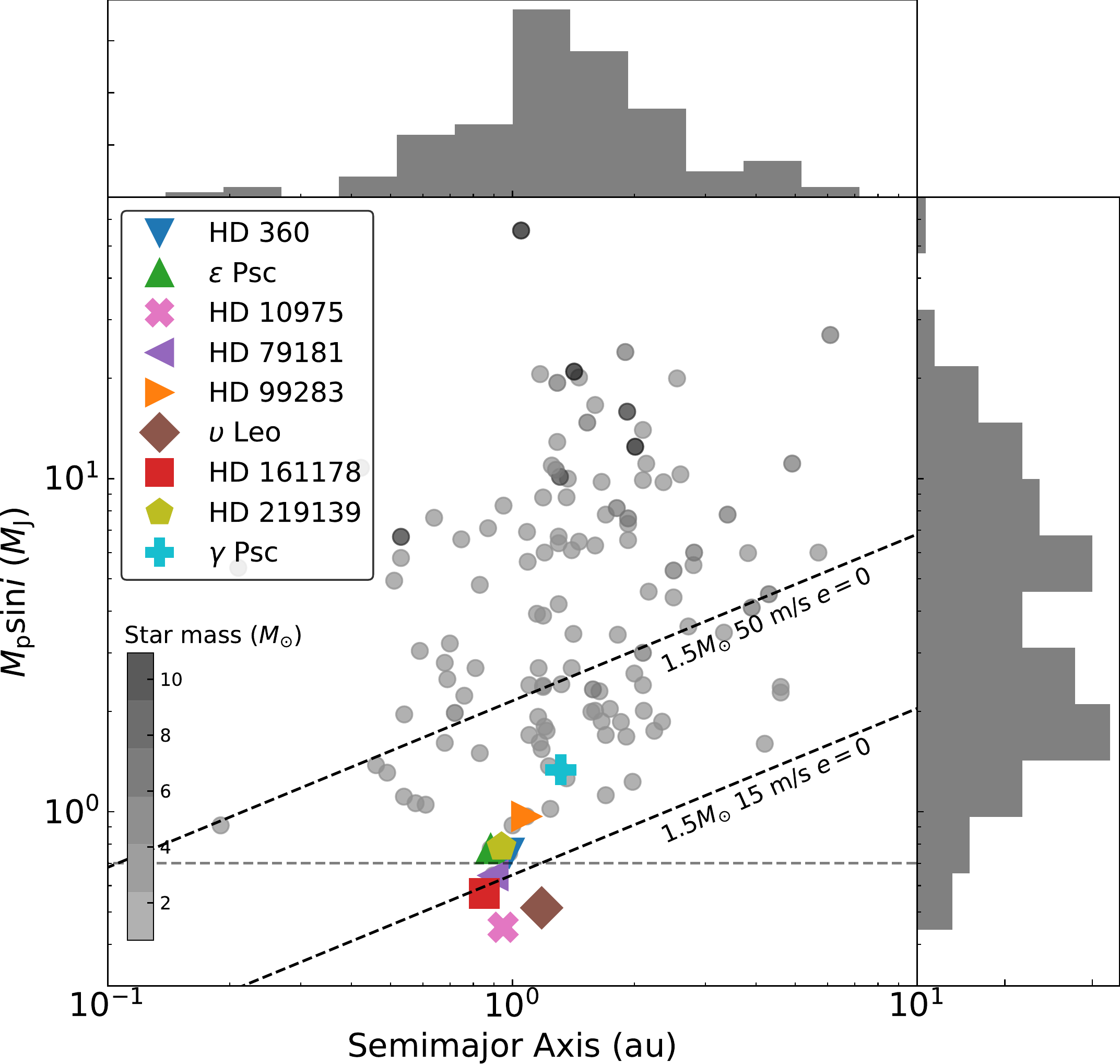} 
 \end{center}
\caption{Minimum mass of planet as a function of semimajor axis. Filled circles are planets around G- and K-type giant stars with their darkness representing the mass of their host stars. The horizontal dashed line represents the minimum mass of $0.7 M_{\rm{J}}$. Two sloped dashed lines respectively show detectabilities of planets around their host stars generating RV semi-amplitude of $15 \rm{m\ s^{-1}}$ and $50 \rm{m\ s^{-1}}$ in circular orbits. (Data are collected on 2019 May 30)}\label{fig:sma_mpsini}
\end{figure}

Among our nine targets, 
HD 360 b ($M_{\rm{p}}\sin{i} = 0.75\  M_{\rm J}$, $a =  0.98\ {\rm au}$), 
the candidate $\epsilon$ Psc b ($M_{\rm{p}}\sin{i} = 0.77\  M_{\rm J}$, $a =  0.87\ {\rm au}$), 
HD 10975 b ($M_{\rm{p}}\sin{i} = 0.45\  M_{\rm J}$, $a =  0.95\ {\rm au}$), 
HD 79181 b ($M_{\rm{p}}\sin{i} = 0.64\  M_{\rm J}$, $a =  0.90\ {\rm au}$),  
$\upsilon$ Leo b ($M_{\rm{p}}\sin{i} = 0.51\  M_{\rm J}$, $a =  1.18\ {\rm au}$),
HD 161178 b ($M_{\rm{p}}\sin{i} = 0.57\  M_{\rm J}$, $a =  0.85\ {\rm au}$), and 
HD 79181 b ($M_{\rm{p}}\sin{i} = 0.78\  M_{\rm J}$, $a =  0.94\ {\rm au}$)
are the least massive planets ($M_{\rm{p}}\sin{i} \lesssim 0.9 M_{\rm J}$) ever discovered around G- and K-type giant stars, including those of both intermediate-mass ($1.5-5 M_{\odot}$) and solar-mass ($0.7-1.5 M_{\odot}$). Before this research, only planetary companions more massive than 0.9 $M_{\rm J}$ were detected around the G- and K-type giant stars: BD+48 738 b ($M_{\rm{p}}\sin{i} = 0.91\  M_{\rm J}$; \cite{Gettel2012}) and 24 Boo b ($M_{\rm{p}}\sin{i} = 0.91\  M_{\rm J}$; \cite{Takarada2018}). Here we did not take transit detected $\it{Kepler}$-91 (\cite{Lillo-Box2014}; \cite{Sato2015}) into account.

Figure \ref{fig:sma_mpsini} shows the minimum mass as a function of semimajor axis. As it is shown in the figure, almost no planet exceeded the detectability of a planet circularly orbiting a star of $1.5\ M_{\odot}$ with a radial velocity semi-amplitude of $15 \ \rm{m\>s^{-1}}$. Pushing the detection limit to a lower level will allow us to directly make a comparison of mass distribution among giant exoplanets between a wider range of distance to the host stars and the mass of their host stars. 

Normally, it is difficult to detect planets with $\lesssim 2 M_{\rm J}$ around G- and K-type giant stars via Doppler measurements because the stellar oscillation raises RV variation to a level of $10-20\ \rm{m\>s^{-1}}$, which means RV signals below $30\ \rm{m\>s^{-1}}$ could be hard to be recognized or detected \citep{Jones2014}. We evaluated the RV jitters induced by stellar oscillation in this work. By using the empirical functions in \citet{Yu2018} and stellar parameters determined in this work, all nine stars have RV amplitudes around or lower than a level of $\sim 10\ \rm{m\>s^{-1}}$. And at the same time, they should be lower than rms of their residuals. \citet{Sato2013} studied the features of RV variations in the G- and K-type giant stars by high cadence observations, showing that solar-like oscillations dominate in these variations with periods of $3-10\ \rm{hr}$. However, unlike dwarf stars, whose p-mode (acoustic pressure mode) oscillation period is around 10 min, the p-mode oscillation of giant stars is hard to be averaged out with longer integration times (e.g., \cite{Mayor2008}, \cite{Dumusque2011}). 

Thus in this study, we further generated simulations in order to explore the detectability of giant planets with various minimum masses around different stars. For each planet mass around each star mass, we generated over 378000 independent RV curves, yet some other conditions, i.e. orbital period, instrumental and stellar jitter, are different.
For the sake of simplicity, we only injected random instrumental RV shifts and RV variation caused by solar-like oscillations to the Keplerian RV curve. 
In each simulation, the instrumental RV shifts were set to be a uniformly distributed random value. The maximum of the instrumental RV shifts in each simulation is also randomly determined with a uniform distribution between 2 and 6 $\rm{m\>s^{-1}}$. We here did not adopt Gaussian noise in instrumental RV shifts, since we have no evidence that the long-term HIDES instrumental RV shift was Gaussian-like according to the instrumental RV stability analysis in Section \ref{sec:ana}.
The jitter caused by solar-like oscillations was set to be Gaussian with boundaries at 3$\sigma$. Empirically, the jitter caused by solar-like oscillations of a G- or K-type giant star is approximately a few to 20 $\rm{m\>s^{-1}}$. We thus set the error of the Gaussian to be a uniformly distributed random number between 6 and 18 $\rm{m\>s^{-1}}$ for each simulation. 
In a single RV curve, we carried out two simulations with time baselines of 10 years (80 data points) and 15 years (120 data points). All input values in the simulations were set empirically according to frequently observed stars in Okayama Planet Search Program. Besides, data points were almost uniformly distributed along the time baseline, and in each week there was at most one observation. Quantified with the K-S test (\cite{Kolmogorov1933}; \cite{Smirnov1948}), we knew our simulated observations were more uniformly distributed than real HIDES observations along the baseline. 
In the simulations, successful recognition of the planet was defined as the case that the strongest signal in GLS indicated the orbital period within an accuracy of 5\% and with a FAP value lower than 0.01\%. 

\begin{figure}
 \begin{center}
  \includegraphics[width=8.0cm]{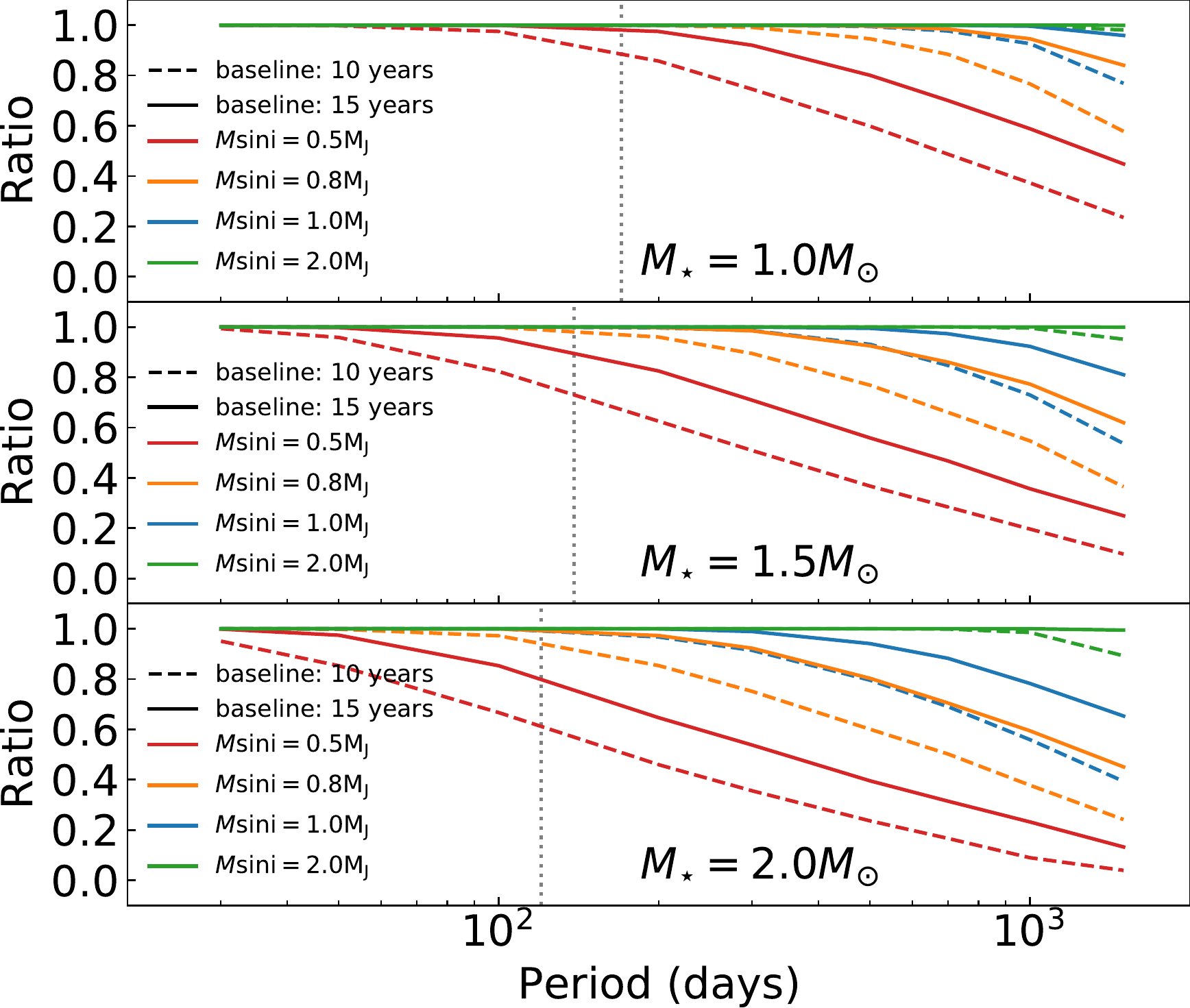} 
 \end{center}
\caption{The ratio of successful detection against simulated orbital periods. From upper panel to lower panel, we respectively simulated stars of $1.0 M_{\odot}$, $1.5 M_{\odot}$, and $2.0 M_{\odot}$. In each panel, planets with minimum masses of $0.5 M_{\rm{J}}$, $0.8 M_{\rm{J}}$, $1.0 M_{\rm{J}}$, and $2.0 M_{\rm{J}}$ are respectively shown in red, orange, blue, and green, and observational baselines of 10 years and 15 years are respectively given in dashed curves and solid curves. The vertical dotted line in each figure refers to 0.6 au, inside which we name it planet desert.}\label{fig:simulation}
\end{figure}
The simulation results are shown in Figure \ref{fig:simulation}. From the figure, it is clear that a planet with minimum mass $2.0 M_{\rm{J}}$ could be easily detected around a star with $2.0 M_{\odot}$ at with orbital period over 1000 days. As a contrast, a planet with a minimum mass of $0.5 M_{\rm{J}}$ orbiting a star with $1.5 M_{\odot}$ at 300 days with 15-year observations has a successful detection probability lower than 0.7. Generally, when the minimum planet mass is higher than $1 M_{\rm{J}}$, it maintains a high successful detection ratio for an orbital period of a few hundred days. Yet, as the minimum planet mass turns to lower than $1 M_{\rm{J}}$, the successful detection ratio sharply decreases. The simulation also legibly tells us planets with shorter orbital periods are much easier to be detected. A planet with a minimum mass of $0.8 M_{\rm{J}}$ orbiting around a star of $1.5 M_{\odot}$ at the orbital period of 800 days only has a successful detection possibility of 0.6 and 0.8 with 10-year and 15-year observational baseline respectively. Here we highlight that, for each simulation, the observational dates are more uniformly distributed than the true HIDES observations along the observational baseline, and on average has more observations on each star. Furthermore, the simulation illustrates that a hypothetical giant planet in the so-called ``planet desert'' can be easily detected. Even if a planet has a minimum mass of $0.5 M_{\rm{J}}$ orbiting a star with $1.5 M_{\odot}$, it can be detected with a probability higher than 0.8 for a 15-year observation. As introduced in Section \ref{sec:intro}, different surveys have targeted thousands of G- and K-type giant stars for many years, yet only few giant planets are detected inside ``planet desert'', whose boundary is around 0.5 au, referring to the orbital period of 100$\sim$200 days. This implies the desert of giant planets around giant stars is like to be true. 
Therefore, in terms of successful detection probability, we consider it is rational that less-massive planets ($M\sin i \lesssim 1 M_{\rm{J}}$) just beyond ``planet desert'' can be more detected, whereas those in wider orbits are not.

As for other studies, \citet{Medina2018} tested simulated p-mode oscillation on subgiant stars, whose p-mode oscillation period is slightly shorter than giant stars and RV variation is around $5-10\ \rm{m\>s^{-1}}$, and they presented possible techniques to detect Neptunian mass planet around subgiant stars. To handle the correlated noise of the star, an efficient method could be Gaussian Process (GP) regression. Different studies (e.g. \cite{Grunblatt2016}; \cite{Grunblatt2017}; \cite{Foreman-Mackey2017}; \cite{Farr2018}) have shown that hidden planets under noise signals, such as oscillation, granulation, or rotation, can be regressed with a Keplerian or transit plus a GP model. 

Several missions e.g. \textit{TESS} and \textit{CHEOPS} are ongoing now and they are opening up a new world to exoplanets. Predicted by \citet{Barclay2018}, 25 \% of the total $\sim 300$ close-in ($P < 85\ \rm days$) planet with radii $R > 4\ R_{\oplus}$ will be detected around evolved intermediate-mass stars in the \textit{TESS} survey. RV follow-ups to these stars with stellar jitter mitigated should provide us more information about planets around these evolved intermediate-mass stars. Even for those non-transiting targets, high cadence photometric observations given by \textit{TESS} will provide us information in asteroseismology and stellar rotation, which will probably help us handle noise in the future RV survey. \\

\subsection{Metallicity of planet-harboring intermediate-mass stars}
\begin{figure}
 \begin{center}
  \includegraphics[width=8.0cm]{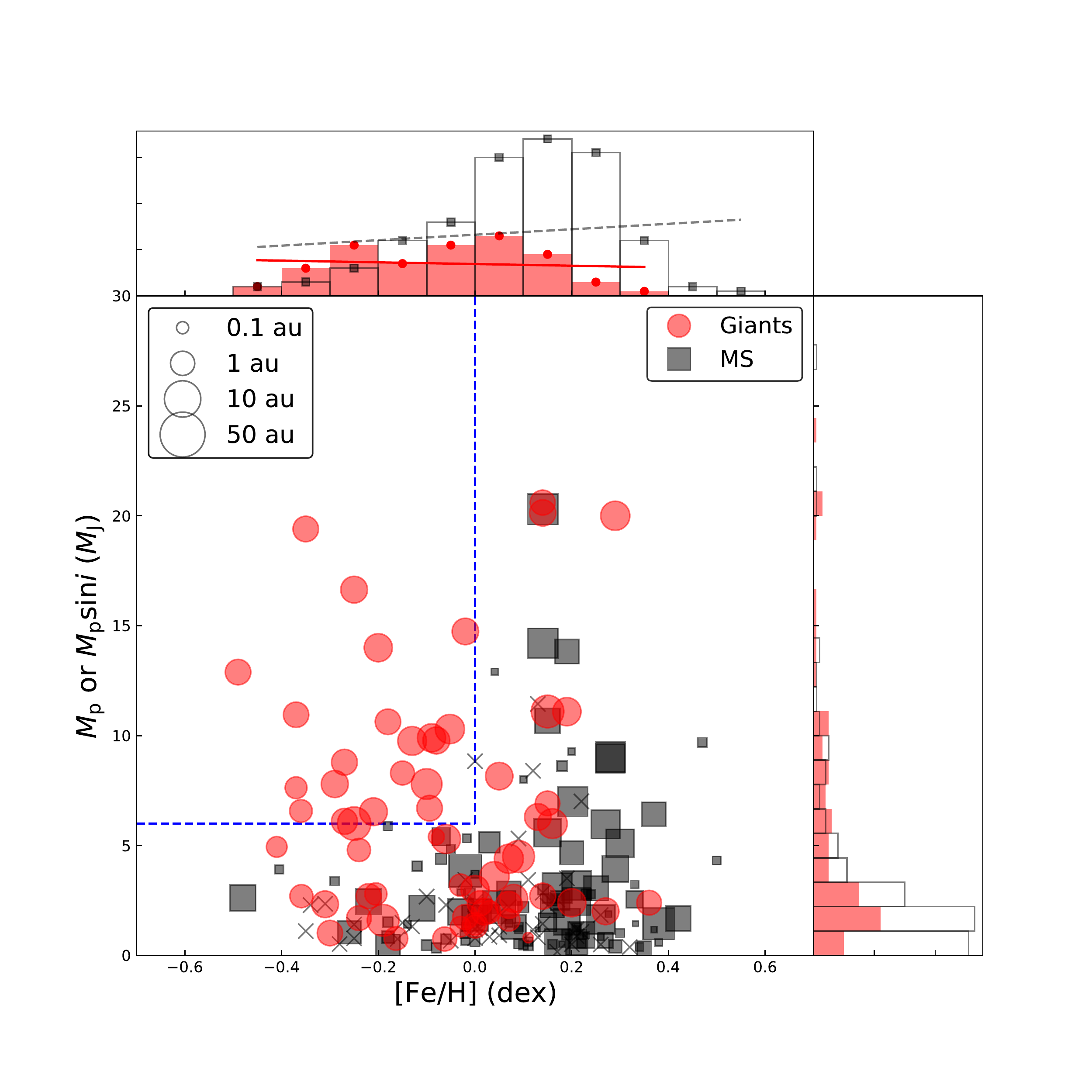} 
 \end{center}
\caption{Mass or minimum mass of giant planets versus metallicity of their host stars with the size of the dots indicating the orbital semimajor axis of the planet. All dots are intermediate-mass planets harboring stars with red circular ones representing giant stars, grey squared ones representing main-sequence stars, and those grey crosses representing main-sequence stars without measurements of orbital semimajor axes of their planets. For both histograms, solid red bars show the counts of giant stars, while the black-framed white bars show the counts of main-sequence stars. Especially, in the horizontal histogram, the solid red line and the black dashed line show the trends of the histograms of giant stars and main-sequence stars respectively. In addition, and only in this figure, we simply define giant stars as stellar radii greater than 5 solar radii. (Data are collected at the end of 2019)}\label{fig:met_mp}
\end{figure}

When saying the occurrence rate of giant planets, many studies showed that they tend to occur around more metal-rich stars including main sequence stars (e.g. \cite{Fischer2005}), and giant stars (e.g. \cite{Reffert2015}). In this research, we mainly focus on the metallicity distribution of intermediate-mass planet-harbouring stars. Figure \ref{fig:met_mp} shows a mass or minimum mass of currently known planets around intermediate-mass stars against the metallicity of these stars. Here we give our discussion under the planetary explanation of the regular RV variations. According to the mass estimation of the prospective planets in this work and the metallicity of their host stars, the nine systems in this work are all located at the most lower left region in this figure. We note that the nine stars are all metal-poor stars, having an average metallicity of -0.25 dex with the richest one of -0.06 dex and the poorest one of -0.62 dex. At present, the most metal-poor planet-harboring intermediate-mass star is BD+20 2457 ($\rm{[Fe/H]}=-1.00$; \cite{Niedzielski2009}), a K-type giant star having two substellar companions. 

In the horizontal histogram of Figure \ref{fig:met_mp}, it is not hard to find a clear increasing trend for the number of detected giant planets around main-sequence stars with the metallicity of their host stars, especially between the range of $\rm{[Fe/H]}=-0.5$ and $\rm{[Fe/H]}= 0.3$. 
In contrast, the distribution seems more uniform for detected giant planets around giant stars, especially in the region between $\rm{[Fe/H]}=-0.4$ and $\rm{[Fe/H]}=0.2$. 
Although the number of detection and occurrence rate does not mean the same, the total detected number seems not as less as it is thought to be. When only saying about the number of detection, the number of planets around metal-poor stars ($\rm [Fe/H]<0$) is slightly larger than the number of those around metal-rich stars ($\rm [Fe/H]>0$). Besides, in the area of $\rm [Fe/H] < 0$ and $M_{\rm{p}}$ or $M_{\rm{p}}\sin i > 6 M_{\rm{J}}$ (upper left region of circled by dashed blue lines), there are almost only giant planets around giant stars at relatively wide orbits ($\gtrsim$ 1 au), and the giant planets in this area contribute nearly half of those around metal-poor giant stars. We notice that our discussion reaches an agreement with an up-to-date research given by \citet{Santos2017}. They indicated that on average, the metallicity of stars with more-massive giant planets ($4 M_{\rm{J}}$) is lower than the metallicity of those with less-massive giant planets (see its definition in section 3 of the related reference), and this is especially for stars more-massive than $1.5 M_{\odot}$.

In the current exploration of giant exoplanets around giant stars, statistically, minimum mass peaks at around $1.5 M_{\rm J}$ and seldom planets were found within mass greater than $3 M_{\rm J}$ and less than $6 M_{\rm J}$ shown by both Figure \ref{fig:sma_mpsini} \& \ref{fig:met_mp}. This gap is close to the distinction between Jupiter-like planets and substellar companions. Moreover, when the distribution is given in logarithm rather than linearly, the valley at around $3.5 M_{\rm J}$ can be more clearly found. This shows a good agreement with the one, a $4 M_{\rm J}$ valley, found by \citet{Santos2017} around whole ``solar-type'' stars. \citet{Santos2017} proposed the boundary as the distinct of two populations of giant planet population, and giant planets on both sides originate from different mechanisms: lighter ones are formed via core-accretion while heavier ones are formed via disk instability, and it could be an explanation to the abundant more massive giant planets around metal-poor stars. \citet{Narang2018} studied candidates at relatively short orbits around M4 type to F0 type main-sequence stars in $Kepler$ DR25 and confirmed this $4 M_{\rm J}$ boundary. 

However, when only taking intermediate-mass main-sequence star into account, the valley seems not to exist from Figure \ref{fig:met_mp}. Here, one possible explanation could be inadequate sampling. Compared to over 800 solar-type giant planet hosts given in \citet{Santos2017}, around 120 intermediate-mass main-sequence stars stands for a minority, which may not correctly reflect the true regimes in mass distribution.  Nonetheless, we can boldly infer that $\sim 4 M_{\rm J}$ valley could be a clue to further investigate planet formation theories. Likewise, it is hard to explain the lack of massive giant planets orbiting metal-poor main-sequence stars in the upper left region in Figure \ref{fig:met_mp}. As these planets are all orbiting stars in their evolved stage, they should have survived when their host stars evolved out of the main-sequence stage. The absence of massive giant planets around metal-poor intermediate-mass stars in their main-sequence stage suggests inadequate sampling could be the best explanation.
In the future, it is necessary to enlarge the samples via different detection methods. Space missions e.g. \textit{TESS} and \textit{CHEOPS} will also offer us new observations to examine today's conjectures. 

\subsection{Summary}
In this research, we have successfully detected regular RV variations in nine G- and K-type giant stars from Okayama Planet Search Program: HD 360, $\epsilon$ Psc, HD 10975, HD 79181, HD 99283, $\upsilon$ Leo, HD 161178, HD 219139, and HD 219615. 
Eight of them are best explained by planets, while $\epsilon$ Psc is classified as a candidate.
The planets will probably end in the engulfment as the stars evolve with time. Among the nine hypothetical planets, HD 360 b is probably accompanied by an outer body indicated by the linear RV long-term trend, and HD 10975 b is a candidate of highly eccentric planets, yet we are not confident about its mechanism. 
Besides, seven of nine planets (including one candidate) are the least massive planets ($M_{\rm{p}}\sin i \lesssim 0.9 M_{\rm{J}}$) discovered around evolved giant stars, which pushing down the detectability of planets around giant stars with RV method. 
In addition, these nine giant stars are all metal-poor stars. By further investigating the metallicity, we confirmed the abundant more-massive giant planets around metal-poor stars and the mass valley around $4 M_{\rm{J}}$, which could be interpreted as the distinct boundary of different giant planet formation mechanisms. 
As a coincidence of variability appears at around 280 days among five stars, we do not fully reject some of periodic RV variations intrinsically come from stars themselves, e.g. LSP or another new type of stellar variation.

%%%%%%%%%%%%%%%%%%%%%%%%%%%%%%%%%%%%%%%

\begin{ack}
This research is based on data collected at the Okayama Astrophysical Observatory (OAO), which is operated by the National Astronomical Observatory of Japan. 
We are grateful to all the staff members of OAO for their support during the observations.
Data of $\tau$ Cet and $\epsilon$ Vir were partially obtained during an engineering time at OAO. We thank the observatory for allowing us to use the data obtained during that time.
We thank students of Tokyo Institute of Technology and Kobe University for their kind help with the observations at OAO. 
B.S. was partially supported by MEXT's program ``Promotion of Environmental Improvement for Independence of Young Researchers" under the Special Coordination Funds for Promoting Science and Technology, and by Grant-in-Aid for Young Scientists (B) 17740106 and 20740101 and Grant-in-Aid for Scientific Research (C) 23540263 from the Japan Society for the Promotion of Science (JSPS). 
H.I. is supported by Grant-in-Aid for Scientific Research (A) 23244038 from JSPS. 
This research has made use of the SIMBAD database, operated at CDS, Strasbourg, France.
This work has made use of data from the European Space Agency (ESA) mission
{\it Gaia} (\url{https://www.cosmos.esa.int/gaia}), processed by the {\it Gaia}
Data Processing and Analysis Consortium (DPAC,
\url{https://www.cosmos.esa.int/web/gaia/dpac/consortium}). Funding for the DPAC
has been provided by national institutions, in particular the institutions
participating in the {\it Gaia} Multilateral Agreement.
This research has made use of the NASA Exoplanet Archive,
which is operated by the California Institute of Technology,
under contract with the National Aeronautics and Space Administration under the Exoplanet Exploration Program.

\end{ack}

\appendix
\section{Effect of IP variability on line profile analysis}\label{sec:bisip}
The spectral line deformation can be either caused by instrumental variation or stellar surface modulation. The rotational velocities of stars in this work are comparable to or lower than the velocity resolution of HIDES, therefore the BIS of CCF (BIS$_{\rm{CCF}}$) might be affected by IP variations. In such a case, we examine the BIS of the mean IP (BIS$_{\rm{CCF}}$) which is derived from RV analysis. The IP is estimated within a different wavelength region as bisector analysis, still, we can expect they share the same variability. To calculate BIS$_{\rm{CCF}}$, we convert pixels to wavelength in each segment as IP is described as a function of pixels. The center of the main Gaussian envelope is set to be the center of the segment. The effect of IP variation on the line profile was evaluated by \citep{Takarada2018} using chromospherically inactive G-type giant star $\epsilon$ Vir. In their work, they reported that the slit mode spectra of $\epsilon$ Vir showed a larger BIS$_{\rm{CCF}}$ scatter in a wider velocity range. They did not find a correlation between BIS$_{\rm{CCF}}$ and RV, while they found a strong correlation between BIS$_{\rm{CCF}}$ and BIS$_{\rm{IP}}$ for both slit mode spectra and fiber mode spectra. Therefore, considering the RV variation e.g. solar-like oscillation and rotational variations, the BIS$_{\rm{CCF}}$ of $\epsilon$ Vir should reflect the extent of line profile deformation caused by IP variation. Here we remind readers that only data acquired after the CCD extension in December 2007 are capable for BIS$_{\rm{CCF}}$ analysis.

\begin{figure*}
 \begin{center}
  \includegraphics[width=15.8cm]{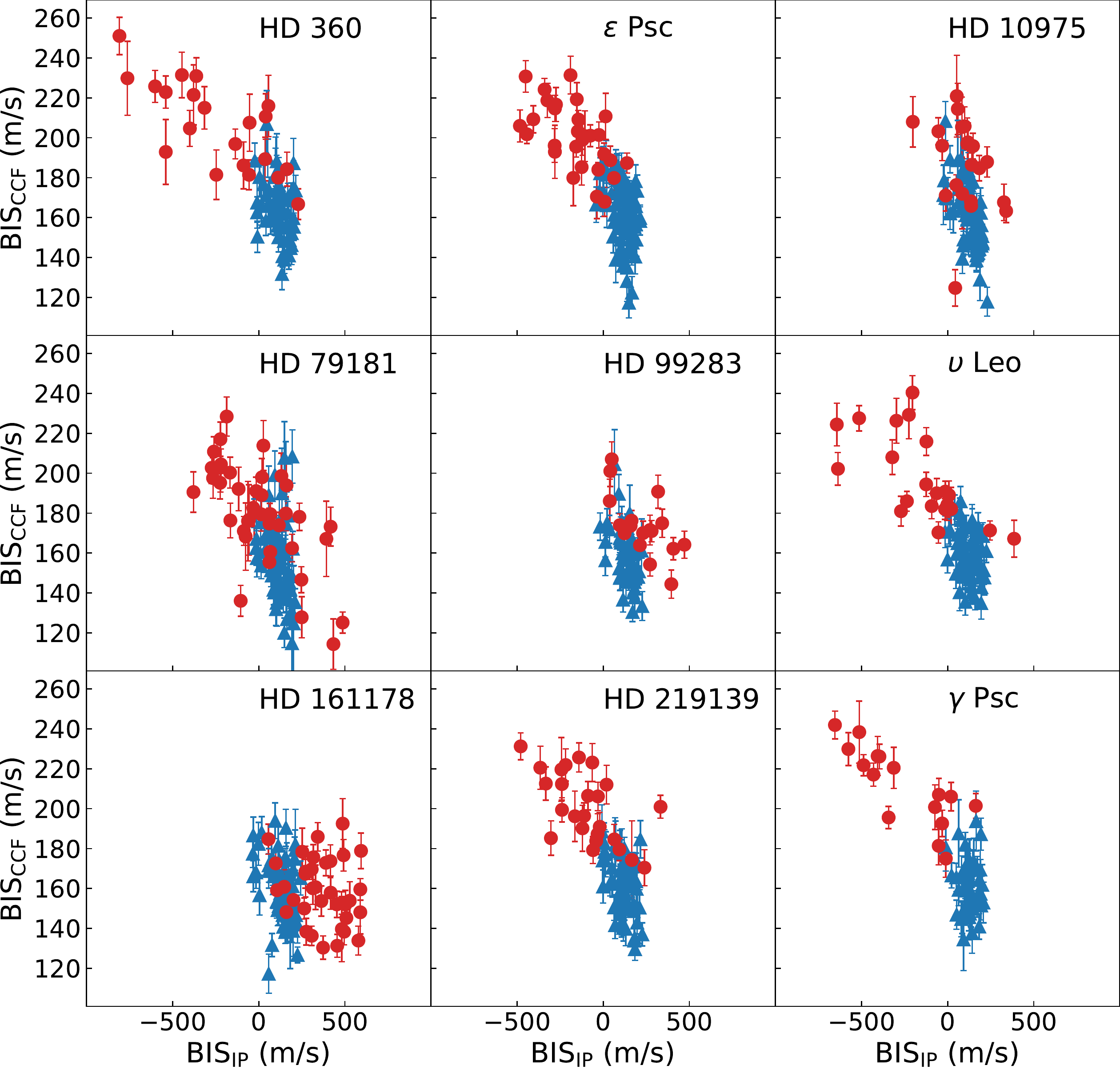} 
 \end{center}
\caption{The relation between BIS$_{\rm{CCF}}$ and BIS$_{\rm{IP}}$. Red points are obtained from HIDES-S and blue points are obtained from HIDES-F.}\label{fig:bis_ip_ccf}
\end{figure*}

\begin{figure}
 \begin{center}
  \includegraphics[width=7.8cm]{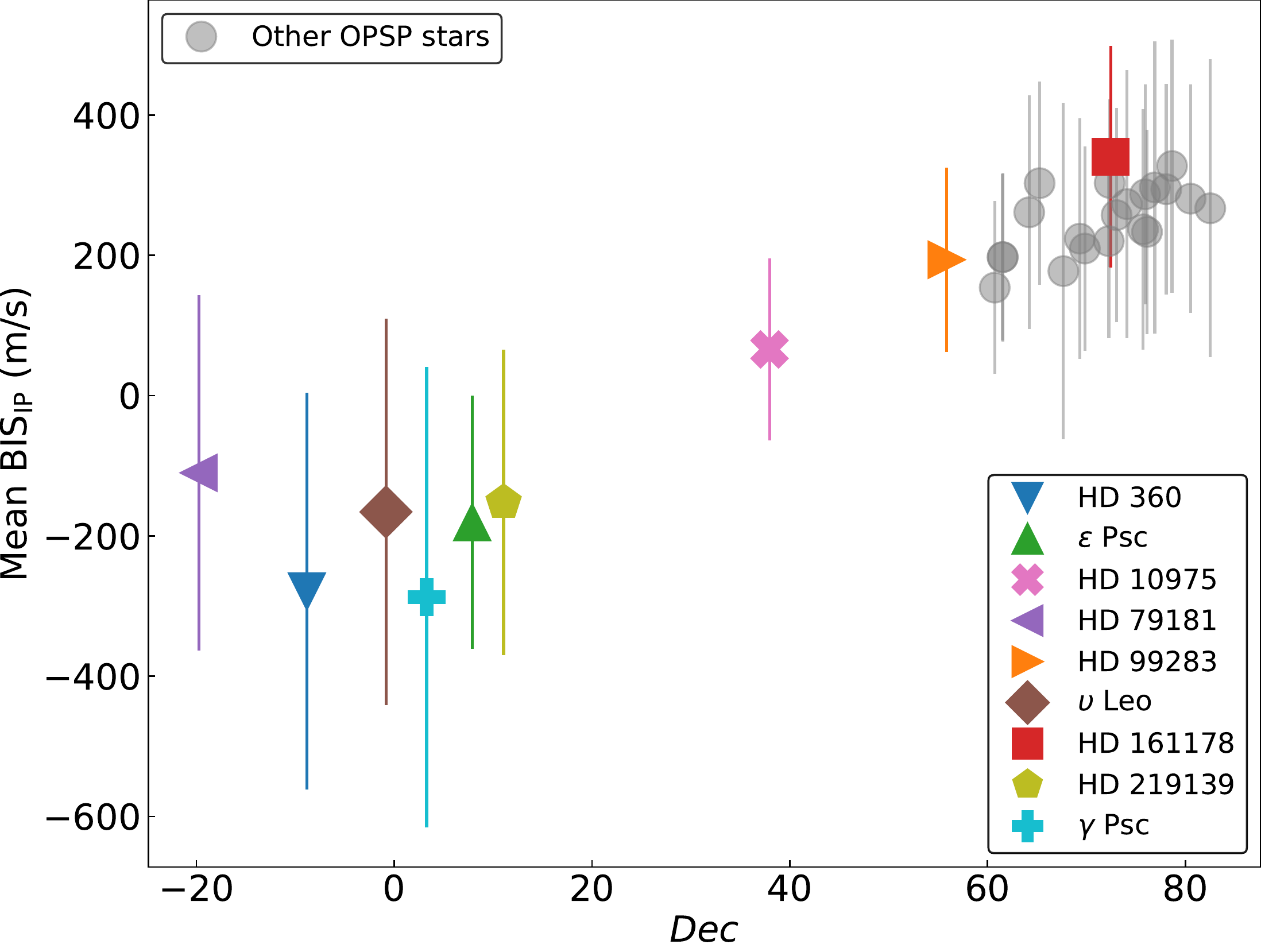} 
 \end{center}
\caption{Mean BIS$_{\rm{IP}}$ measured from slit mode spectra against stars' declination. In the figure, the colored dots in different shapes refer to nine stars in this study, and the circular grey dots refer to other stars with $Dec >$ \timeform{+60D} in Okayama Planet Search Program. The errorbars indicate the rms scatters of BIS$_{\rm{IP}}$.}\label{fig:bis_ip_dec}
\end{figure}

In this work, BIS$_{\rm{IP}}$ of all nine stars in this work are calculated, and the relation between BIS$_{\rm{CCF}}$ and BIS$_{\rm{IP}}$ is shown in Figure \ref{fig:bis_ip_ccf}. 
Generally, BIS$_{\rm{IP}}$ of fiber mode spectra show higher stability with mean BIS$_{\rm{IP}}$ around $100.0 \rm{m\ s^{-1}}$ and lower rms scatter.
In most cases, including HD 360, $\epsilon$ Psc, HD 79181, $\upsilon$ Leo, HD 219139, and $\gamma$ Psc, BIS$_{\rm{CCF}}$ of slit mode spectra are generally correlated with BIS$_{\rm{IP}}$ (correlation coefficient $|r| \gtrsim 0.6$), and the BIS$_{\rm{CCF}}$ of slit mode spectra are on average greater than BIS$_{\rm{CCF}}$ of fiber mode spectra. 
Therefore, the IP variability is a more dominant factor than line profile in the BIS$_{\rm{CCF}}$ variations, and slit mode spectra suffer from IP variation to a greater extent. 
While in the case of HD 161178, we do not find strong correlation between BIS$_{\rm{CCF}}$ and BIS$_{\rm{IP}}$ (correlation coefficient $|r| < 0.23$). 
Furthermore, the rms of BIS$_{\rm{CCF}}$ for both observation modes are almost the same ($16.2 \rm{m\ s^{-1}}$ and $15.0 \rm{m\ s^{-1}}$ for slit mode and fiber mode respectively). This suggests that stellar modulation is the dominant factor in BIS$_{\rm{CCF}}$ rather than BIS$_{\rm{IP}}$ for HD 161178. 
Also, we notice that, for HD 161178, the BIS$_{\rm{IP}}$ of slit mode spectra is on average larger and more positive than BIS$_{\rm{IP}}$ of fiber mode spectra compared to other stars. 
As the star is located very close to the north celestial pole on the sky ($RA$ \timeform{17h37m08.88s}, $Dec$ \timeform{+72D27'20.86''}), we infer it is related to the deformation of the stellar image dependent on the direction to which the telescope points. Therefore, we calculate the mean BIS$_{\rm{IP}}$ of slit mode spectra of our nine stars and the mean BIS$_{\rm{IP}}$ of slit mode spectra of other 20 stars with $Dec >$ \timeform{+60D} in Okayama Planet Search Program. Consequently, stars with higher declination tend to have positive and larger BIS$_{\rm{IP}}$ values compared to stars located around the celestial equator (Figure \ref{fig:bis_ip_dec}), which verifies our inference of origin of the stellar image deformation.
In the rest cases, including, HD 10975 and HD 99283, the correlation may not be convincing enough since there are fewer slit mode observations that we can simultaneously obtain both BIS$_{\rm{CCF}}$ and BIS$_{\rm{IP}}$. 

\begin{figure*}
 \begin{center}
  \includegraphics[width=16.0cm]{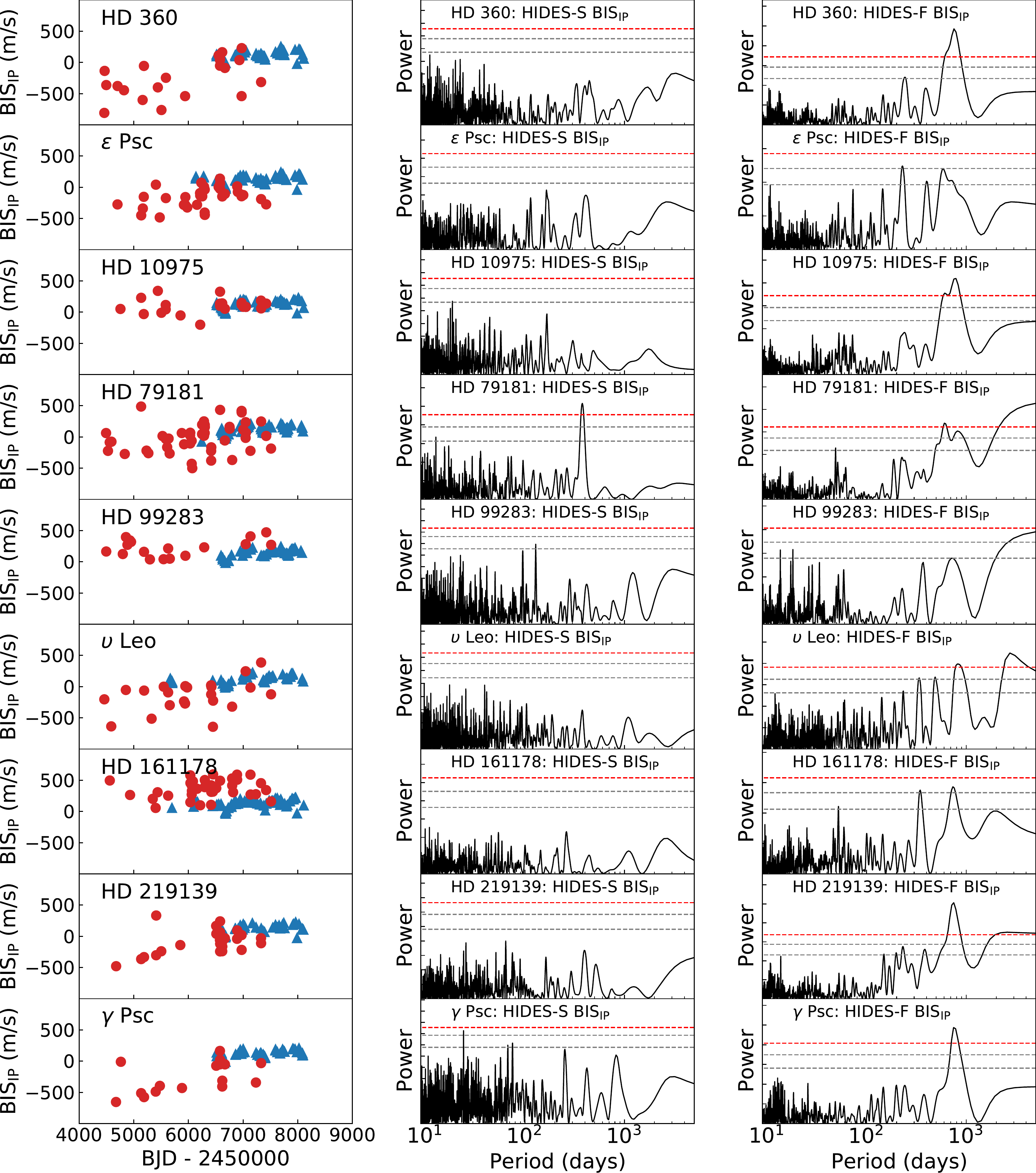} 
 \end{center}
\caption{Left column: BIS$_{\rm{IP}}$ variation with time after Dec. 2007, where red dots are obtained from slit mode spectra (HIDES-S) and blue dots are obtained from fiber mode spectra (HIDES-F). Middle column: GLS of BIS$_{\rm{IP}}$ obtained from HIDES-S. Right column, GLS of BIS$_{\rm{IP}}$ obtained from HIDES-F.}\label{fig:bis_ip_gls}
\end{figure*}

To investigate the IP variation with time, we performed GLS on BIS$_{\rm{IP}}$ (Figure \ref{fig:bis_ip_gls}). For BIS$_{\rm{IP}}$ of fiber mode spectra, the periodograms show signals at around 600 to 800 days for all the stars. In the case of $\epsilon$ Psc, a signal around 400 days might be an explanation for the FWHM regular variation mentioned in Section \ref{sec:res}.
This implies the regular IP variation of fiber mode spectra. For BIS$_{\rm{IP}}$ of slit mode spectra, we can by-eye find a general acceleration trend. From periodograms, the IP variation of slit mode spectra seems not to be as regular as the IP variation of fiber mode spectra. However, it might be on account of fewer observation numbers taken by HIDES-S. 
Peculiarly in the case of HD 79181, the periodogram shows a significant signal at around 377 days which is close to 1 year, yet other stars do not show similar variability. As it is more frequently observed than other stars, we infer it might be the true annually instrumental variation. 

\section{Instrumental RV stability}\label{sec:inststable}
\begin{figure*}
 \begin{center}
  \includegraphics[width=16cm]{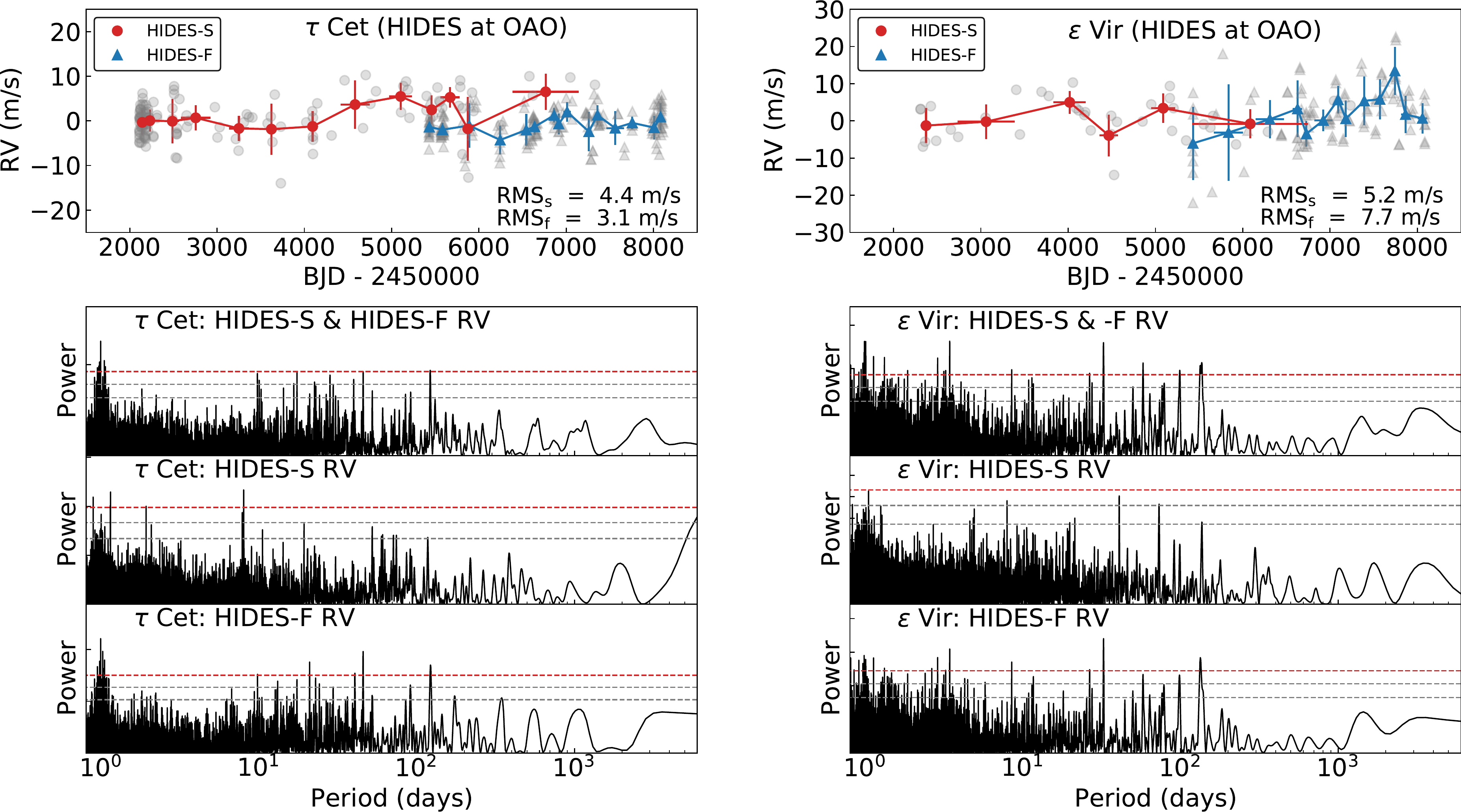} 
 \end{center}
\caption{RV stability of HIDES. The upper panels show the observations to two RV standard stars: $\tau$ Cet and $\epsilon$ Vir. Color dots are binned observations while grey dots in the backgrounds are original observations. Red circular dots indicates HIDES-S observations while blue triangle dots indicates HIDES-F observations.
The lower panels show the Generalized Lomb-Scargle (GLS) periodogram of the original RV time series. In GLS periodograms, horizontal lines represent 10\%, 1\%, and 0.1\% FAP level from bottom to top respectively. }\label{fig:inst}
\end{figure*}
Two RV standard stars were adopted to examine the instrumental stability in RV measurements for this study (Figure \ref{fig:inst}). $\tau$ Cet (HD 10700, G8 V, 0.8$M_{\odot}$), known to have stable RV among solar-type stars, was set to be one of the RV standard stars in this work. For both observation modes of HIDES, the typical exposure time was set to be 5min. The root mean square of HIDES-S was $4.4\ \rm{m\>s^{-1}}$, and the root mean square of HIDES-F was $3.2\ \rm{m\>s^{-1}}$. $\epsilon$ Vir (HD 113226, G8 III, 2.5$M_{\odot}$), one of the most RV-stable Late-G giant stars among targets in the Okayama Planet Search Program \citep{Sato2005}, was set as a comparison RV standard star. Its stellar mass was re-estimated in the same way as nine target stars in this study. For both observation modes of HIDES, the typical exposure time was set to be 5min. The root mean square of HIDES-S was $5.2\ \rm{m\>s^{-1}}$. As for the root mean square of HIDES-F, we obtained $7.7\ \rm{m\>s^{-1}}$ for all exposures, and $6.7\ \rm{m\>s^{-1}}$ for exposures after 2012. Although on average the scatter is larger than $\tau$ Cet, it is likely due to the solar-like oscillation. In order to investigate the long-term variation of the instrument, we firstly binned neighboring observations, and each bin was set to be three months at least. If there were less than five observations in a single bin, we combined the next neighboring bin until there were at least five observations in the bin. We calculated the average of time and RV of observations in each bin, and by our educated eyes, we found slight RV variation with time for both stars. Comparing both RV standard stars, $\tau$ Cet showed betters RV stability than $\epsilon$ Vir. In order to have quantified measurements on long-term variations, we checked periodicity (method: Section \ref{sec:period}) in the time series of both two RV standard stars. Consequently, we did not find regular variations over 150 days. Therefore, we could rule out regular instrumental variations longer than this period. 

%%%%%%%%%%%%%%%%%%%%%%%%%%%%%%%%%%%%%%%%%%%%%%%%%%%%%%%%%%%%%%%%%%%%%%%%%%%%%%%%%%%%%%%%%%%%%%%%%%%%%%%%%%%%%%%%%%%%%%%%
\section{Individual stars}\label{sec:stars}

%\newpage
%%%%%%%%%%%%%%%%%%%%%%%%%%%%%%%%%%%%%%%%%%%%%%%%%%%%%%%%%%%%%%%%%%%%%%%%%%%%%%%%%%%%%%%%%%%%%%%%%%%%%%%%%%%%%%%%%%%%%%%%
\subsection{HD 360}
\begin{table}
\tbl{Radial Velocities of HD 360 }{%
\begin{tabular}{lccc}
\hline\hline
BJD & Radial Velocity & Uncertainty & Observation \\
$(-2450000)$ & ($\rm{m\ s^{-1}}$) & ($\rm{m\ s^{-1}}$) & Mode \\
\hline
$3028.8938$ & $-67.4$ & $4.5$ & HIDES-S  \\ 
$3285.1626$ & $-67.8$ & $3.9$ & HIDES-S  \\ 
$3362.9977$ & $-23.3$ & $4.3$ & HIDES-S  \\ 
$3401.9169$ & $-11.0$ & $6.2$ & HIDES-S  \\ 
$3579.2258$ & $-80.5$ & $4.0$ & HIDES-S  \\ 
... &... &... &... \\
\hline
\end{tabular}}
\begin{tabnote}
\hangindent6pt\noindent
\hbox to6pt{\footnotemark[$*$]\hss}\unskip% 
Only the first five sets of RV data are listed. A complete data listing including RV, FWHM, BIS, and S/N will be available online as supplementary after the publication.
\end{tabnote}
\label{RV:HD360}
\end{table}

\begin{figure*}
 \begin{center}
 \includegraphics[width=16.0cm]{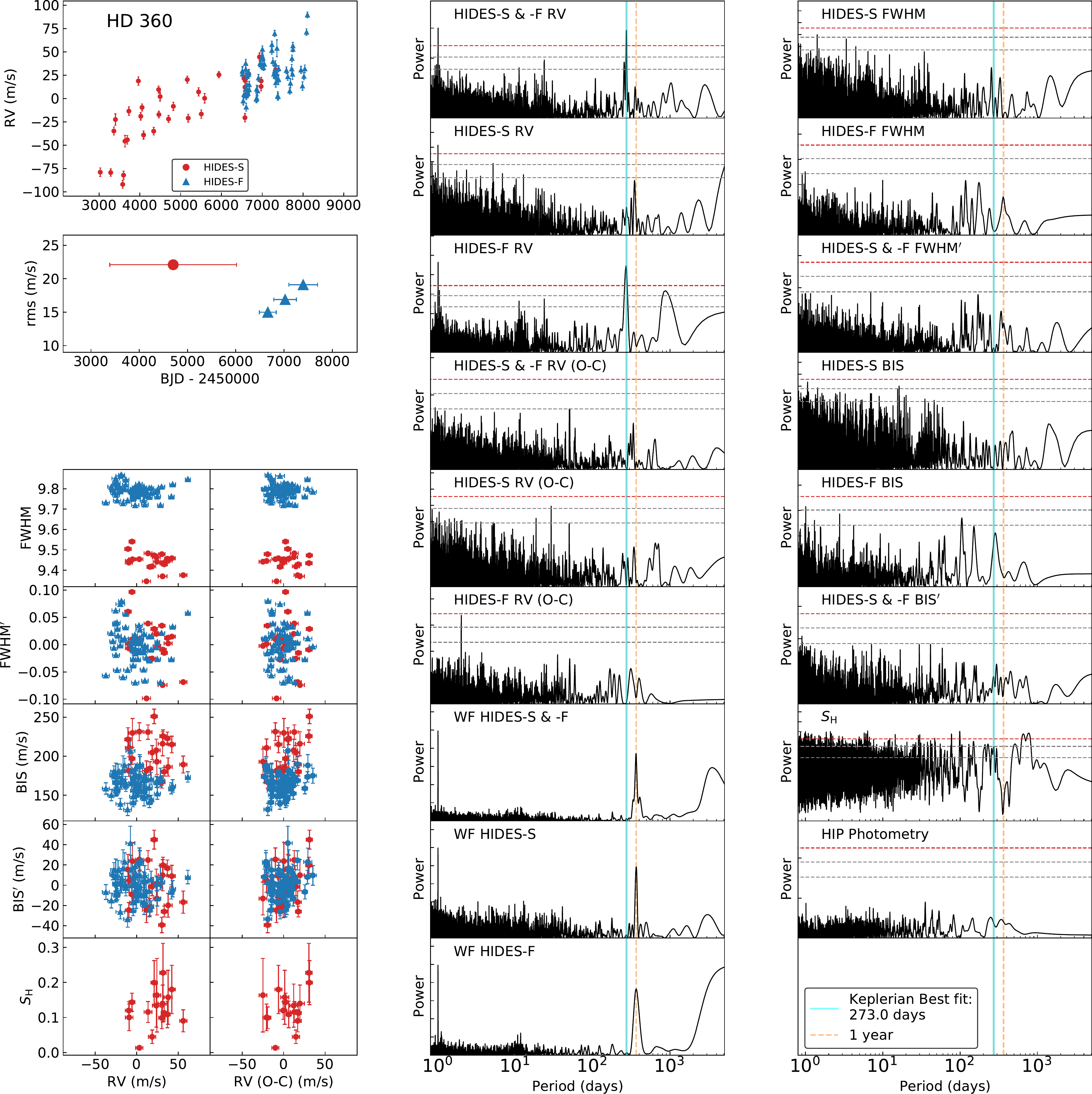}
 \end{center}
\caption{
Summary figure of HD 360. 
Left column from top to bottom: 
RV time series; rms of RVs varying by time; FWHM, FWHM$^{\prime}$, BIS, BIS$^{\prime}$, and Ca \emissiontype{II} H index $S_{\rm{H}}$ respectively against RVs and RV residuals.
Middle column from top to bottom:
GLS periodogram of full RV data, HIDES-S RV data, HIDES-F RV data, full RV residuals, HIDES-S RV residuals, and HIDES-F RV residuals; window function of full RV data, HIDES-S data, and HIDES-F data.
Right column from top to bottom:
GLS periodogram of HIDES-S FWHM, HIDES-F FWHM, FWHM$^{\prime}$, HIDES-S BIS, HIDES-F BIS, BIS$^{\prime}$, Ca \emissiontype{II} H index $S_{\rm{H}}$, and Hipparcos photometry. In GLS periodograms, the horizontal lines represent 10\%, 1\%, and 0.1\% FAP levels from bottom to top. The vertical cyan solid line indicates the best-fitted period from the Keplerian model, and the vertical orange dashed line indicates 1 year. 
}\label{fig:HD360_rv_all}
\end{figure*}

\begin{figure}
 \begin{center}
  \includegraphics[width=8.0cm]{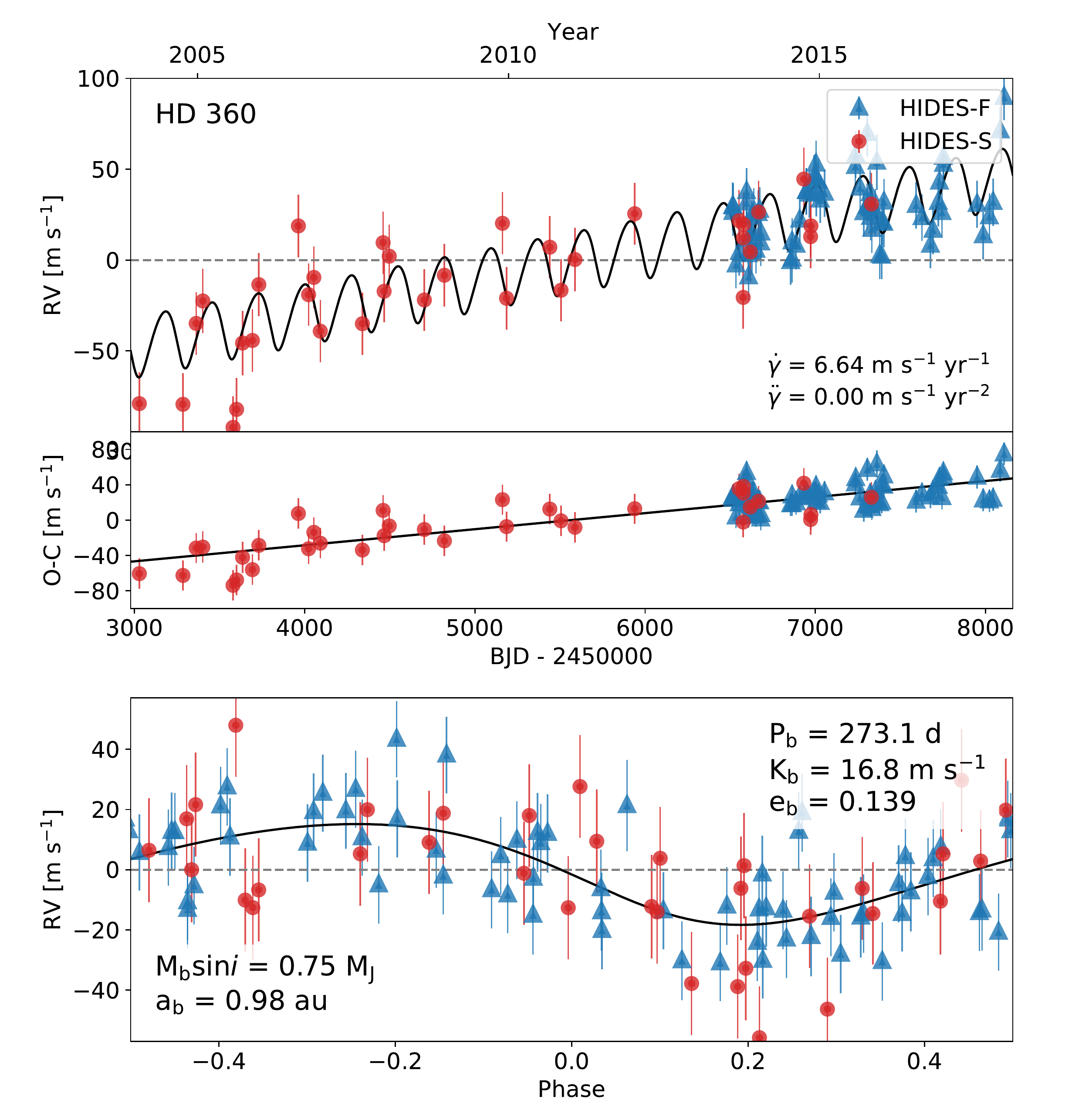} 
 \end{center}
\caption{Orbital solution of HD 360. HIDES-S data are shown in red and HIDES-F data are shown in blue Top: Best fit single Keplerian curve in the full time span with fitted RV offsets between instruments and fitted jitters included in the errorbars. Mid: RV Residuals to the best fit. Bottom: phase-folded of Top panel.}\label{fig:HD360_curve}
\end{figure}

We collected a total of 102 data including 35 taken by HIDES-S and 67 taken by HIDES-F between 2004 January and 2017 December. The RV data are shown in Figure \ref{fig:HD360_rv_all} and the data are listed in Table \ref{RV:HD360}. For the full RV data, the GLS periodogram shows one significant peak at 273 days with FAP lower than $0.1 \%$. We did not find significant periodicity with GLS in spectral line profiles or Ca \emissiontype{II} H index time series, and we did not find RV strongly correlating with either spectral line profiles or Ca \emissiontype{II} H index ($|r| < 0.4$, where $r$ is the Pearson's correlation coefficient\footnote{Since we only lay emphasis on whether two sets of data correlate with each other or not, but not how they correlate with each other, we here use the absolute value of correlation coefficient $r$ to illustrate the correlation in this work.}, and $r$ hereafter is likewise defined.) We also calculated a periodogram for Hipparcos photometry found no significant periodicity.

We adopted a single Keplerian curve to fit the data and obtained orbital parameters for the companion of $P = 273.1_{-0.8}^{+1.6}\ {\rm days}$, $K =16.8_{-2.6}^{+1.8}\ \rm{m\>s^{-1}}$, $e = 0.139_{-0.104}^{+0.108}$. The rms scatter of the residuals to the Keplerian fit is $14.8\ \rm{m\>s^{-1}}$. We did not find any significant periodicity in the residuals, and we did not find line profiles or Ca \emissiontype{II} H index strongly correlating with RV residuals ($|r| < 0.4$). Adopting a stellar mass of $M = 1.69\ M_{\odot}$, we obtained a minimum mass $M_{\rm{p}}\sin{i} = 0.75_{-0.15}^{+0.12}\  M_{\rm J}$ and a semimajor axis $a = 0.98_{-0.03}^{+0.11}\ {\rm au}$ for the companion. The phase-folded RV curve is shown in the lower panel in Figure \ref{fig:HD360_curve}.

As for HD 360, it shows a significant long-term linear trend, which strongly suggests an outer companion surviving around the host star. 
The acceleration, which means the slope of the RV trend, can be used for estimate the minimum of dynamical mass of the outer body \citep{Winn2009}. By adopting the acceleration of $\dot{\gamma} = 0.018_{-0.002}^{+0.002}\ \rm{m\>s^{-1}\>d^{-1}}$, we obtained:
\begin{equation}
\label{eq:dvdt}
\frac{M_{\rm c}\sin i_{\rm c}}{a_{\rm c}^{2}} \sim \frac{\dot{\gamma}}{G} = 0.037_{-0.004}^{+0.004}\  M_{\rm J}\ \rm{au}^{-2}
\end{equation}
where $\dot{\gamma}$ is the RV acceleration, $M_{\rm c}$ is the mass of the outer companion, $i_{\rm c}$ is the orbital inclination, $G$ is the gravitational constant, and $a_{\rm c}$ is the semimajor axis of the outer companion. 
%%%%%%%%%%%%%%%%%%%%%%%%%%%%%%%%%%%%%%%%%%%%%%%%%%%%%%%%%%%%%%%%%%%%%%%%%%%%%%%%%%%%%%%%%%%%%%%%%%%%%%%%%%%%%%%%%%%%%%%%

%\newpage
%%%%%%%%%%%%%%%%%%%%%%%%%%%%%%%%%%%%%%%%%%%%%%%%%%%%%%%%%%%%%%%%%%%%%%%%%%%%%%%%%%%%%%%%%%%%%%%%%%%%%%%%%%%%%%%%%%%%%%%%
\subsection{$\epsilon$ Psc}
\begin{table}
\tbl{Radial Velocities of $\epsilon$ Psc }{%
\begin{tabular}{lccc}
\hline\hline
BJD & Radial Velocity & Uncertainty & Observation \\
$(-2450000)$ & ($\rm{m\ s^{-1}}$) & ($\rm{m\ s^{-1}}$) & Mode \\
\hline
$2311.9252$ & $-22.9$ & $4.2$ & HIDES-S  \\ 
$2488.2308$ & $10.5$ & $4.4$ & HIDES-S  \\ 
$2507.2159$ & $-10.1$ & $4.3$ & HIDES-S  \\ 
$2542.0624$ & $-16.7$ & $3.7$ & HIDES-S  \\ 
$2652.9970$ & $-12.0$ & $4.6$ & HIDES-S  \\ 
... & ... & ... & ... \\
\hline
\end{tabular}}
\begin{tabnote}
\hangindent6pt\noindent
\hbox to6pt{\footnotemark[$*$]\hss}\unskip% 
Only the first five sets of RV data are listed. A complete data listing including RV, FWHM, BIS, and S/N will be available online as supplementary after the publication.
\end{tabnote}
\label{RV:HD6186}
\end{table}

\begin{figure*}
 \begin{center}
 \includegraphics[width=16.0cm]{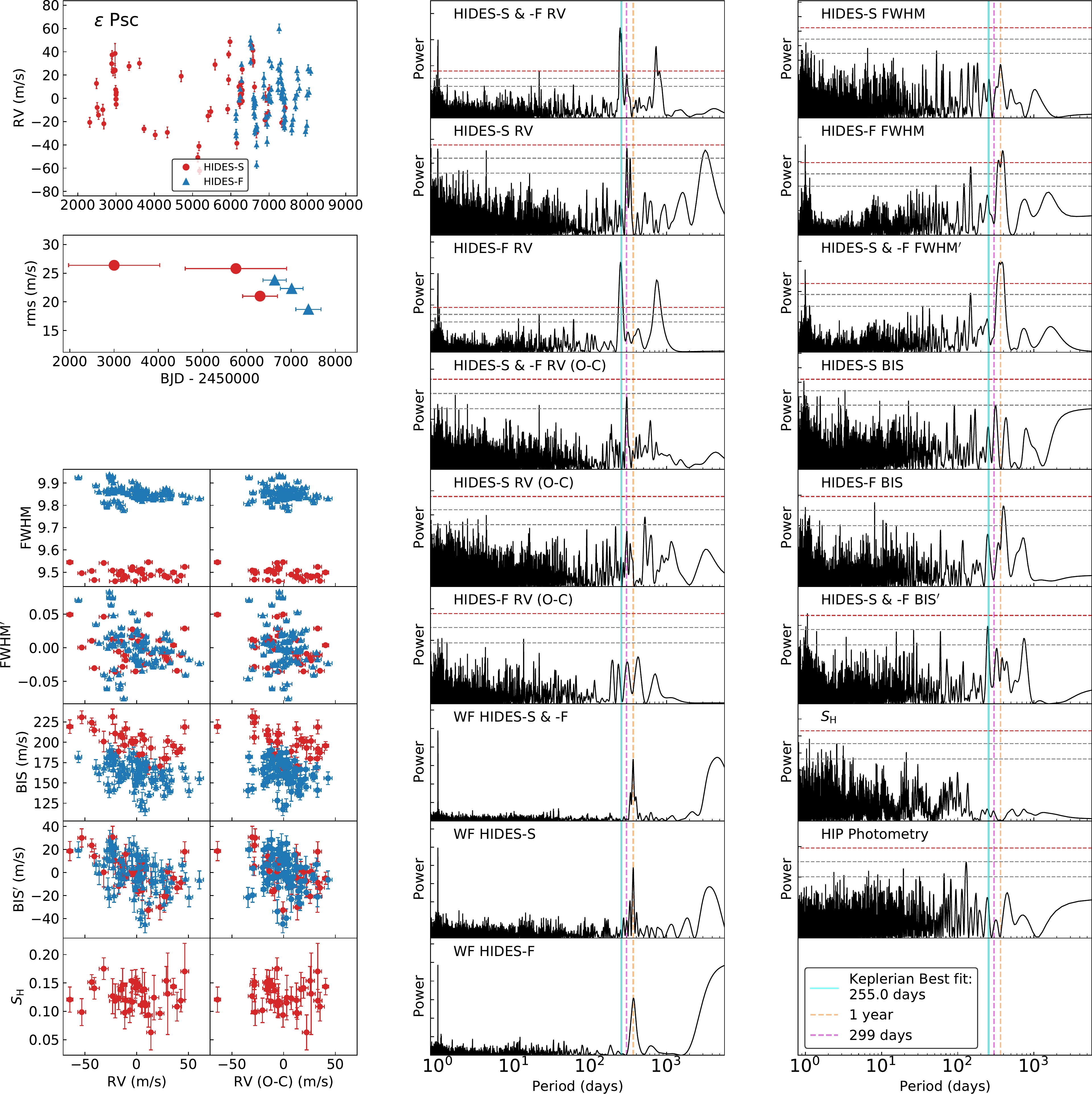}
 \end{center}
\caption{
Summary figure of $\epsilon$ Psc. 
Left column from top to bottom: 
RV time series; rms of RVs varying by time; FWHM, FWHM$^{\prime}$, BIS, BIS$^{\prime}$, and Ca \emissiontype{II} H index $S_{\rm{H}}$ respectively against RVs and RV residuals.
Middle column from top to bottom:
GLS periodogram of full RV data, HIDES-S RV data, HIDES-F RV data, full RV residuals, HIDES-S RV residuals, and HIDES-F RV residuals; window function of full RV data, HIDES-S data, and HIDES-F data.
Right column from top to bottom:
GLS periodogram of HIDES-S FWHM, HIDES-F FWHM, FWHM$^{\prime}$, HIDES-S BIS, HIDES-F BIS, BIS$^{\prime}$, Ca \emissiontype{II} H index $S_{\rm{H}}$, and Hipparcos photometry. In GLS periodograms, the horizontal lines represent 10\%, 1\%, and 0.1\% FAP levels from bottom to top. The vertical cyan solid line indicates the best-fitted period from the Keplerian model, the vertical orange dashed line indicates 1 year, and the vertical pink dashed line indicates the interested 299-day period. 
}\label{fig:HD6186_rv_all}
\end{figure*}

\begin{figure}
 \begin{center}
  \includegraphics[width=8.0cm]{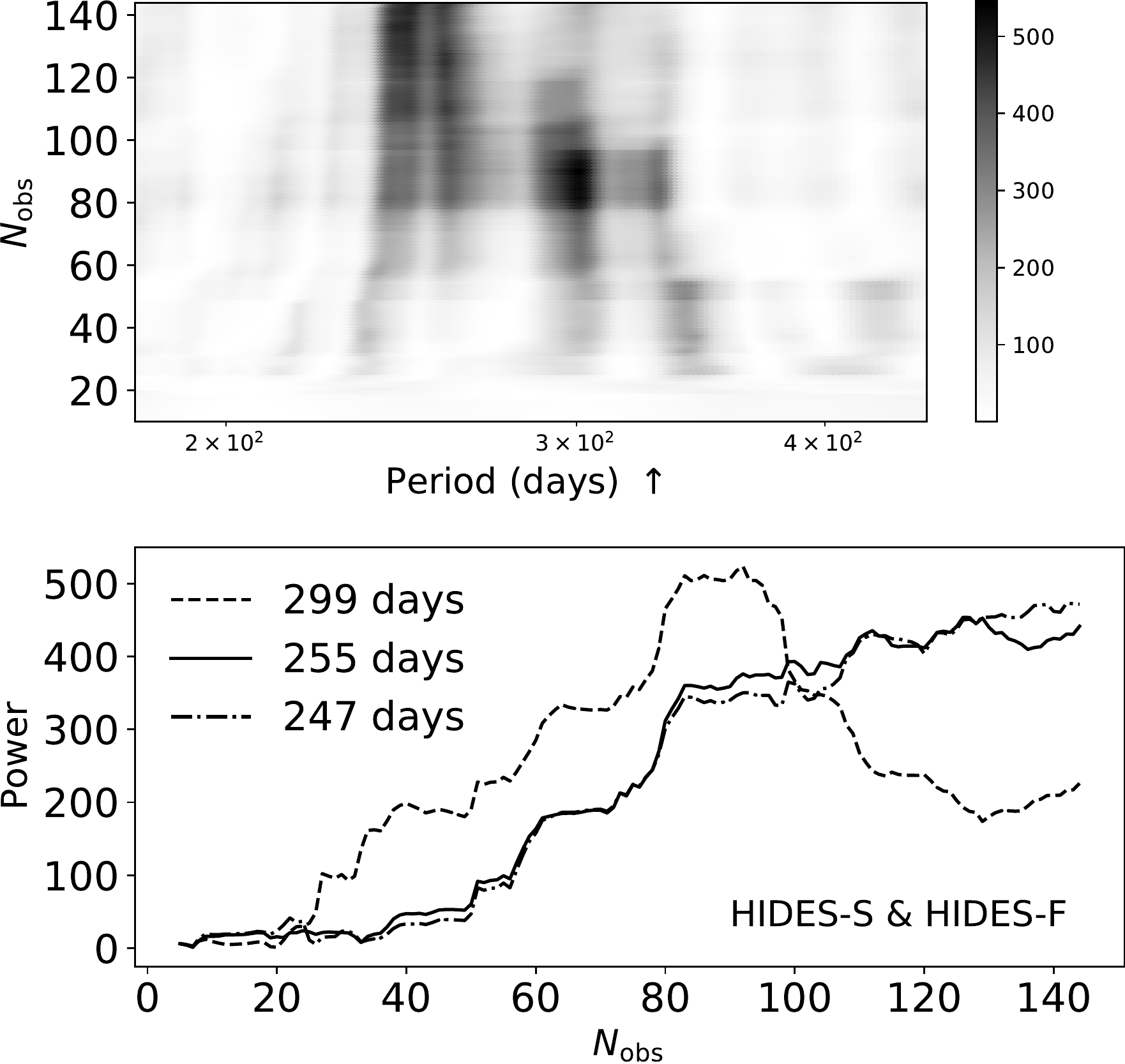}
 \end{center}
\caption{Stacked Bayesian Generalized Lomb-Scargle periodogram of the full data of $\epsilon$ Psc. The upper panel shows the power variation with time and the number of data points. The color bar illustrates the strength of power. The lower panel shows the power of interested periods varying with time. The legend of 299 days marked with a dashed line, the legend of 255 days marked with a solid line, and the legend of 247 days marked with dot-and-dash line respectively represent the interested periods in the GLS periodogram of RV data.
}\label{fig:HD6186_sbgls}
\end{figure}

\begin{figure}
 \begin{center}
  \includegraphics[width=8.0cm]{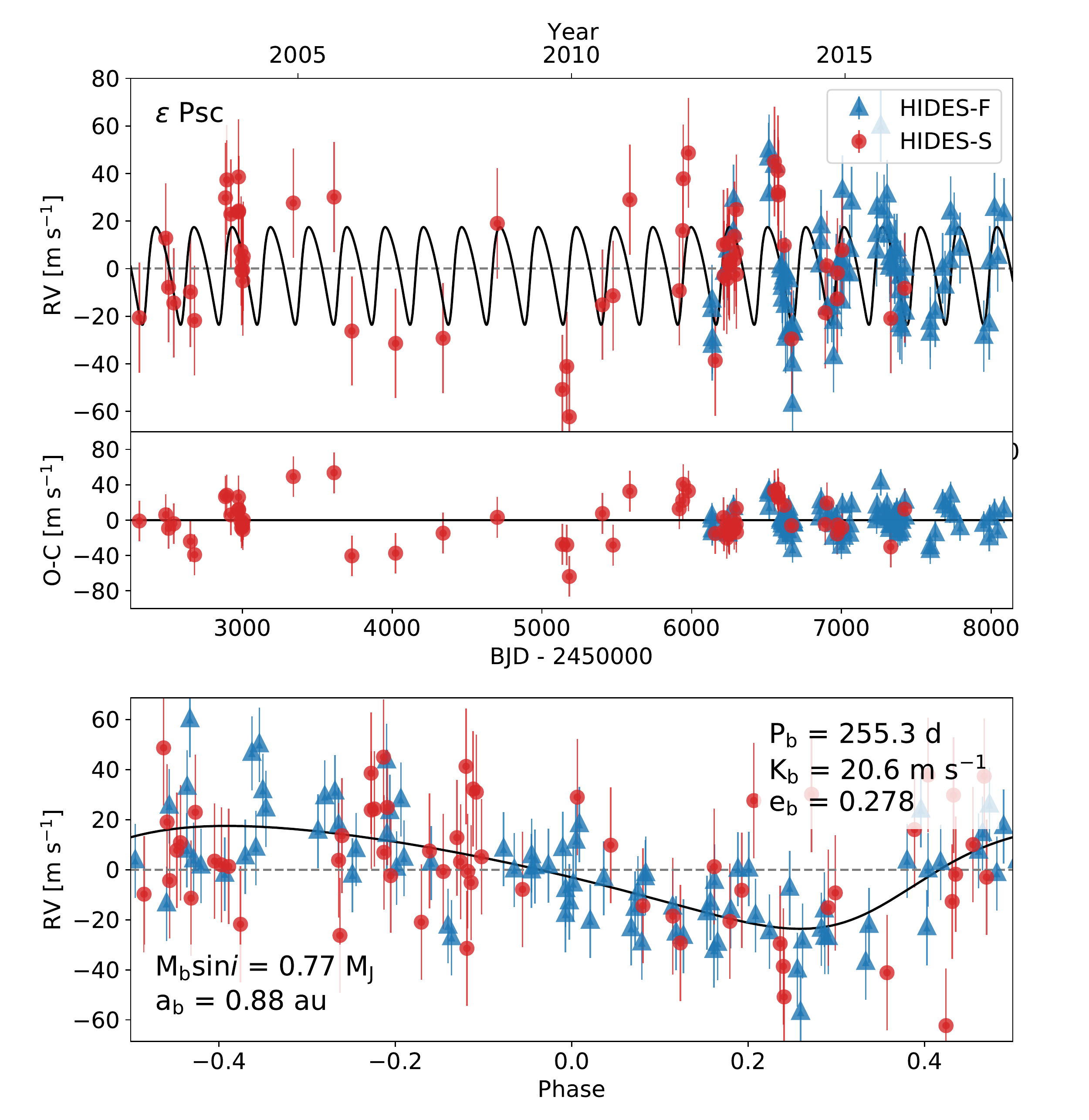} 
 \end{center}
\caption{Orbital solution of $\epsilon$ Psc. HIDES-S data are shown in red and HIDES-F data are shown in blue Top: Best fit single Keplerian curve in the full time span with fitted RV offsets between instruments and fitted jitters included in the errorbars. Mid: RV Residuals to the best fit. Bottom: phase-folded of Top panel.}\label{fig:HD6186_curve}
\end{figure}

We collected a total of 144 data including 60 taken by HIDES-S and 84 taken by HIDES-F between 2002 February and 2017 November. The RV data are shown in Figure \ref{fig:HD6186_rv_all} and the data are listed in Table \ref{RV:HD6186}. For the full RV data, the GLS periodogram shows two significant peaks with a FAP less than $0.1 \%$ including one peak at the period of 247 days ($\sim 2/3$ years) and another peak at a period of 729 days ($\sim 2$ years). The window function shows one peak at 1 year and another one peak over 4000 days. The 1yr window terribly caused an aliasing effect to the true period (noticing $1/2 = 1/(2/3) - 1/1$). Since the power of the 247-day signal is significantly higher than the 729-day signal, we then confirmed that the 247-day signal should be the true period. The other long-term window is the result of sparse observations in the middle of the HIDES-S baseline. 
However, it is distinct that the main peak in the full data GLS is almost dominated by fiber mode data by comparing the GLS periodogram of full data and GLS periodogram of each instrument.
We performed SBGLS to investigate the power of different periods varying by the number of data. In Figure \ref{fig:HD6186_sbgls}, it is clear that the power of the 299-day period sharply decreases after 90 observations, which implies the instability of this variation, while the approximately 250-day signals show a more stable increasing power trend.  
%Its aliasing signal reflects in the branch peak of 247-day signal. 
As for the activity indicator, we did not find significant periodicity with GLS in Ca \emissiontype{II} H index time series, and we did not find RV strongly correlating with fiber mode spectral line profiles or Ca \emissiontype{II} H index ($|r| \lesssim 0.4$).
The FWHM of fiber mode spectra might vary regularly with a period of around 1-year.
We also found a difference in BIS between slit mode and fiber mode spectra.
The difference of BIS between slit mode and fiber mode spectra should be similarly caused by IP difference. However, we found RV moderately correlate with BIS of slit mode spectra ($|r| = 0.6$) and BIS$^{\prime}$ ($|r| = 0.5$).
In the GLS of the BIS time series, we noticed there was a similar periodicity as RV presents for both individual instruments. Therefore, it should be considered that the RV variation may be caused by the star intrinsically.
As for Hipparcos photometry, we did not find any significant periodicity.

We adopted a single Keplerian curve to fit the data and obtained orbital parameters for the companion of $P = 255.4_{-1.4}^{+2.1}\ {\rm days}$, $K =20.6_{-2.9}^{+2.1}\ \rm{m\>s^{-1}}$, $e = 0.057_{-0.030}^{+0.159}$. The rms scatter of the residuals to the Keplerian fit is $18.9\ \rm{m\>s^{-1}}$. 
Adopting a stellar mass of $M = 1.41\ M_{\odot}$, we obtained a minimum mass $M_{\rm{p}}\sin{i} = 0.77_{-0.10}^{+0.14}\  M_{\rm J}$ and a semimajor axis $a = 0.88_{-0.10}^{+0.11}\ {\rm au}$ for the companion. The phase-folded RV curve is shown in the lower panel in Figure \ref{fig:HD6186_curve}.
The fitted extra noise of HIDES-S is relatively large (larger than semi-amplitude) with a value of $22.7\ \rm{m\>s^{-1}}$. 
The periodicity of 299 days still remains in the full-data residuals although the power of this signal is not high enough (FAP $> 1\%$). 
Here, we generated mock data in order to test if the 299-day signal could be a fake signal which is mimicked by the  true period and noise. 
We used the true observation window of HIDES-S and the best-fit Keplerian model of $\epsilon$ Psc, and we added random Gaussian noise to the RV time series with the RV jitter derived from the Keplerian fitting. 
There was a total of 100000 mock time series generated, and GLS periodograms were performed to find the periods in the noisy mocked time series. 
Consequently, 18\% of the simulations failed in recognizing the true period ($\pm 5\%$ as a criterion), and fake periods could distribute everywhere in the period phase.
In such a case, the 299-day period could be a variation either true or triggered by noise. In other words, it is hard to affirm if the period varies with time. Although there is no more accessible observation indicating the RV variation of $\epsilon$ Psc is intrinsically triggered by the star, in consideration of moderate correlation and similar periods between RV and BIS, we could only label $\epsilon$ Psc as a planet-harboring candidate.
%%%%%%%%%%%%%%%%%%%%%%%%%%%%%%%%%%%%%%%%%%%%%%%%%%%%%%%%%%%%%%%%%%%%%%%%%%%%%%%%%%%%%%%%%%%%%%%%%%%%%%%%%%%%%%%%%%%%%%%%

%\newpage
%%%%%%%%%%%%%%%%%%%%%%%%%%%%%%%%%%%%%%%%%%%%%%%%%%%%%%%%%%%%%%%%%%%%%%%%%%%%%%%%%%%%%%%%%%%%%%%%%%%%%%%%%%%%%%%%%%%%%%%%
\subsection{HD 10975}
\begin{table}
\tbl{Radial Velocities of HD 10975 }{%
\begin{tabular}{lccc}
\hline\hline
BJD & Radial Velocity & Uncertainty & Observation \\
$(-2450000)$ & ($\rm{m\ s^{-1}}$) & ($\rm{m\ s^{-1}}$) & Mode \\
\hline
$3026.0042$ & $-22.6$ & $4.1$ & HIDES-S  \\ 
$3245.2826$ & $-9.5$ & $4.9$ & HIDES-S  \\ 
$3333.1050$ & $14.7$ & $2.8$ & HIDES-S  \\ 
$3401.9937$ & $13.0$ & $5.1$ & HIDES-S  \\ 
$3608.3144$ & $-3.7$ & $3.5$ & HIDES-S  \\ 
... & ... & ... & ... \\
\hline
\end{tabular}}
\begin{tabnote}
\hangindent6pt\noindent
\hbox to6pt{\footnotemark[$*$]\hss}\unskip% 
Only the first five sets of RV data are listed. A complete data listing including RV, FWHM, BIS, and S/N will be available online as supplementary after the publication.
\end{tabnote}
\label{RV:HD10975}
\end{table}

\begin{figure*}
 \begin{center}
 \includegraphics[width=16.0cm]{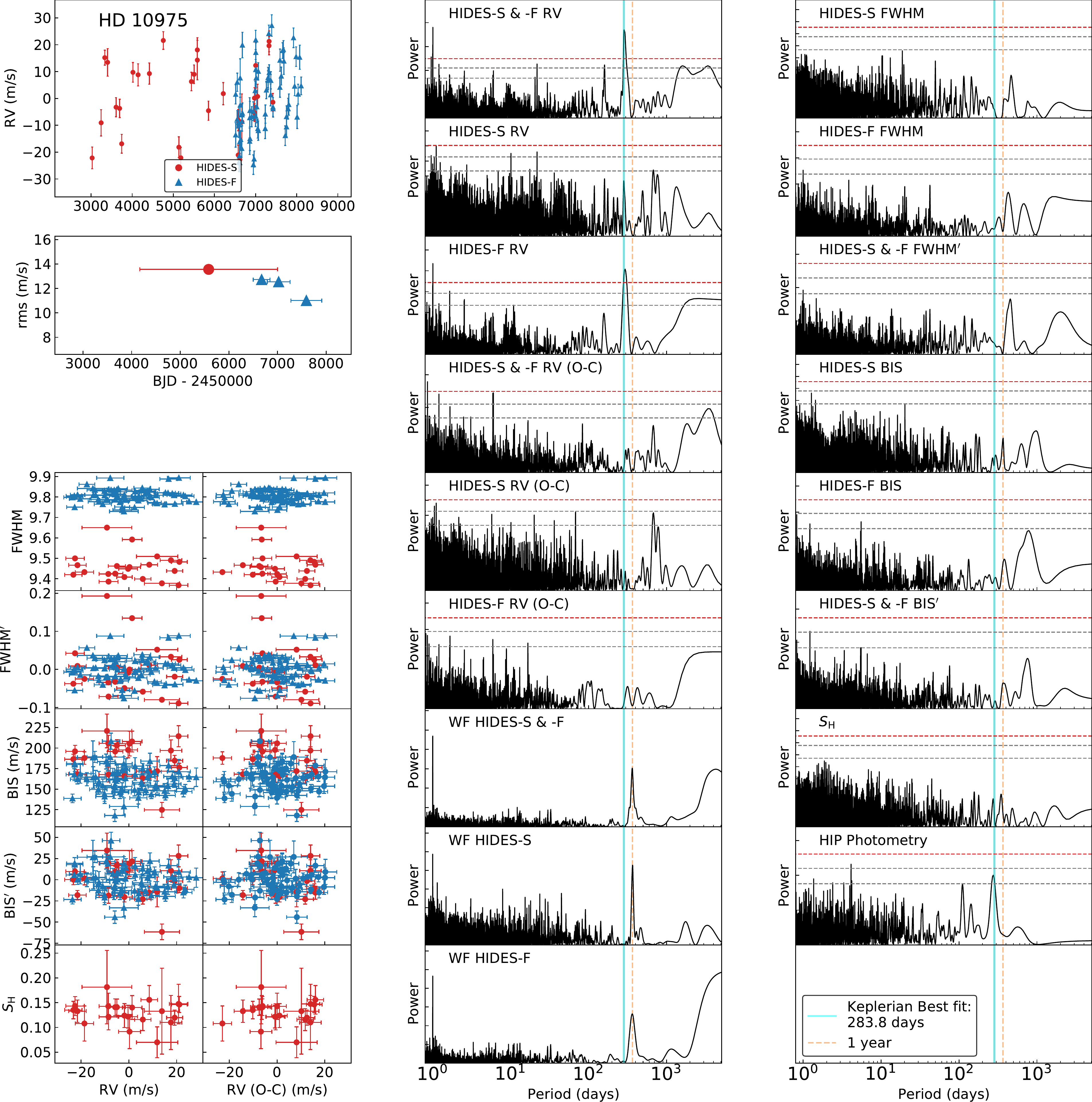}
 \end{center}
\caption{
Summary figure of HD 10975. 
Left column from top to bottom: 
RV time series; rms of RVs varying by time; FWHM, FWHM$^{\prime}$, BIS, BIS$^{\prime}$, and Ca \emissiontype{II} H index $S_{\rm{H}}$ respectively against RVs and RV residuals.
Middle column from top to bottom:
GLS periodogram of full RV data, HIDES-S RV data, HIDES-F RV data, full RV residuals, HIDES-S RV residuals, and HIDES-F RV residuals; window function of full RV data, HIDES-S data, and HIDES-F data.
Right column from top to bottom:
GLS periodogram of HIDES-S FWHM, HIDES-F FWHM, FWHM$^{\prime}$, HIDES-S BIS, HIDES-F BIS, BIS$^{\prime}$, Ca \emissiontype{II} H index $S_{\rm{H}}$, and Hipparcos photometry. In GLS periodograms, the horizontal lines represent 10\%, 1\%, and 0.1\% FAP levels from bottom to top. The vertical cyan solid line indicates the best-fitted period from the Keplerian model, and the vertical orange dashed line indicates 1 year. 
}\label{fig:HD10975_rv_all}
\end{figure*}

\begin{figure}
 \begin{center}
  \includegraphics[width=8.0cm]{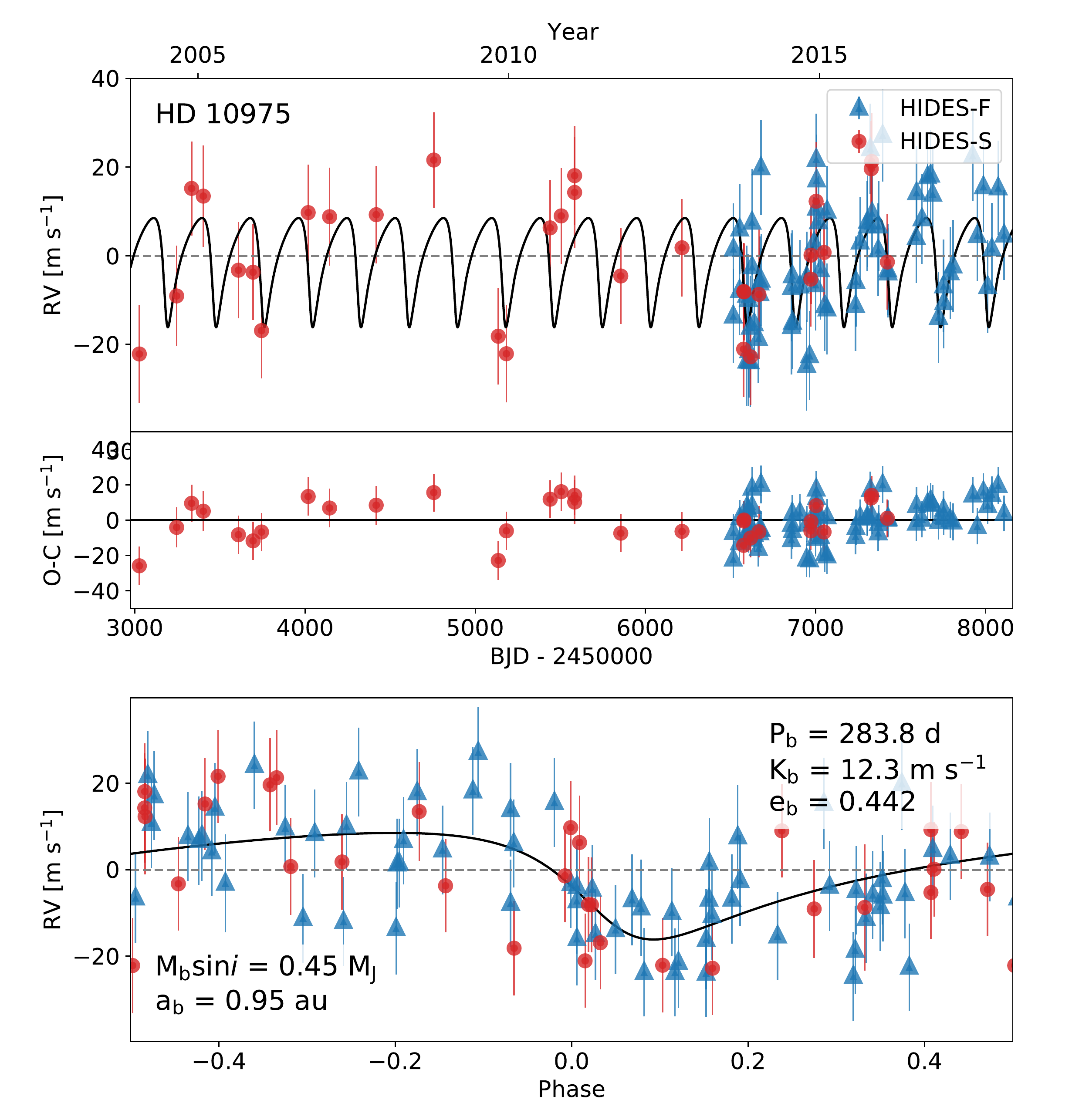} 
 \end{center}
\caption{Orbital solution of HD 10975. HIDES-S data are shown in red and HIDES-F data are shown in blue Top: Best fit single Keplerian curve in the full time span with fitted RV offsets between instruments and fitted jitters included in the errorbars. Mid: RV Residuals to the best fit. Bottom: phase-folded of Top panel. }\label{fig:HD10975_curve}
\end{figure}

We collected a total of 99 data including 32 taken by HIDES-S and 67 taken by HIDES-F between 2004 January and 2017 December. The RV data are shown in Figure \ref{fig:HD10975_rv_all} and the data are listed in Table \ref{RV:HD10975}. For full data, GLS shows a significant peak at the period of 286 days with a FAP of less than $0.1 \%$. We did not find significant periodicity with GLS in Ca \emissiontype{II} H index time series, and we did not find RV strongly correlating with either spectral line profiles or Ca \emissiontype{II} H index ($|r| < 0.2$). We also calculated a periodogram for Hipparcos photometry. We could see one weak signal with a similar RV period in the GLS periodogram, yet it is not significant enough (FAP $> 1\% $).

We adopted a single Keplerian curve to fit the data and obtained orbital parameters for the companion of $P = 283.8_{-0.1}^{+12.7} \ {\rm days}$, $K =12.3_{-3.5}^{+0.2} \ \rm{m\>s^{-1}}$, $e = 0.442_{-0.389}^{+0.040}$. The rms scatter of the residuals to the Keplerian fit is $10.7\ \rm{m\>s^{-1}}$. We did not find any significant periodicity in the residuals, and we did not find line profiles or Ca \emissiontype{II} H index strongly correlating with RV residuals ($|r| < 0.3$). Adopting a stellar mass of $M = 1.41\ M_{\odot}$, we obtained a minimum mass $M_{\rm{p}}\sin{i} = 0.45_{-0.15}^{+0.06}\  M_{\rm J}$ and a semimajor axis $a = 0.95_{-0.04}^{+0.06}\ {\rm au}$ for the companion. The phase-folded RV curve is shown in the lower panel in Figure \ref{fig:HD10975_curve}.
%%%%%%%%%%%%%%%%%%%%%%%%%%%%%%%%%%%%%%%%%%%%%%%%%%%%%%%%%%%%%%%%%%%%%%%%%%%%%%%%%%%%%%%%%%%%%%%%%%%%%%%%%%%%%%%%%%%%%%%%

%\newpage
%%%%%%%%%%%%%%%%%%%%%%%%%%%%%%%%%%%%%%%%%%%%%%%%%%%%%%%%%%%%%%%%%%%%%%%%%%%%%%%%%%%%%%%%%%%%%%%%%%%%%%%%%%%%%%%%%%%%%%%%
\subsection{HD 79181}
\begin{table}
\tbl{Radial Velocities of HD 79181 }{%
\begin{tabular}{lccc}
\hline\hline
BJD & Radial Velocity & Uncertainty & Observation \\
$(-2450000)$ & ($\rm{m\ s^{-1}}$) & ($\rm{m\ s^{-1}}$) & Mode \\
\hline
$1987.0079$ & $-28.7$ & $6.5$ & HIDES-S  \\ 
$2016.0188$ & $-11.9$ & $6.8$ & HIDES-S  \\ 
$2041.9640$ & $12.2$ & $6.0$ & HIDES-S  \\ 
$2272.3077$ & $-26.5$ & $5.9$ & HIDES-S  \\ 
$2282.2919$ & $-13.1$ & $6.5$ & HIDES-S  \\ 
... & ... & ... & ...  \\
\hline
\end{tabular}}
\begin{tabnote}
\hangindent6pt\noindent
\hbox to6pt{\footnotemark[$*$]\hss}\unskip% 
Only the first five sets of RV data are listed. A complete data listing including RV, FWHM, BIS, and S/N will be available online as supplementary after the publication.
\end{tabnote}
\label{RV:HD79181}
\end{table}

\begin{figure*}
 \begin{center}
  \includegraphics[width=16.0cm]{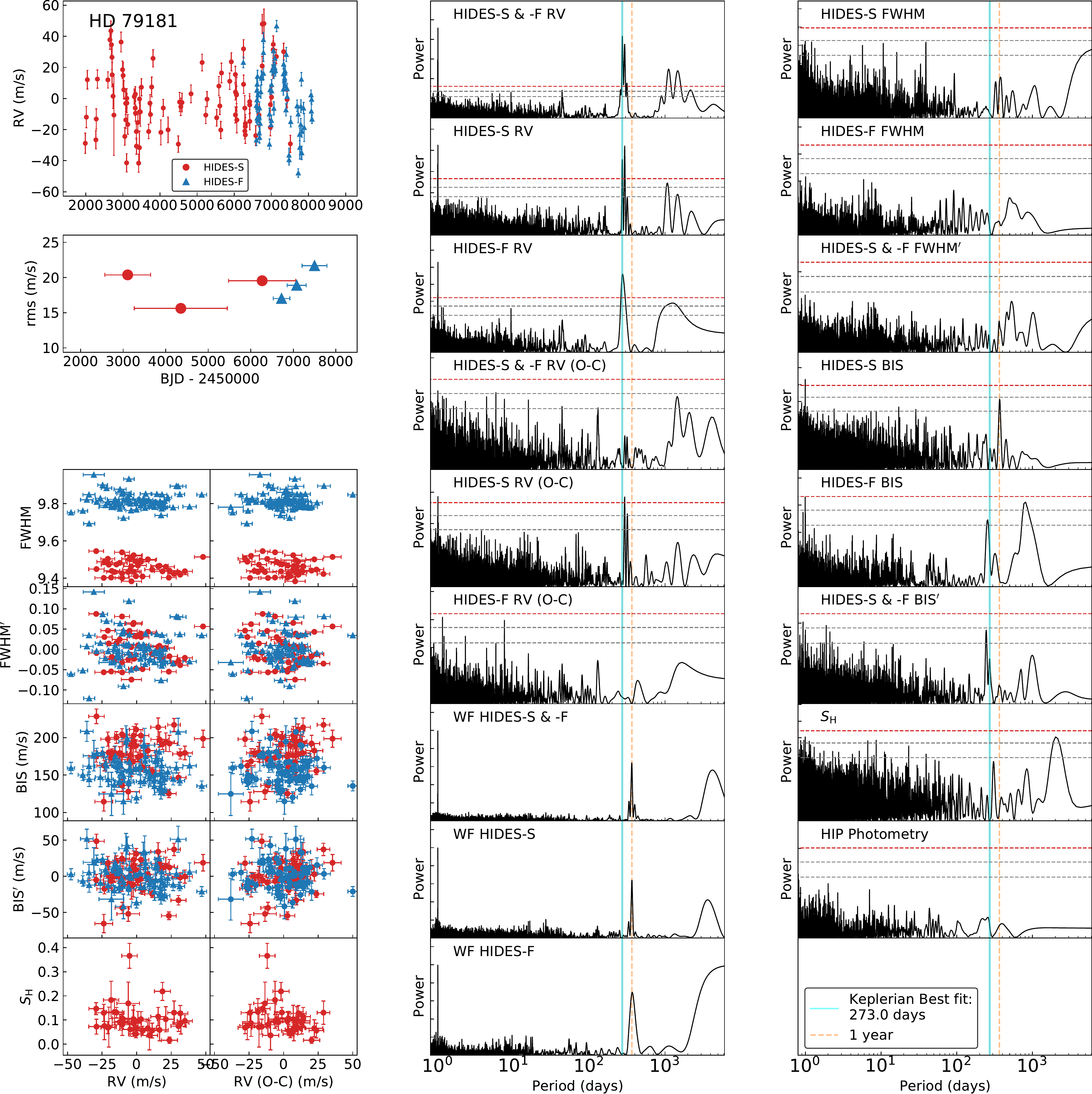}
 \end{center}
\caption{
Summary figure of HD 79181. 
Left column from top to bottom: 
RV time series; rms of RVs varying by time; FWHM, FWHM$^{\prime}$, BIS, BIS$^{\prime}$, and Ca \emissiontype{II} H index $S_{\rm{H}}$ respectively against RVs and RV residuals.
Middle column from top to bottom:
GLS periodogram of full RV data, HIDES-S RV data, HIDES-F RV data, full RV residuals, HIDES-S RV residuals, and HIDES-F RV residuals; window function of full RV data, HIDES-S data, and HIDES-F data.
Right column from top to bottom:
GLS periodogram of HIDES-S FWHM, HIDES-F FWHM, FWHM$^{\prime}$, HIDES-S BIS, HIDES-F BIS, BIS$^{\prime}$, Ca \emissiontype{II} H index $S_{\rm{H}}$, and Hipparcos photometry. In GLS periodograms, the horizontal lines represent 10\%, 1\%, and 0.1\% FAP levels from bottom to top. The vertical cyan solid line indicates the best-fitted period from the Keplerian model, and the vertical orange dashed line indicates 1 year. 
}\label{fig:HD79181_rv_all}
\end{figure*}

\begin{figure}
 \begin{center}
  \includegraphics[width=8.0cm]{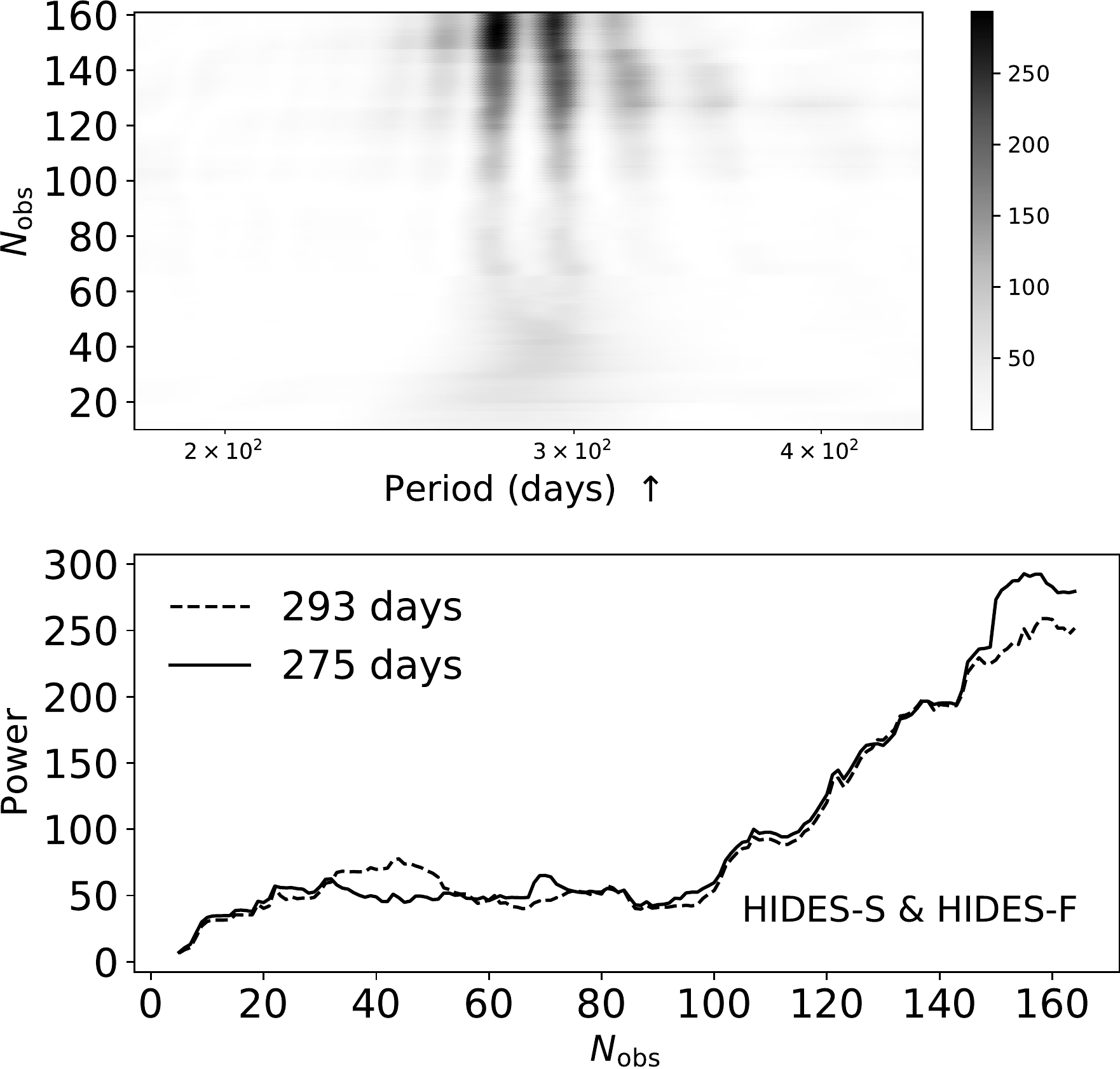}
 \end{center}
\caption{Stacked Bayesian Generalized Lomb-Scargle periodogram of the full data of HD 79181. The upper panel shows the power variation with time and the number of data points. The color bar illustrates the strength of power. The lower panel shows the power of interested periods varying with time. The legend of 293 days marked with a dashed line and the legend of 275 days marked with solid line respectively represent the interested periods of 293-day signal and 275-day signal in the GLS periodogram of RV data. 
}\label{fig:HD79181_sbgls}
\end{figure}

\begin{figure}
 \begin{center}
  \includegraphics[width=8.0cm]{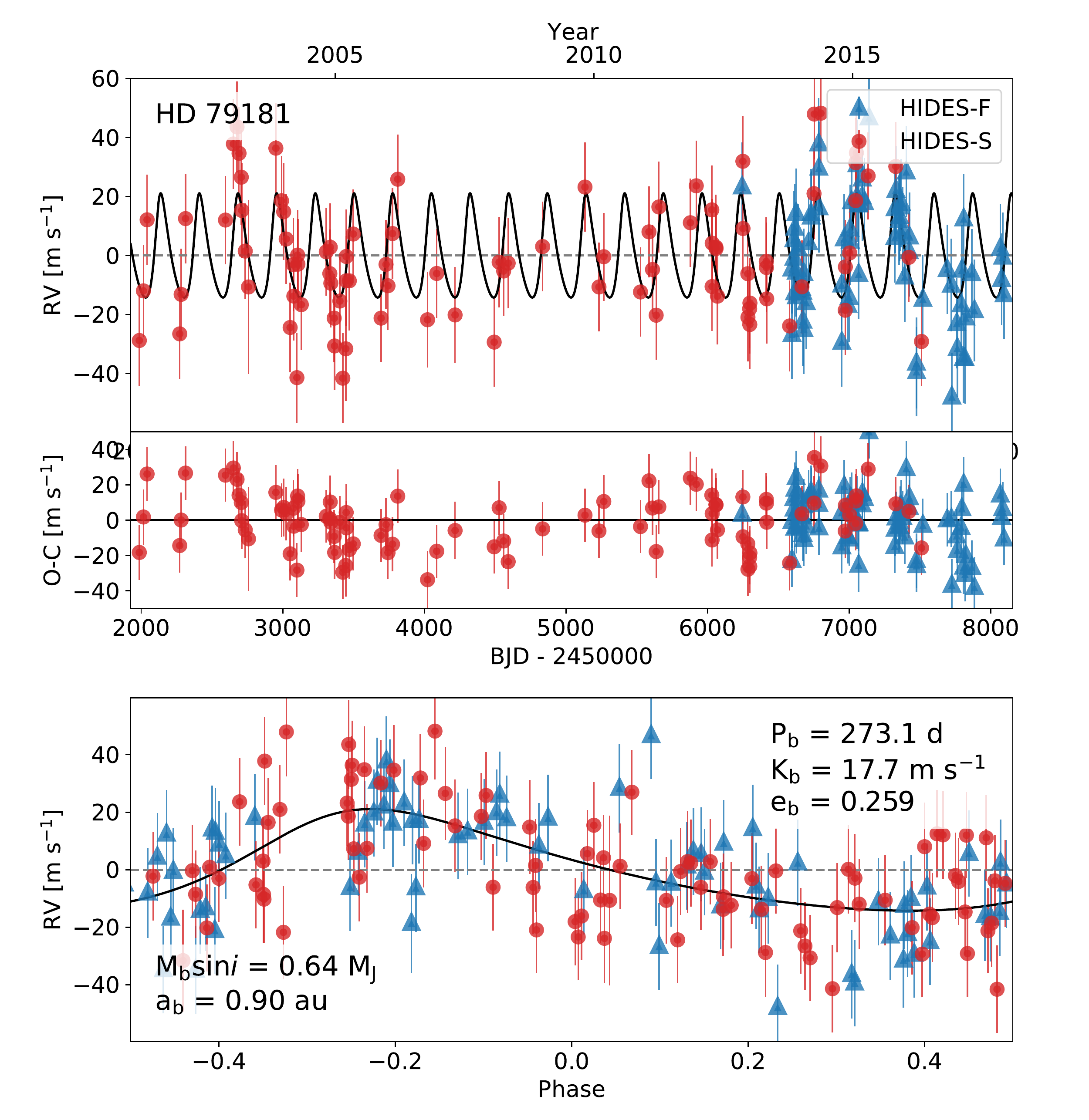} 
 \end{center}
\caption{Orbital solution of HD 79181. HIDES-S data are shown in red and HIDES-F data are shown in blue Top: Best fit single Keplerian curve in the full time span with fitted RV offsets between instruments and fitted jitters included in the errorbars. Mid: RV Residuals to the best fit. Bottom: phase-folded of Top panel.}\label{fig:HD79181_curve}
\end{figure}

This star is firstly reported in the result of the first three years Okayama Planet Search Program, yet no planet was detected at that time \citep{Sato2005}. This time, we collected a total of 164 data including 92 taken by HIDES-S and 72 taken by HIDES-F between 2001 March and 2017 December. The RV data are shown in Figure \ref{fig:HD79181_rv_all} and the data are listed in Table \ref{RV:HD79181}. For full data, the Generalized Lomb-Scargle periodogram shows four significant peaks with a FAP of less than $0.1 \%$ including peaks at periods of 275, 293, 1119, 1477 days.
The window function shows one peak at 1 year and another one peak over 4000 days. The 1yr window terribly causes an aliasing effect in the period analysis. Recognizing the relations of ($1/1119 \simeq 1/275 - 1/365$) and ($1/1477 \simeq 1/293 - 1/365$), we, therefore,  confirmed that periods of 1119 and 1477 days should be the aliases of 275 and 293 days respectively. The other long-term window is the result of sparse observations in the middle of the HIDES-S baseline and it also causes aliasing on the true signal. Either 275-day or 293-day is the alias of the other one. Since full RV data and HIDES-S RV data show different stronger signals, and HIDES-F RV data show only one peak, therefore we performed SBGLS to investigate the signal closer to the true period in the periodogram.
Consequently, SBGLS shows that significant periodicity appears after 100 observations, and the power of 275 days becomes apparently stronger than the power of 293 days after gathering 145 observations (Figure \ref{fig:HD79181_sbgls}). 
There is one signal at around one year in the periodograms of BIS for slit mode spectra. According to the IP discussion in Section Appendix \ref{sec:bisip}, we confirm they are regular instrumental variations. There are another two signals at 254 days and 812 days in the periodograms of BIS for fiber mode spectra and BIS$^{\prime}$ for both spectra. 
Recognizing ($1/254 \simeq 1/812 + 1/365$, we, therefore, knew the 254-day signal should be the alias. 
We did not find significant periodicity with GLS in Ca \emissiontype{II} H index time series, and we did not find RV strongly correlating with either spectral line profiles or Ca \emissiontype{II} H index ($|r| \lesssim 0.2$).

We adopted a single Keplerian curve to fit the data and obtained orbital parameters for the companion of $P = 273.1_{-0.4}^{+1.3}\ {\rm days}$, $K =17.7_{-2.2}^{+1.6}\ \rm{m\>s^{-1}}$, $e = 0.259_{-0.206}^{+0.067} $.
The rms scatter of the residuals to the Keplerian fit is $15.5\ \rm{m\>s^{-1}}$. 
We find a relatively significant signal at 292 days in the residuals of data taken by only HIDES-S. Since the period of this regular variation is the almost same as the one detected from the original RV taken by HIDES-S, it implies that the 293-day variation cannot be fully subtracted. 
In such a case, we infer that sparse observation of HIDES-S could be account for the inaccuracy in searching for the true period. 
As for the activity indicators, we did not find line profiles or Ca \emissiontype{II} H index strongly correlating with RV residuals ($|r| < 0.3$). In this case, it is confident to say that the period of 273 day, rather than 293 days, is the true orbital period. Adopting a stellar mass of $M = 1.28\ M_{\odot}$, we obtained a minimum mass $M_{\rm{p}}\sin{i} = 0.64_{-0.16}^{+0.06}\  M_{\rm J}$ and a semimajor axis $a = 0.90_{-0.08}^{+0.07}\ {\rm au}$ for the companion. The phase-folded RV curve is shown in the lower panel in Figure \ref{fig:HD79181_curve}.
%%%%%%%%%%%%%%%%%%%%%%%%%%%%%%%%%%%%%%%%%%%%%%%%%%%%%%%%%%%%%%%%%%%%%%%%%%%%%%%%%%%%%%%%%%%%%%%%%%%%%%%%%%%%%%%%%%%%%%%%

%\newpage
%%%%%%%%%%%%%%%%%%%%%%%%%%%%%%%%%%%%%%%%%%%%%%%%%%%%%%%%%%%%%%%%%%%%%%%%%%%%%%%%%%%%%%%%%%%%%%%%%%%%%%%%%%%%%%%%%%%%%%%%
\subsection{HD 99283}
\begin{table}
\tbl{Radial Velocities of HD 99283 }{%
\begin{tabular}{lccc}
\hline\hline
BJD & Radial Velocity & Uncertainty & Observation \\
$(-2450000)$ & ($\rm{m\ s^{-1}}$) & ($\rm{m\ s^{-1}}$) & Mode \\
\hline
$3030.2885$ & $1.3$ & $3.8$ & HIDES-S  \\ 
$3105.1390$ & $-1.3$ & $4.2$ & HIDES-S  \\ 
$3333.2632$ & $-4.4$ & $3.1$ & HIDES-S  \\ 
$3428.1573$ & $-16.8$ & $3.8$ & HIDES-S  \\ 
$3498.0961$ & $5.5$ & $4.1$ & HIDES-S  \\ 
... & ... & ... & ... \\
\hline
\end{tabular}}
\begin{tabnote}
\hangindent6pt\noindent
\hbox to6pt{\footnotemark[$*$]\hss}\unskip% 
Only the first five sets of RV data are listed. A complete data listing including RV, FWHM, BIS, and S/N will be available online as supplementary after the publication.
\end{tabnote}
\label{RV:HD99283}
\end{table}

\begin{figure*}
 \begin{center}
 \includegraphics[width=16.0cm]{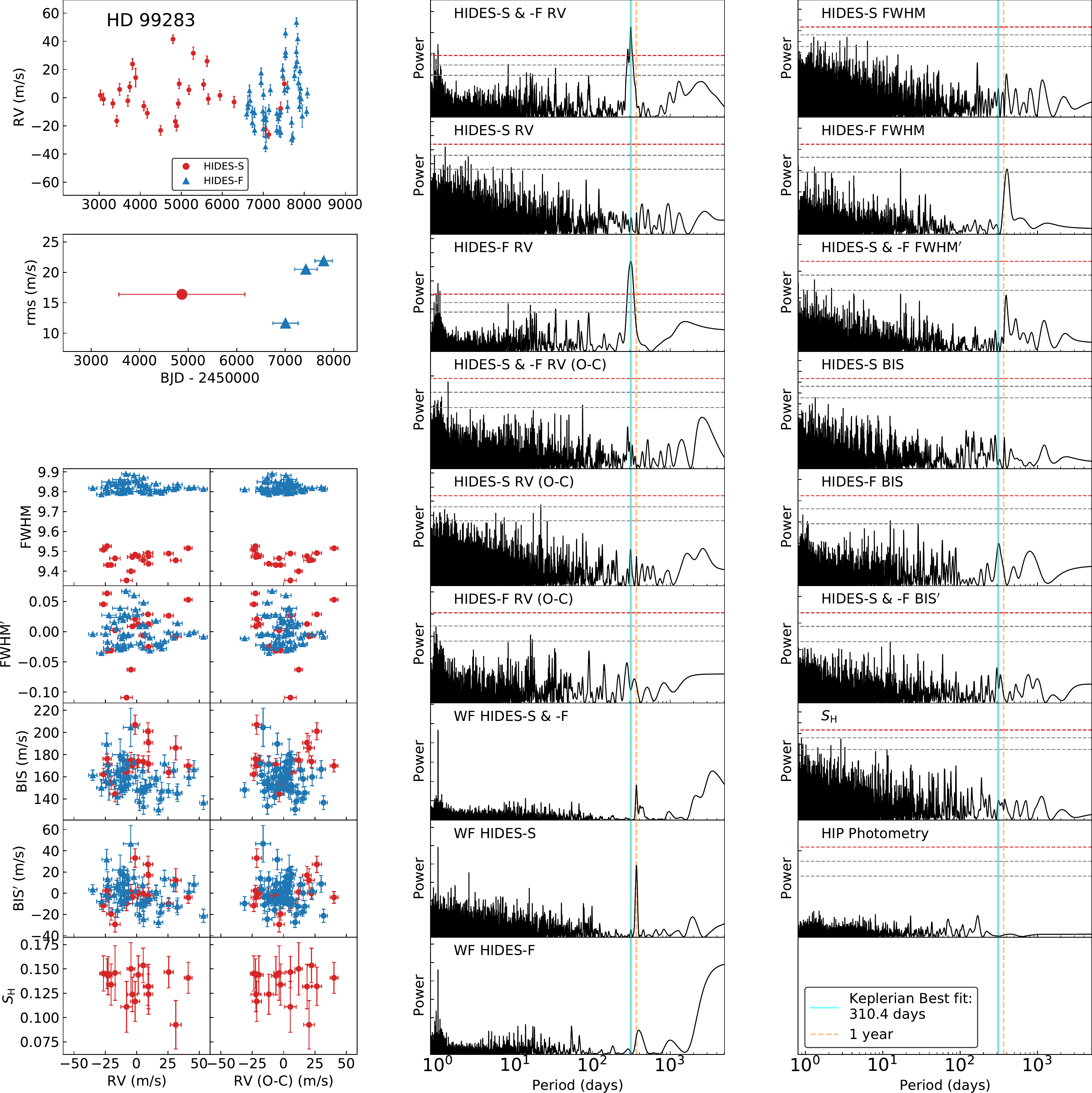}
 \end{center}
\caption{
Summary figure of HD 99283. 
Left column from top to bottom: 
RV time series; rms of RVs varying by time; FWHM, FWHM$^{\prime}$, BIS, BIS$^{\prime}$, and Ca \emissiontype{II} H index $S_{\rm{H}}$ respectively against RVs and RV residuals.
Middle column from top to bottom:
GLS periodogram of full RV data, HIDES-S RV data, HIDES-F RV data, full RV residuals, HIDES-S RV residuals, and HIDES-F RV residuals; window function of full RV data, HIDES-S data, and HIDES-F data.
Right column from top to bottom:
GLS periodogram of HIDES-S FWHM, HIDES-F FWHM, FWHM$^{\prime}$, HIDES-S BIS, HIDES-F BIS, BIS$^{\prime}$, Ca \emissiontype{II} H index $S_{\rm{H}}$, and Hipparcos photometry. In GLS periodograms, the horizontal lines represent 10\%, 1\%, and 0.1\% FAP level from bottom to top. The vertical cyan solid line indicates the best-fitted period from the Keplerian model, and the vertical orange dashed line indicates 1 year. 
}\label{fig:HD99283_rv_all}
\end{figure*}

\begin{figure}
 \begin{center}
  \includegraphics[width=8.0cm]{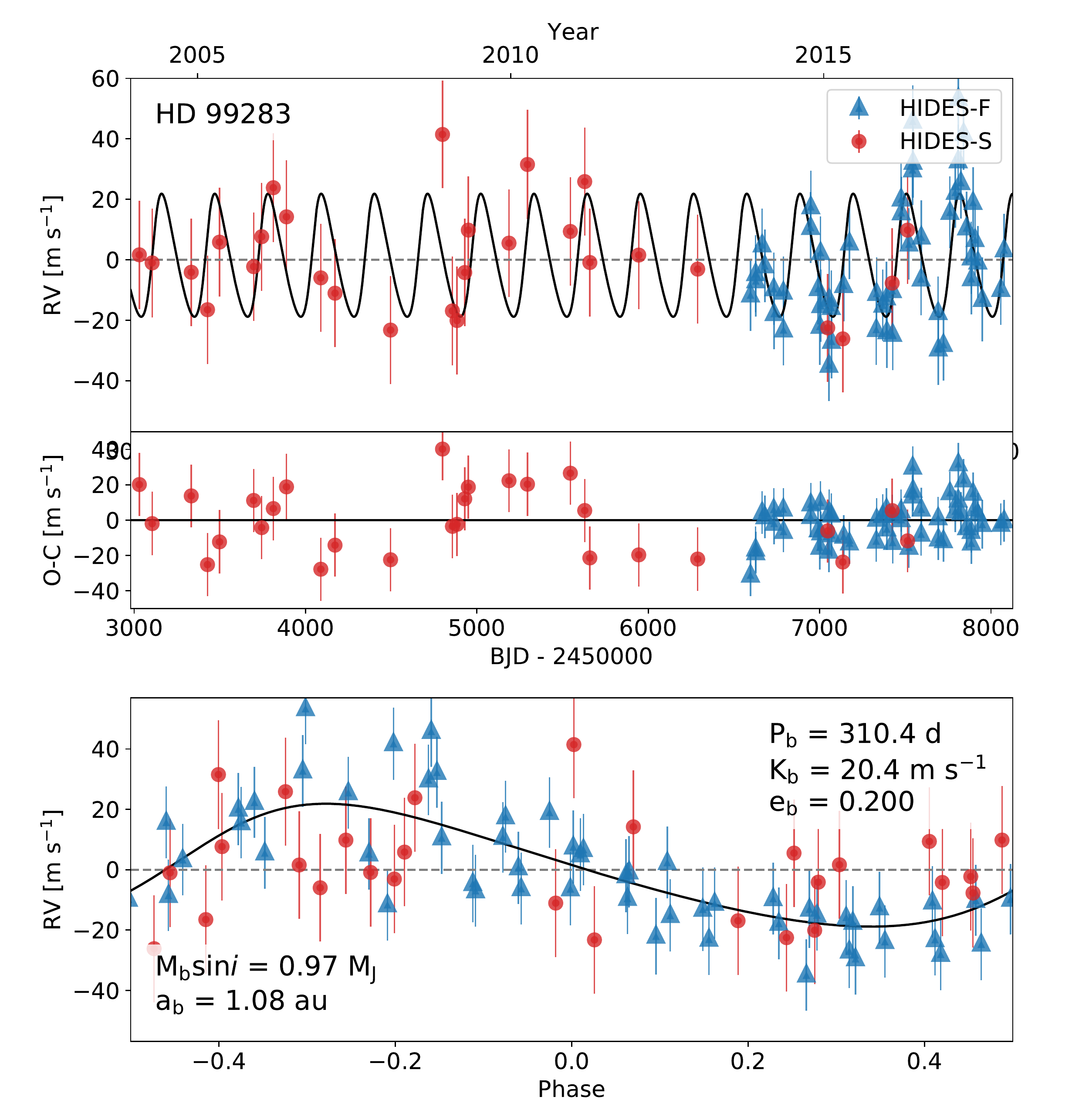} 
 \end{center}
\caption{Orbital solution of HD 99283. HIDES-S data are shown in red and HIDES-F data are shown in blue Top: Best fit single Keplerian curve in the full time span with fitted RV offsets between instruments and fitted jitters included in the errorbars. Mid: RV Residuals to the best fit. Bottom: phase-folded of Top panel.}\label{fig:HD99283_curve}
\end{figure}

We collected a total of 82 data including 28 taken by HIDES-S and 54 taken by HIDES-F between 2004 January and 2017 November. The RV data are shown in Figure \ref{fig:HD99283_rv_all} and the data are listed in Table \ref{RV:HD99283}. The GLS shows a significant peak at the period of 310 days with a FAP of less than $0.1 \%$. We did not find significant periodicity with GLS in Ca \emissiontype{II} H index time series, and we did not find RV strongly correlating with either spectral line profiles or Ca \emissiontype{II} H index ($|r| < 0.3$).

We adopted a single Keplerian curve to fit the data and obtained orbital parameters for the companion of $P = 310.4_{-1.7}^{+5.2} \ {\rm days}$, $K =20.4_{-3.5}^{+1.9} \ \rm{m\>s^{-1}}$, $e = 0.200_{-0.146}^{+0.102} $. The rms scatter of the residuals to the Keplerian fit is $14.4 \ \rm{m\>s^{-1}}$. We did not find any significant periodicity in the residuals, and we did not find line profiles or Ca \emissiontype{II} H index strongly correlating with RV residuals ($|r| < 0.2$). Adopting a stellar mass of $M = 1.76\ M_{\odot}$, we obtained a minimum mass $M_{\rm{p}}\sin{i} = 0.97_{-0.25}^{+0.06}\  M_{\rm J}$ and a semimajor axis $a = 1.08_{-0.05}^{+0.07}\ {\rm au}$ for the companion. Besides, we present RV scatters varying with time in Figure \ref{fig:HD99283_rv_all}. From the figure, we can find the first scatter measurement of HIDES-F is apparently lower than the other three measurements. Combining with the result from Keplerian fitting, we know that the data included by the first scatter measurement of HIDES-F is mainly located on the second half in the orbital phase, which refers to the phase between 0 and 0.5 in the bottom panel in Figure \ref{fig:HD99283_curve}. For the other three scatter measurements, by considering solar-like oscillations and instrumental jitter, we believe the scatter of rms is acceptable as these rms measurements locate around 20 $\rm{m\>s^{-1}}$.
%%%%%%%%%%%%%%%%%%%%%%%%%%%%%%%%%%%%%%%%%%%%%%%%%%%%%%%%%%%%%%%%%%%%%%%%%%%%%%%%%%%%%%%%%%%%%%%%%%%%%%%%%%%%%%%%%%%%%%%%

%\newpage
%%%%%%%%%%%%%%%%%%%%%%%%%%%%%%%%%%%%%%%%%%%%%%%%%%%%%%%%%%%%%%%%%%%%%%%%%%%%%%%%%%%%%%%%%%%%%%%%%%%%%%%%%%%%%%%%%%%%%%%%
\subsection{$\upsilon$ Leo}
\begin{table}
\tbl{Radial Velocities of $\upsilon$ Leo}{%
\begin{tabular}{lccc}
\hline\hline
BJD & Radial Velocity & Uncertainty & Observation \\
$(-2450000)$ & ($\rm{m\ s^{-1}}$) & ($\rm{m\ s^{-1}}$) & Mode \\
\hline
$1987.2705$ & $-6.1$ & $4.1$ & HIDES-S  \\ 
$2016.1508$ & $26.6$ & $5.4$ & HIDES-S  \\ 
$2033.0790$ & $0.3$ & $4.0$ & HIDES-S  \\ 
$2043.0306$ & $26.5$ & $3.6$ & HIDES-S  \\ 
$2271.2775$ & $-23.4$ & $3.6$ & HIDES-S  \\ 
... & ... & ... & ... \\
\hline
\end{tabular}}
\begin{tabnote}
\hangindent6pt\noindent
\hbox to6pt{\footnotemark[$*$]\hss}\unskip% 
Only the first five sets of RV data are listed. A complete data listing including RV, FWHM, BIS, and S/N will be available online as supplementary after the publication.
\end{tabnote}
\label{RV:HD100920}
\end{table}

\begin{figure*}
 \begin{center}
 \includegraphics[width=16.0cm]{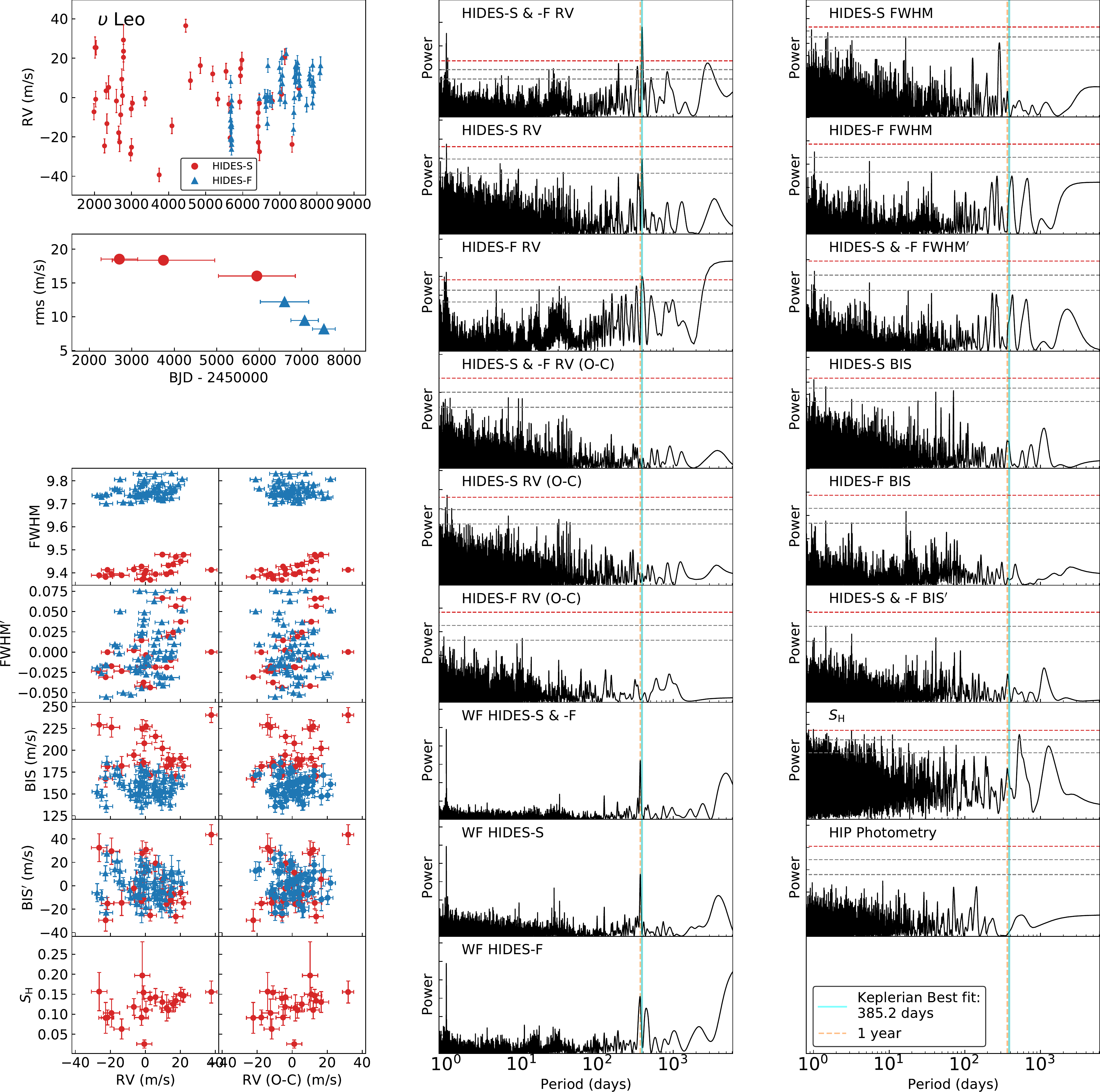}
 \end{center}
\caption{
Summary figure of $\upsilon$ Leo. 
Left column from top to bottom: 
RV time series; rms of RVs varying by time; FWHM, FWHM$^{\prime}$, BIS, BIS$^{\prime}$, and Ca \emissiontype{II} H index $S_{\rm{H}}$ respectively against RVs and RV residuals.
Middle column from top to bottom:
GLS periodogram of full RV data, HIDES-S RV data, HIDES-F RV data, full RV residuals, HIDES-S RV residuals, and HIDES-F RV residuals; window function of full RV data, HIDES-S data, and HIDES-F data.
Right column from top to bottom:
GLS periodogram of HIDES-S FWHM, HIDES-F FWHM, FWHM$^{\prime}$, HIDES-S BIS, HIDES-F BIS, BIS$^{\prime}$, Ca \emissiontype{II} H index $S_{\rm{H}}$, and Hipparcos photometry. In GLS periodograms, the horizontal lines represent 10\%, 1\%, and 0.1\% FAP level from bottom to top. The vertical cyan solid line indicates the best-fitted period from the Keplerian model, and the vertical orange dashed line indicates 1 year. 
}\label{fig:HD100920_rv_all}
\end{figure*}

\begin{figure}
 \begin{center}
  \includegraphics[width=8.0cm]{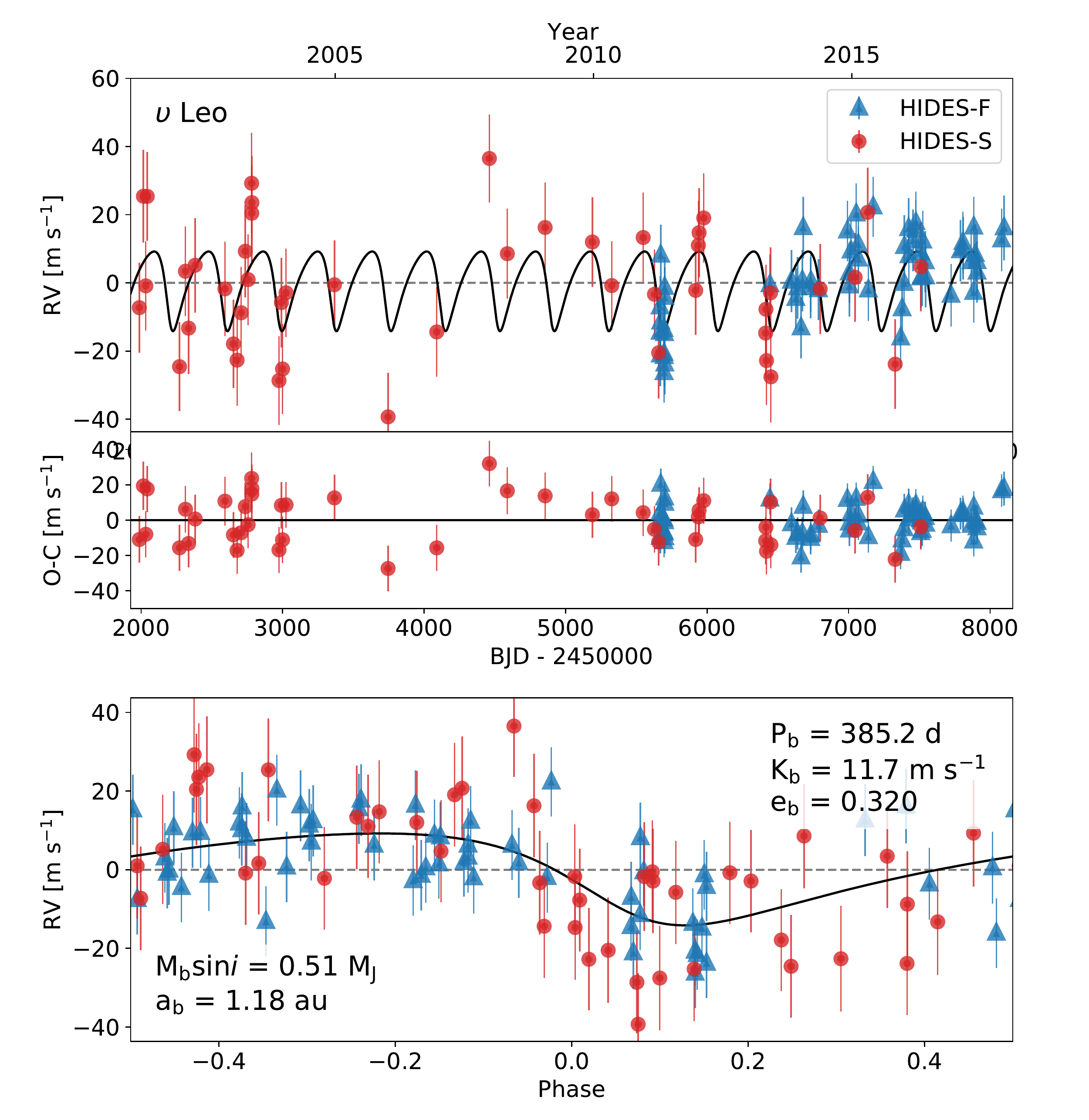} 
 \end{center}
\caption{Orbital solution of $\upsilon$ Leo. HIDES-S data are shown in red and HIDES-F data are shown in blue Top: Best fit single Keplerian curve in the full time span with fitted RV offsets between instruments and fitted jitters included in the errorbars. Mid: RV Residuals to the best fit. Bottom: phase-folded of Top panel.}\label{fig:HD100920_curve}
\end{figure}

We collected a total of 105 data including 46 taken by HIDES-S and 59 taken by HIDES-F between 2001 December and 2017 November. The RV data are shown in Figure \ref{fig:HD100920_rv_all} and the data are listed in Table \ref{RV:HD100920}. The GLS shows a significant peak at the period of 387 days with a FAP of less than $0.1 \%$. 
Although a signal at 527 days derived from Ca \emissiontype{II} H index time series might indicate the rotational period, we did not find RV strongly correlating with either spectral line profiles or Ca \emissiontype{II} H index ($|r| < 0.5$). We did not find RV correlate with line profiles. 

We adopted a single Keplerian curve to fit the data and obtained orbital parameters for the companion of $P = 385.2_{-1.3}^{+2.8} \ {\rm days}$, $K =11.7_{-3.5}^{+1.9} \ \rm{m\>s^{-1}}$, $e = 0.320_{-0.218}^{+0.134} $. The rms scatter of the residuals to the Keplerian fit is $11.2 \ \rm{m\>s^{-1}}$. The fitted extra noise of HIDES-S is relatively large (larger than semi-amplitude) with a value of $12.5\ \rm{m\>s^{-1}}$. We did not find any significant periodicity in the residuals, and we did not find line profiles or Ca \emissiontype{II} H index strongly correlating with RV residuals ($|r| < 0.5$). Adopting a stellar mass of $M = 1.48\ M_{\odot}$, we obtained a minimum mass $M_{\rm{p}}\sin{i} = 0.51_{-0.25}^{+0.06}\  M_{\rm J}$ and a semimajor axis $a = 1.18_{-0.05}^{+0.07}\ {\rm au}$ for the companion.
Besides, we present RV scatters varying with time in Figure \ref{fig:HD100920_rv_all}. We find that the scatters of HIDES-S data are lower than the ones of HIDES-F data. 
However, we also notice that the data included in the second and third scatter measurements of HIDES-F observations are mainly located at the first half of the orbital phase (phase between -0.5 and 0 in the bottom panel of Figure \ref{fig:HD100920_curve}). In such a case, we could know that the low scatter of HIDES-F data was due to the observational window along the observation baseline. 
%%%%%%%%%%%%%%%%%%%%%%%%%%%%%%%%%%%%%%%%%%%%%%%%%%%%%%%%%%%%%%%%%%%%%%%%%%%%%%%%%%%%%%%%%%%%%%%%%%%%%%%%%%%%%%%%%%%%%%%%

%\newpage
%%%%%%%%%%%%%%%%%%%%%%%%%%%%%%%%%%%%%%%%%%%%%%%%%%%%%%%%%%%%%%%%%%%%%%%%%%%%%%%%%%%%%%%%%%%%%%%%%%%%%%%%%%%%%%%%%%%%%%%%
\subsection{HD 161178}
\begin{table}
\tbl{Radial Velocities of HD 161178 }{%
\begin{tabular}{lccc}
\hline\hline
BJD & Radial Velocity & Uncertainty & Observation \\
$(-2450000)$ & ($\rm{m\ s^{-1}}$) & ($\rm{m\ s^{-1}}$) & Mode \\
\hline
$1987.3376$ & $38.3$ & $4.3$ & HIDES-S  \\ 
$2016.2007$ & $35.7$ & $7.2$ & HIDES-S  \\ 
$2018.2127$ & $20.9$ & $2.9$ & HIDES-S  \\ 
$2036.1503$ & $20.3$ & $4.7$ & HIDES-S  \\ 
$2041.2751$ & $5.7$ & $2.9$ & HIDES-S  \\ 
... & ... & ... & ... \\
\hline
\end{tabular}}
\begin{tabnote}
\hangindent6pt\noindent
\hbox to6pt{\footnotemark[$*$]\hss}\unskip% 
Only the first five sets of RV data are listed. A complete data listing including RV, FWHM, BIS, and S/N will be available online as supplementary after the publication.
\end{tabnote}
\label{RV:HD161178}
\end{table}

\begin{figure*}
 \begin{center}
 \includegraphics[width=16.0cm]{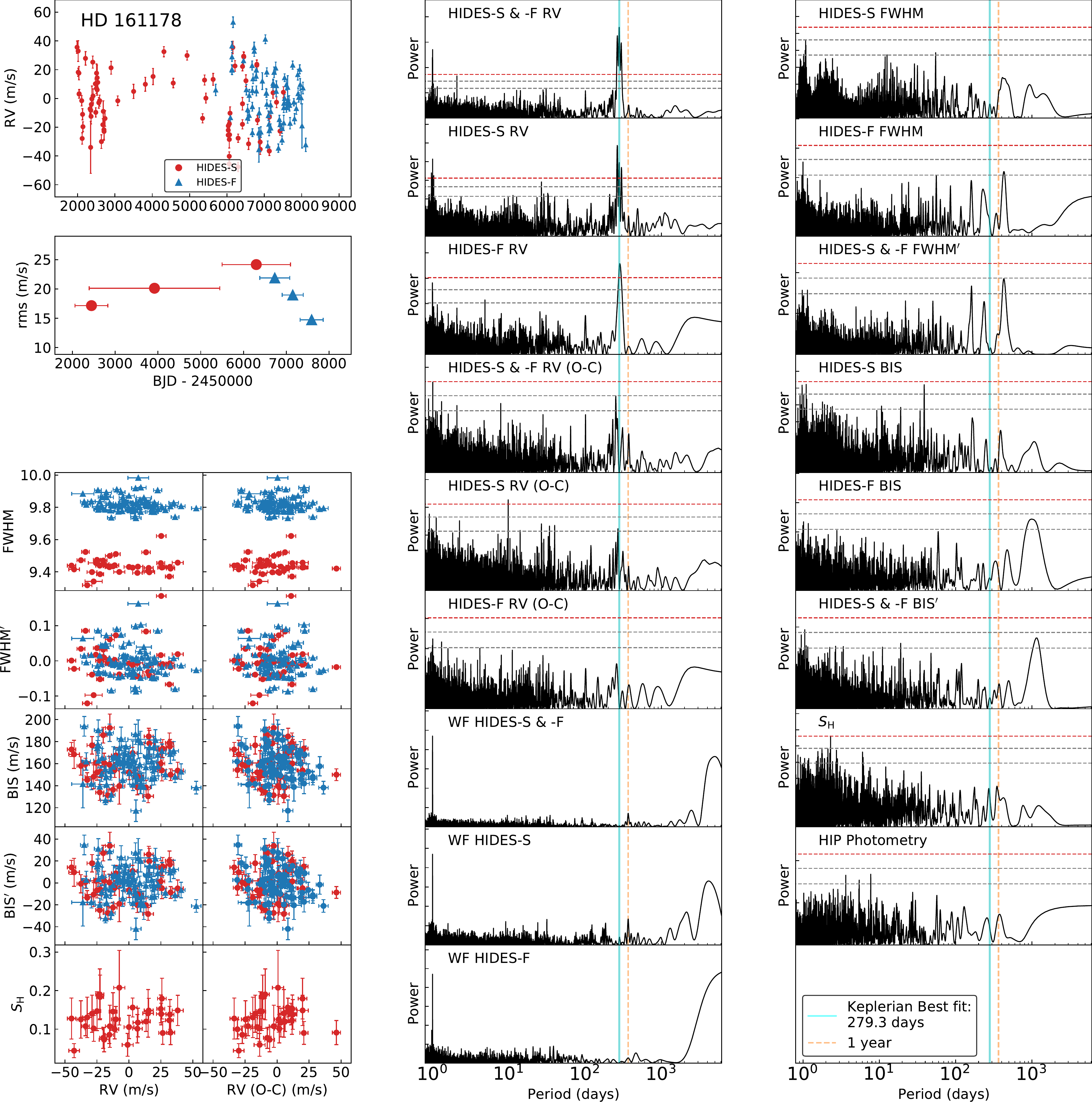}
 \end{center}
\caption{
Summary figure of HD 161178. 
Left column from top to bottom: 
RV time series; rms of RVs varying by time; FWHM, FWHM$^{\prime}$, BIS, BIS$^{\prime}$, and Ca \emissiontype{II} H index $S_{\rm{H}}$ respectively against RVs and RV residuals.
Middle column from top to bottom:
GLS periodogram of full RV data, HIDES-S RV data, HIDES-F RV data, full RV residuals, HIDES-S RV residuals, and HIDES-F RV residuals; window function of full RV data, HIDES-S data, and HIDES-F data.
Right column from top to bottom:
GLS periodogram of HIDES-S FWHM, HIDES-F FWHM, FWHM$^{\prime}$, HIDES-S BIS, HIDES-F BIS, BIS$^{\prime}$, Ca \emissiontype{II} H index $S_{\rm{H}}$, and Hipparcos photometry. In GLS periodograms, the horizontal lines represent 10\%, 1\%, and 0.1\% FAP levels from bottom to top. The vertical cyan solid line indicates the best-fitted period from the Keplerian model, and the vertical orange dashed line indicates 1 year. 
}\label{fig:HD161178_rv_all}
\end{figure*}

\begin{figure}
 \begin{center}
  \includegraphics[width=8.0cm]{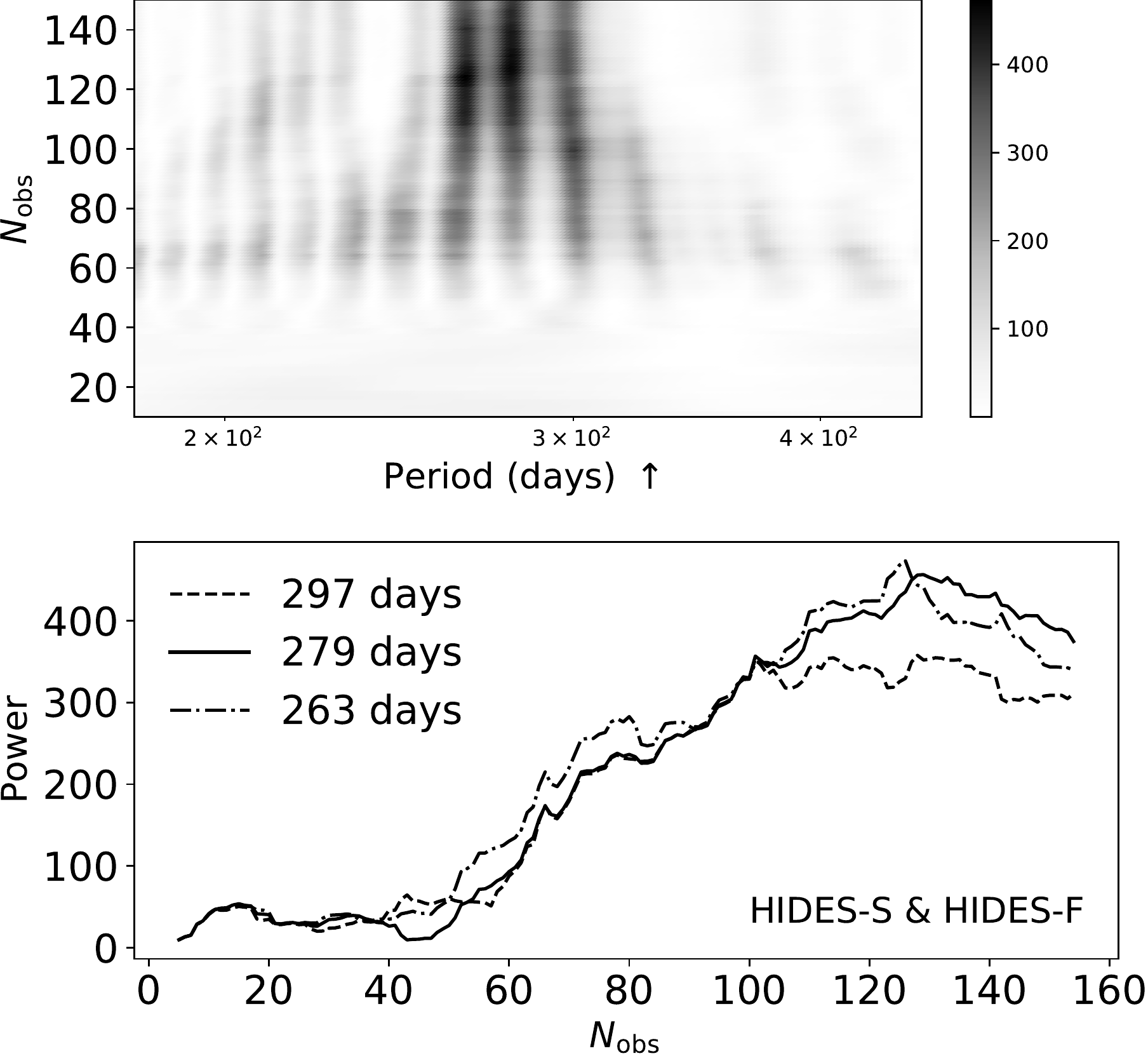} 
 \end{center}
\caption{
Stacked Bayesian Generalized Lomb-Scargle periodogram of the full data of HD 161178. The upper panel shows the power variation with time and the number of data points. The color bar illustrates the strength of power. The lower panel shows the power of interested periods varying with time. The legend of 297 days marked with a dashed line, the legend of 279 days marked with a solid line, and the legend of 263 days marked with dot-and-dash line respectively represent the interested periods of 297-day signal, 279-day signal, and 275-day signal in the GLS periodogram of RV data.
}\label{fig:HD161178_sbgls}
\end{figure}

\begin{figure}
 \begin{center}
  \includegraphics[width=8.0cm]{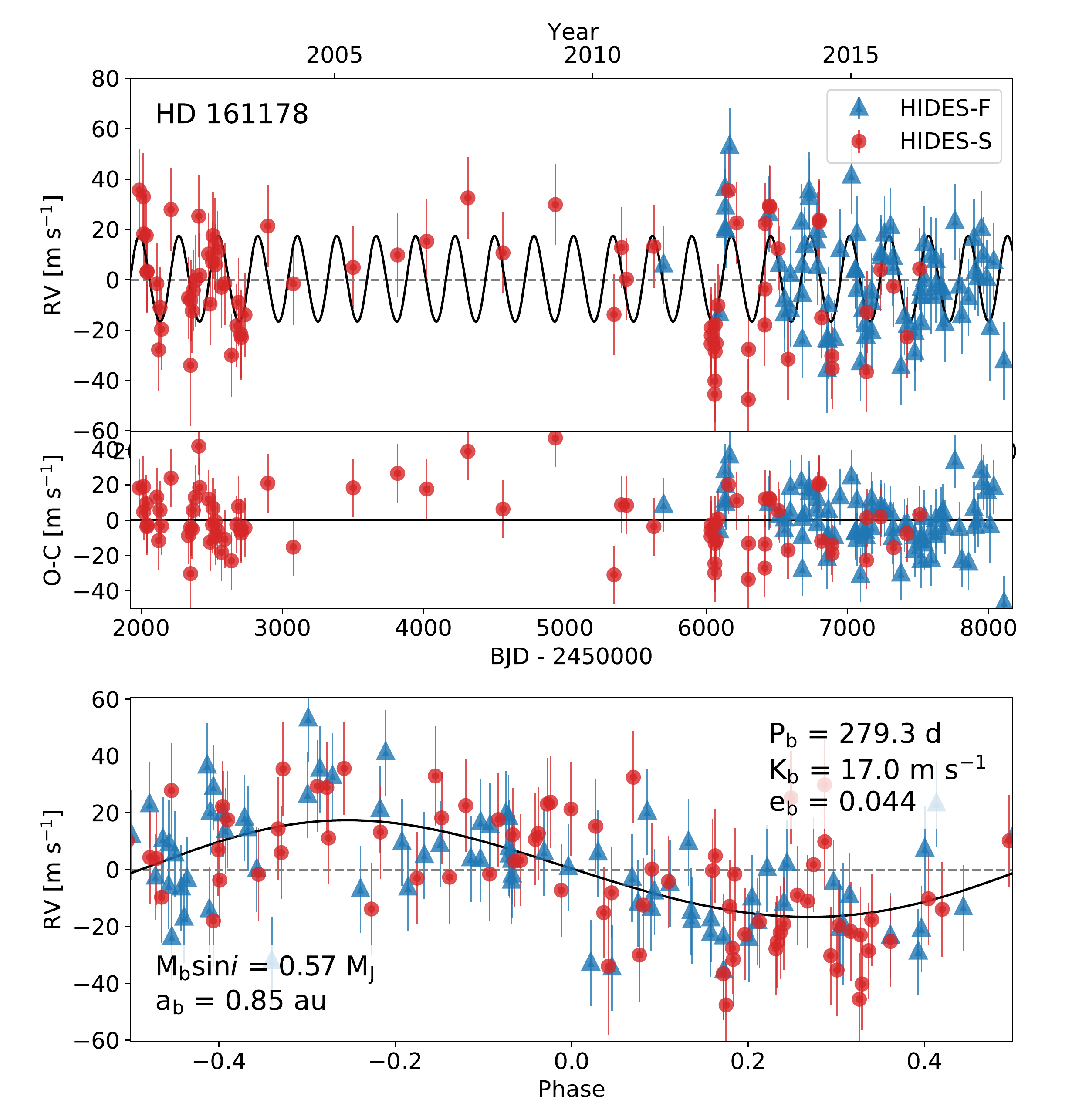} 
 \end{center}
\caption{Orbital solution of HD 161178. HIDES-S data are shown in red and HIDES-F data are shown in blue Top: Best fit single Keplerian curve in the full time span with fitted RV offsets between instruments and fitted jitters included in the errorbars. Mid: RV Residuals to the best fit. Bottom: phase-folded of Top panel.}\label{fig:HD161178_curve}
\end{figure}

This star is firstly reported in the result of the first three years Okayama Planet Search Program, yet no planet was detected at that time \citep{Sato2005}. 
This time, we collected a total of 154 data including 77 taken by HIDES-S and 77 taken by HIDES-F between 2001 March and 2017 December. The RV data are shown in Figure \ref{fig:HD161178_rv_all} and the data are listed in Table \ref{RV:HD161178}. The generalized Lomb-Scargle periodogram shows three significant peaks with a FAP of less than $0.1 \%$ including peaks at periods of 263, 279, and 297 days for full data. This is the aliasing effect by a strong window at around 6000 days. This window function is the result of sparse observation in the middle of the HIDES-S baseline. Since full RV data and HIDES-S RV data show different strongest signals, and HIDES-F RV data show only one peak in the periodogram,  therefore we performed SBGLS to investigate the signal in the neighboring period region. Consequently, the SBGLS periodogram shows that the power of 279 days becomes the strongest one after 125 observations (Figure \ref{fig:HD161178_sbgls}). Here, we also notice a slight power decreases for all these three peaks after about 125 observations. This could be probably introduced by jitters, and it calls for our attention in future follow-up observations. 
Furthermore, we did not find significant periodicity with GLS in line profiles and Ca \emissiontype{II} H index time series, and we did not find RV strongly correlating with either spectral line profiles or Ca \emissiontype{II} H index ($|r| < 0.2$). 

We adopted a single Keplerian curve to fit the data and obtained orbital parameters for the companion of $P = 279.3_{-0.8}^{+1.0} \ {\rm days}$, $K =17.0_{-0.8}^{+1.0}\ \rm{m\>s^{-1}}$, $e = 0.044_{-0.019}^{+0.132}$. The rms scatter of the residuals to the Keplerian fit is $16.0\ \rm{m\>s^{-1}}$. we did not find any significant periodicity in the residuals. In this case, it is confident to say that the period of 279 days, rather than 269 or 297 days, is the true orbital period, and we did not find line profiles or Ca \emissiontype{II} H index strongly correlating with RV residuals ($|r| < 0.2$).
By adopting a stellar mass of $M = 1.05\ M_{\odot}$, we obtained a minimum mass $M_{\rm{p}}\sin{i} = 0.57_{-0.16}^{+0.02}\  M_{\rm J}$ and a semimajor axis $a = 0.85_{-0.07}^{+0.05}\ {\rm au}$ for the companion. The phase-folded RV curve is shown in the lower panel in Figure \ref{fig:HD161178_curve}.
%%%%%%%%%%%%%%%%%%%%%%%%%%%%%%%%%%%%%%%%%%%%%%%%%%%%%%%%%%%%%%%%%%%%%%%%%%%%%%%%%%%%%%%%%%%%%%%%%%%%%%%%%%%%%%%%%%%%%%%%

%\newpage
%%%%%%%%%%%%%%%%%%%%%%%%%%%%%%%%%%%%%%%%%%%%%%%%%%%%%%%%%%%%%%%%%%%%%%%%%%%%%%%%%%%%%%%%%%%%%%%%%%%%%%%%%%%%%%%%%%%%%%%%
\subsection{HD 219139}
\begin{table}
\tbl{Radial Velocities of HD 219139 }{%
\begin{tabular}{lccc}
\hline\hline
BJD & Radial Velocity & Uncertainty & Observation \\
$(-2450000)$ & ($\rm{m\ s^{-1}}$) & ($\rm{m\ s^{-1}}$) & Mode \\
\hline
$3025.8918$ & $9.7$ & $4.5$ & HIDES-S  \\ 
$3245.2039$ & $8.6$ & $3.7$ & HIDES-S  \\ 
$3332.0253$ & $26.8$ & $3.4$ & HIDES-S  \\ 
$3582.2343$ & $-23.9$ & $6.7$ & HIDES-S  \\ 
$3656.1926$ & $-12.6$ & $3.6$ & HIDES-S  \\ 
... & ... & ... & ... \\
\hline
\end{tabular}}
\begin{tabnote}
\hangindent6pt\noindent
\hbox to6pt{\footnotemark[$*$]\hss}\unskip% 
Only the first five sets of RV data are listed. A complete data listing including RV, FWHM, BIS, and S/N will be available online as supplementary after the publication.
\end{tabnote}
\label{RV:HD219139}
\end{table}

\begin{figure*}
 \begin{center}
 \includegraphics[width=16.0cm]{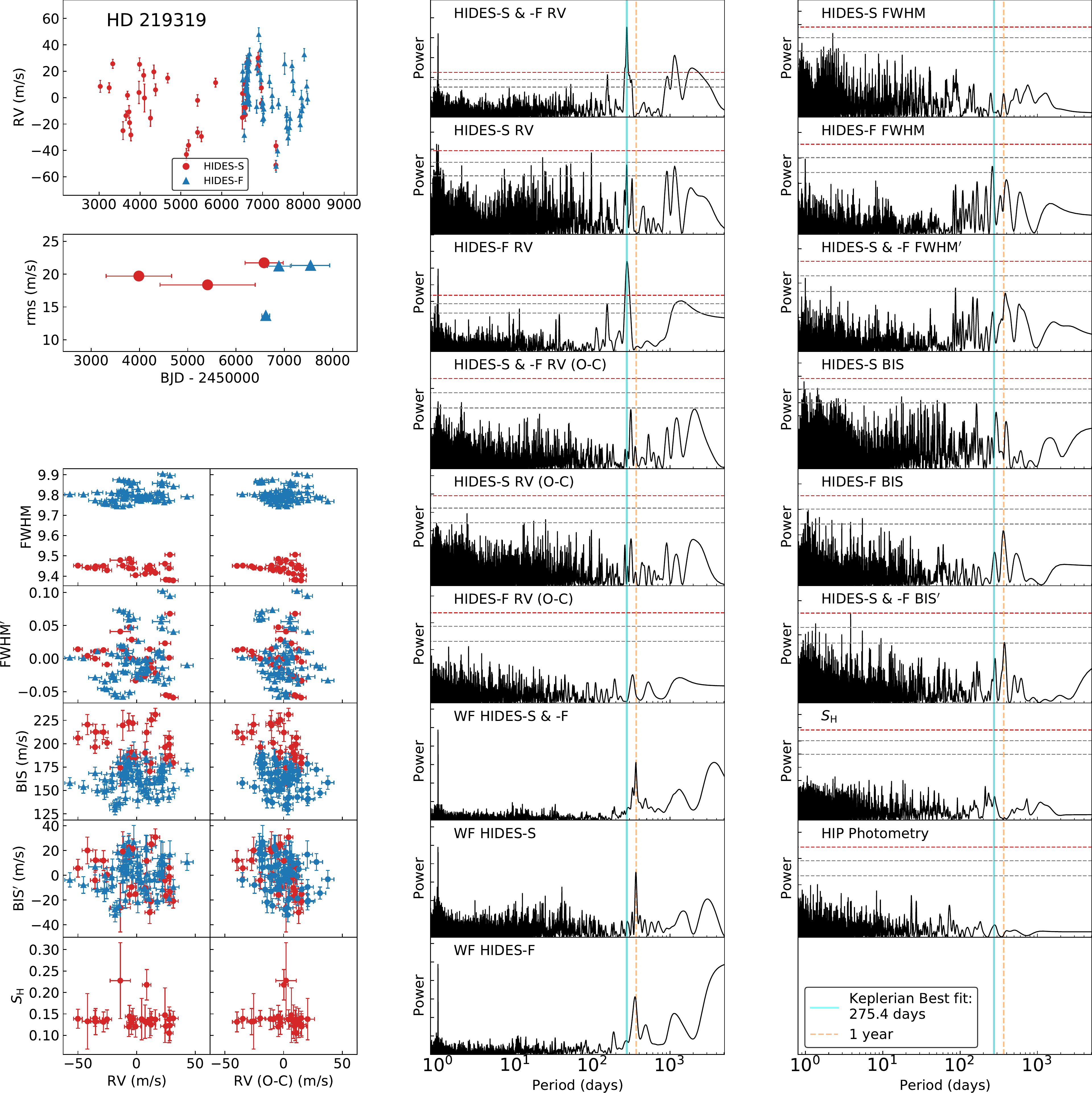}
 \end{center}
\caption{
Summary figure of HD219139. 
Left column from top to bottom: 
RV time series; rms of RVs varying by time; FWHM, FWHM$^{\prime}$, BIS, BIS$^{\prime}$, and Ca \emissiontype{II} H index $S_{\rm{H}}$ respectively against RVs and RV residuals.
Middle column from top to bottom:
GLS periodogram of full RV data, HIDES-S RV data, HIDES-F RV data, full RV residuals, HIDES-S RV residuals, and HIDES-F RV residuals; window function of full RV data, HIDES-S data, and HIDES-F data.
Right column from top to bottom:
GLS periodogram of HIDES-S FWHM, HIDES-F FWHM, FWHM$^{\prime}$, HIDES-S BIS, HIDES-F BIS, BIS$^{\prime}$, Ca \emissiontype{II} H index $S_{\rm{H}}$, and Hipparcos photometry. In GLS periodograms, the horizontal lines represent 10\%, 1\%, and 0.1\% FAP level from bottom to top. The vertical cyan solid line indicates the best-fitted period from the Keplerian model, and the vertical orange dashed line indicates 1 year. 
}\label{fig:HD219139_rv_all}
\end{figure*}

\begin{figure}
 \begin{center}
  \includegraphics[width=8.0cm]{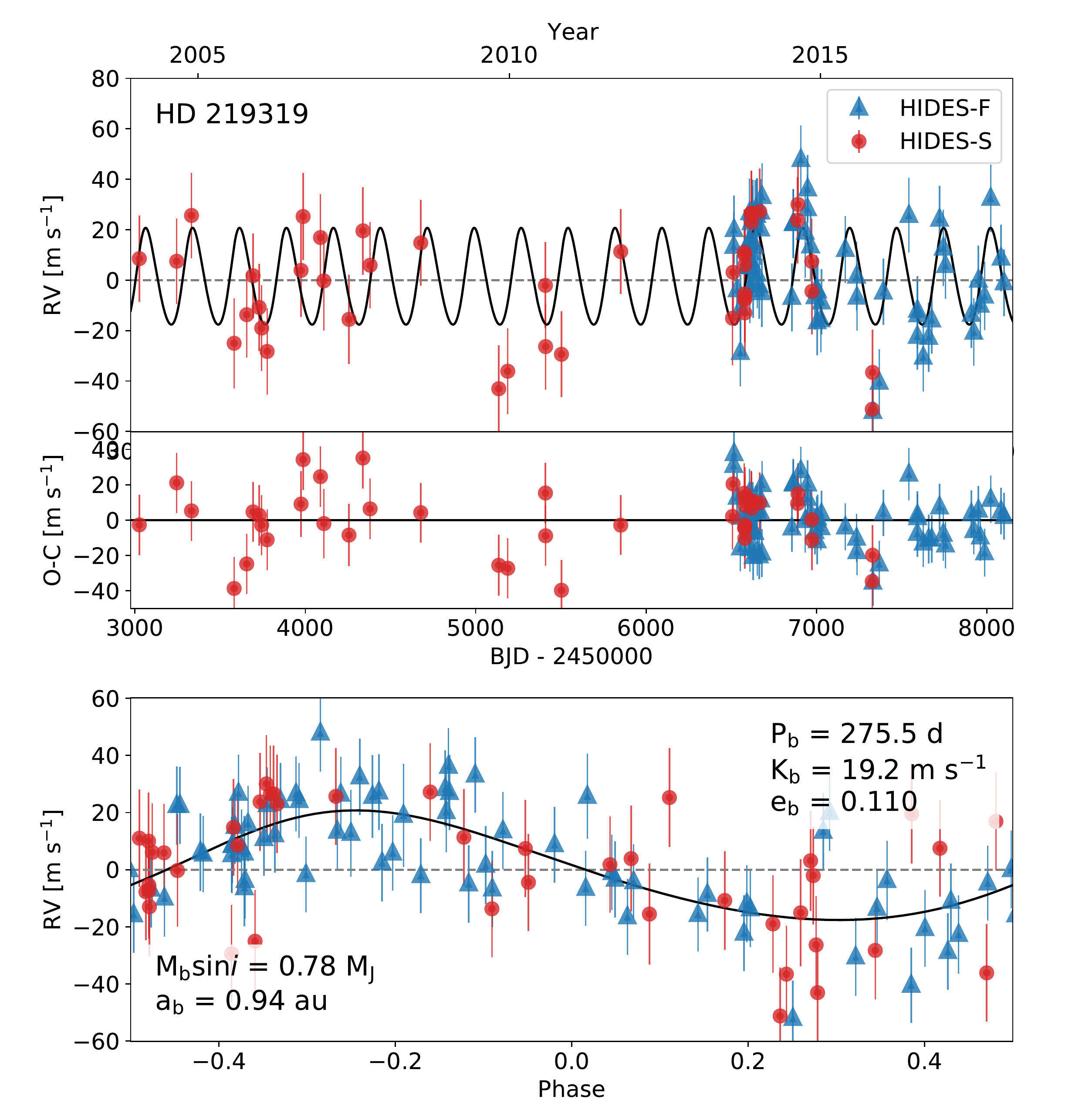} 
 \end{center}
\caption{Orbital solution of HD 219139. HIDES-S data are shown in red and HIDES-F data are shown in blue Top: Best fit single Keplerian curve in the full time span with fitted RV offsets between instruments and fitted jitters included in the errorbars. Mid: RV Residuals to the best fit. Bottom: phase-folded of Top panel.}\label{fig:HD219139_curve}
\end{figure}

We collected a total of 111 data including 42 taken by HIDES-S and 69 taken by HIDES-F between 2004 January and 2017 December. The RV data are shown in Figure \ref{fig:HD219139_rv_all} and the data are listed in Table \ref{RV:HD219139}. The GLS shows a significant peak at the period of 276 days with a FAP of less than $0.1 \%$, and another significant signal locates at 1153 days. Recognizing the relations of ($1/1153 \simeq 1/276 - 1/365$), we are aware that the 1153-day signal is the alias caused by the 1-year window function. For further investigation of periodicity, we did not find significant periodicity with GLS in line profiles or Ca \emissiontype{II} H index time series, and we did not find RV strongly correlating with either spectral line profiles or Ca \emissiontype{II} H index ($|r| < 0.3$). 

We adopted a single Keplerian curve to fit the data and obtained orbital parameters for the companion of $P = 275.5_{-1.0}^{+2.3} \ {\rm days}$, $K =19.2_{-2.6}^{+1.8} \ \rm{m\>s^{-1}}$, $e = 0.110_{-0.081}^{+0.087} $. The rms scatter of the residuals to the Keplerian fit is $15.1 \ \rm{m\>s^{-1}}$. We did not find any significant periodicity in the residuals, and we did not find line profiles or Ca \emissiontype{II} H index strongly correlating with RV residuals ($|r| < 0.5$). Adopting a stellar mass of $M = 1.46\ M_{\odot}$, we obtained a minimum mass $M_{\rm{p}}\sin{i} = 0.78_{-0.20}^{+0.05}\  M_{\rm J}$ and a semimajor axis $a = 0.94_{-0.05}^{+0.06}\ {\rm au}$ for the companion. Besides, we present RV scatters varying with time in Figure \ref{fig:HD219139_rv_all}. 
Combining with the result from Keplerian fitting, we know that the data included by the first scatter measurement of HIDES-F is mainly located on the first half in the orbital phase (phase between -0.5 and 0 in the lower panel in Figure \ref{fig:HD219139_curve}). For these other scatter measurements, by considering solar-like oscillations and instrumental jitter, we believe the scatter of rms is acceptable as these rms measurements locate around 20 $\rm{m\>s^{-1}}$.
%%%%%%%%%%%%%%%%%%%%%%%%%%%%%%%%%%%%%%%%%%%%%%%%%%%%%%%%%%%%%%%%%%%%%%%%%%%%%%%%%%%%%%%%%%%%%%%%%%%%%%%%%%%%%%%%%%%%%%%%

%\newpage
%%%%%%%%%%%%%%%%%%%%%%%%%%%%%%%%%%%%%%%%%%%%%%%%%%%%%%%%%%%%%%%%%%%%%%%%%%%%%%%%%%%%%%%%%%%%%%%%%%%%%%%%%%%%%%%%%%%%%%%%
\subsection{$\gamma$ Psc}
\begin{table}
\tbl{Radial Velocities of $\gamma$ Psc}{%
\begin{tabular}{lccc}
\hline\hline
BJD & Radial Velocity & Uncertainty & Observation \\
$(-2450000)$ & ($\rm{m\ s^{-1}}$) & ($\rm{m\ s^{-1}}$) & Mode \\
\hline
$2489.2149$ & $-1.4$ & $4.9$ & HIDES-S  \\ 
$2509.1015$ & $-2.8$ & $4.4$ & HIDES-S  \\ 
$2541.1795$ & $12.2$ & $4.3$ & HIDES-S  \\ 
$2570.0815$ & $-2.5$ & $4.2$ & HIDES-S  \\ 
$2652.9006$ & $-20.8$ & $4.0$ & HIDES-S  \\ 
... & ... & ... & ... \\
\hline
\end{tabular}}
\begin{tabnote}
\hangindent6pt\noindent
\hbox to6pt{\footnotemark[$*$]\hss}\unskip% 
Only the first five sets of RV data are listed. A complete data listing including RV, FWHM, BIS, and S/N will be available online as supplementary after the publication.
\end{tabnote}
\label{RV:HD219615}
\end{table}

\begin{figure*}
 \begin{center}
 \includegraphics[width=16.0cm]{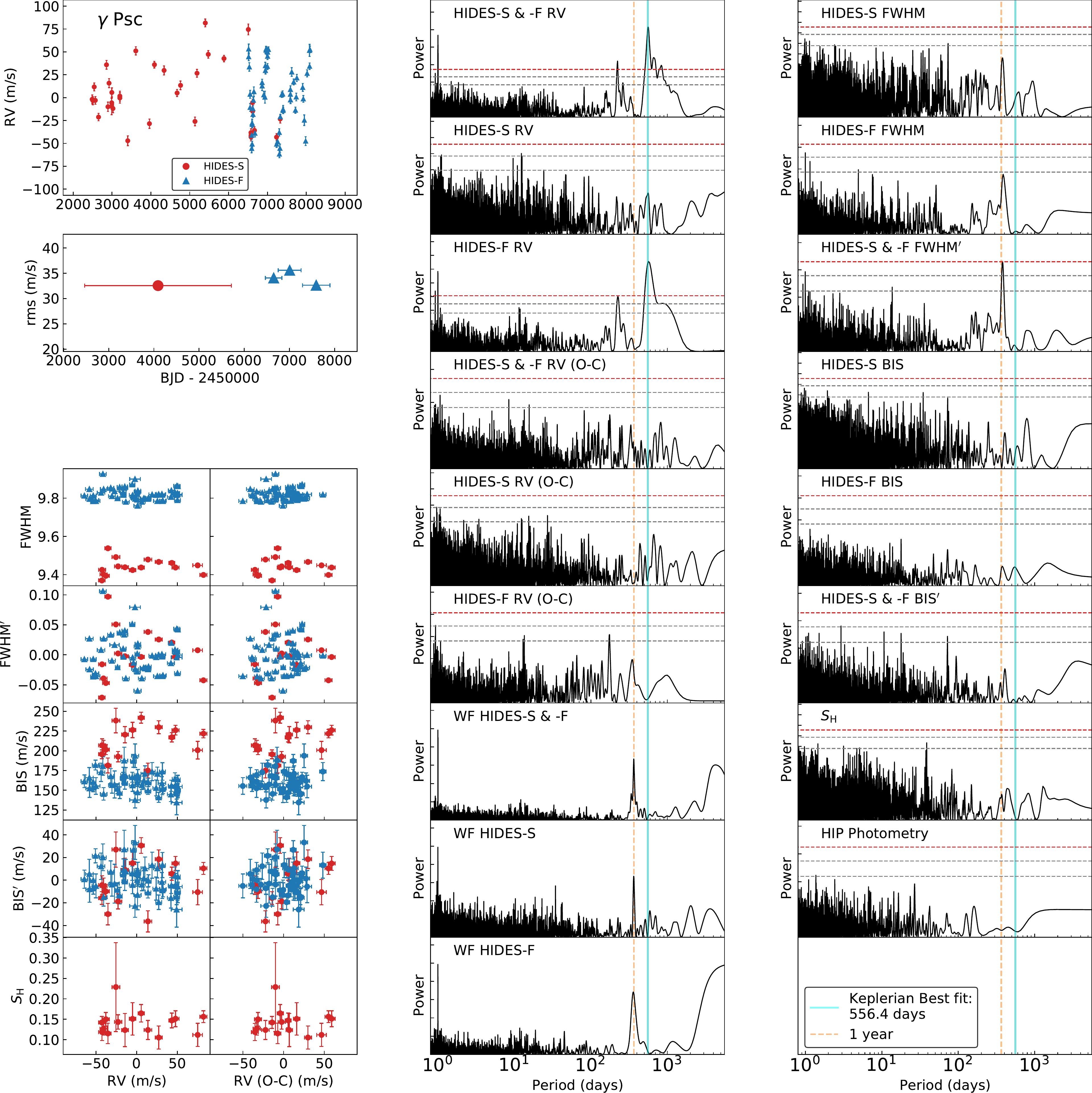}
 \end{center}
\caption{
Summary figure of $\gamma$ Psc. 
Left column from top to bottom: 
RV time series; rms of RVs varying by time; FWHM, FWHM$^{\prime}$, BIS, BIS$^{\prime}$, and Ca \emissiontype{II} H index $S_{\rm{H}}$ respectively against RVs and RV residuals.
Middle column from top to bottom:
GLS periodogram of full RV data, HIDES-S RV data, HIDES-F RV data, full RV residuals, HIDES-S RV residuals, and HIDES-F RV residuals; window function of full RV data, HIDES-S data, and HIDES-F data.
Right column from top to bottom:
GLS periodogram of HIDES-S FWHM, HIDES-F FWHM, FWHM$^{\prime}$, HIDES-S BIS, HIDES-F BIS, BIS$^{\prime}$, Ca \emissiontype{II} H index $S_{\rm{H}}$, and Hipparcos photometry. In GLS periodograms, the horizontal lines represent 10\%, 1\%, and 0.1\% FAP level from bottom to top. The vertical cyan solid line indicates the best-fitted period from the Keplerian model, and the vertical orange dashed line indicates 1 year.   

}\label{fig:HD219615_rv_all}
\end{figure*}

\begin{figure}
 \begin{center}
  \includegraphics[width=8.0cm]{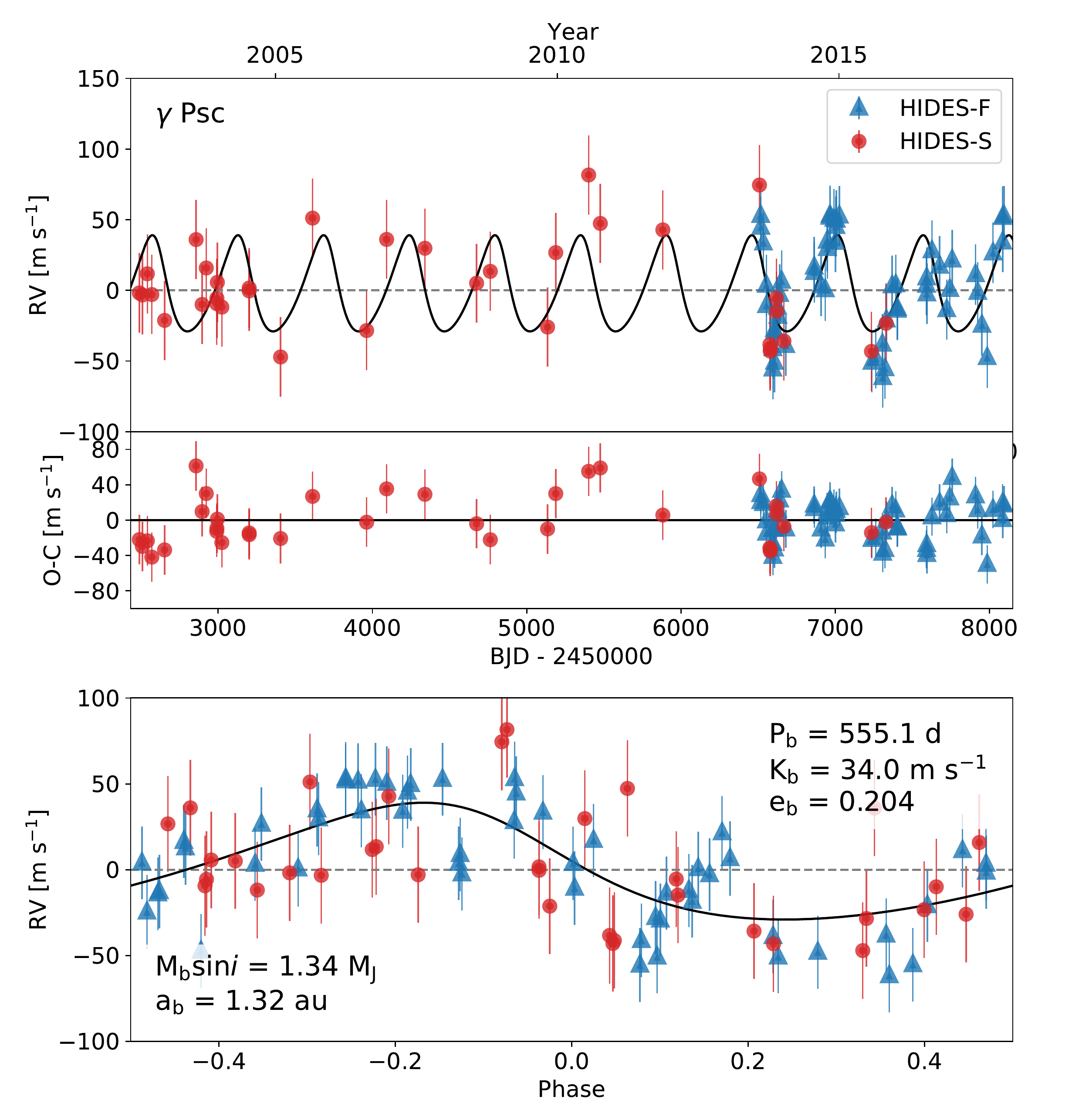} 
 \end{center}
\caption{Orbital solution of $\gamma$ Psc. HIDES-S data are shown in red and HIDES-F data are shown in blue. Top: Best fit single Keplerian curve in the full time span with fitted RV offsets between instruments and fitted jitters included in the errorbars. Mid: RV Residuals to the best fit. Bottom: phase-folded of Top panel.}\label{fig:HD219615_curve}
\end{figure}

We collected a total of 89 data including 34 taken by HIDES-S and 55 taken by HIDES-F between 2002 February and 2017 December. The RV data are shown in Figure \ref{fig:HD219615_rv_all} and the data are listed in Table \ref{RV:HD219615}. The GLS shows a significant peak at the period of 561 days with a FAP of less than $0.1 \%$, and another significant signal locates at 224 days. Recognizing the relations of ($1/224 \simeq 1/561 + 1/365$), we know that the 224-day signal is the alias caused by a 1-year window function. For further investigation of periodicity, we did not find significant periodicity with GLS in line profiles and Ca \emissiontype{II} H index time series, and we did not find RV strongly correlating with either spectral line profiles or Ca \emissiontype{II} H index ($|r| < 0.3$). 

We adopted a single Keplerian curve to fit the data and obtained orbital parameters for the companion of $P = 555.1_{-2.5}^{+6.0} \ {\rm days}$, $K =34.0_{-5.0}^{+3.9} \ \rm{m\>s^{-1}}$, $e = 0.204_{-0.141}^{+0.114} $. The rms scatter of the residuals to the Keplerian fit is $24.6 \ \rm{m\>s^{-1}}$. We did not find any significant periodicity in the residuals, and we did not find line profiles or Ca \emissiontype{II} H index strongly correlating with RV residuals ($|r| < 0.4$). Adopting a stellar mass of $M = 0.99\ M_{\odot}$, we obtained a minimum mass $M_{\rm{p}}\sin{i} = 1.34_{-0.31}^{+0.02}\  M_{\rm J}$ and a semimajor axis $a = 1.32_{-0.08}^{+0.05}\ {\rm au}$ for the companion. The phase-folded RV curve is shown in the lower panel in Figure \ref{fig:HD219615_curve}.
%%%%%%%%%%%%%%%%%%%%%%%%%%%%%%%%%%%%%%%%%%%


\begin{thebibliography}{}
% Journals(e.g. A\&A,ApJ,AJ,NMRAS,PASP ...)
% Authors, Year, Journal, Vol#, Page#
% Journal Title Abbreviation >> http://www.asj.or.jp/pasj/Jabb.html
\bibitem[Adam{\'o}w et al.(2018)]{Adamow2018} 
Adam{\'o}w, M., Niedzielski, A., Kowalik, K., et al.\ 2018, \aap, 613, A47.
\bibitem[Adam{\'o}w et al.(2012)]{Adamow2012} 
Adam{\'o}w, M., Niedzielski, A., Villaver, E., Nowak, G., \& Wolszczan, A.\ 2012, \apjl, 754, L15.
\bibitem[Alibert et al.(2011)]{Alibert2011} 
Alibert, Y., Mordasini, C., \& Benz, W.\ 2011, \aap, 526, A63.
\bibitem[Andrews et al.(2013)]{Andrews2013} 
Andrews, S.~M., Rosenfeld, K.~A., Kraus, A.~L., et al.\ 2013, \apj, 771, 129. 
%\bibitem[Aoki \& Hełminiak (2014)]{Aoki2014}
%Aoki, W., \& Hełminiak, K.\ Data reduction of echelle spectra with IRAF version 1.2
\bibitem[Arenou, Grenon, \& Gomez(1992)]{Arenou1992} 
Arenou, F., Grenon, M., \& Gomez, A.\ 1992, \aap, 258, 104.
\bibitem[Baluev(2008)]{Baluev2008} 
Baluev, R.~V.\ 2008, \mnras, 385, 1279.
\bibitem[Baranne et al.(1996)]{Baranne1996} 
Baranne, A., Queloz, D., Mayor, M., et al.\ 1996, \aaps, 119, 373
\bibitem[Barclay, Pepper, \& Quintana(2018)]{Barclay2018} 
Barclay, T., Pepper, J., \& Quintana, E.~V.\ 2018, \apjs, 239, 2.
\bibitem[Berger et al.(2020)]{Berger2020} 
Berger, T.~A., Huber, D., van Saders, J.~L., et al.\ 2020, \aj, 159, 280. 
\bibitem[Bovy et al.(2016)]{Bovy2016} 
\bibitem[Borucki et al.(2012)]{Borucki2012} 
Borucki, W.~J., Koch, D.~G., Batalha, N., et al.\ 2012, \apj, 745, 120. 
Bovy, J., Rix, H.-W., Green, G.~M., et al.\ 2016, \apj, 818, 130.
\bibitem[Butler et al.(1996)]{Butler1996} 
Butler, R.~P., Marcy, G.~W., Williams, E., et al.\ 1996, \pasp, 108, 500.
\bibitem[Cameron(1978)]{Cameron1978} 
Cameron, A.~G.~W.\ 1978, Moon and Planets, 18, 5.
\bibitem[Choi et al.(2016)]{Choi2016} 
Choi, J., Dotter, A., Conroy, C., et al.\ 2016, The Astrophysical Journal, 823, 102.
\bibitem[Conroy et al.(in prep.)]{ConroyInPrep} 
Conroy et al.\ In preparation.
\bibitem[Dall et al.(2006)]{Dall2006} 
Dall, T.~H., Santos, N.~C., Arentoft, T., et al.\ 2006, \aap, 454, 341. doi:10.1051/0004-6361:20065021
\bibitem[Dotter(2016)]{Dotter2016} 
Dotter, A.\ 2016, The Astrophysical Journal Supplement Series, 222, 8.
\bibitem[D{\"o}llinger et al.(2007)]{Dollinger2007} 
D{\"o}llinger, M.~P., Hatzes, A.~P., Pasquini, L., et al.\ 2007, \aap, 472, 649.
\bibitem[D{\"o}llinger \& Hartmann(2021)]{Dollinger2021} 
D{\"o}llinger, M.~P. \& Hartmann, M.\ 2021, \apjs, 256, 10. 
\bibitem[Dumusque et al.(2011)]{Dumusque2011} 
Dumusque, X., Udry, S., Lovis, C., Santos, N.~C., \& Monteiro, M.~J.~P.~F.~G.\ 2011, \aap, 525, A140.
\bibitem[Duncan et al.(1991)]{Duncan1991} 
Duncan, D.~K., Vaughan, A.~H., Wilson, O.~C., et al.\ 1991, \apjs, 76, 383.
\bibitem[Eastman et al.(2010)]{Eastman2010} 
Eastman, J., Siverd, R., \& Gaudi, B.~S.\ 2010, \pasp, 122, 935. 
\bibitem[ESA(1997)]{ESA1997} 
ESA\ 1997, ESA Special Publication, 1200
\bibitem[Farr et al.(2018)]{Farr2018} 
Farr, W.~M., Pope, B.~J.~S., Davies, G.~R., et al.\ 2018, \apjl, 865, L20.
\bibitem[Fischer \& Valenti(2005)]{Fischer2005} 
Fischer, D.~A., \& Valenti, J.\ 2005, \apj, 622, 1102
\bibitem[Ford(2006)]{Ford2006} 
Ford, E.~B.\ 2006, \apj, 642, 505.
\bibitem[Foreman-Mackey, Agol, Ambikasaran, \& Angus(2017)]{Foreman-Mackey2017} 
Foreman-Mackey, D., Agol, E., Ambikasaran, S., \& Angus, R.\ 2017, \aj, 154, 220.
\bibitem[Foreman-Mackey, Hogg, Lang, \& Goodman(2013)]{Foreman-Mackey2013} 
Foreman-Mackey, D., Hogg, D.~W., Lang, D., \& Goodman, J.\ 2013, \pasp, 125, 306.
\bibitem[Frink et al.(2001)]{Frink2001} 
Frink, S., Quirrenbach, A., Fischer, D., R{\"o}ser, S., \& Schilbach, E.\ 2001, \pasp, 113, 173.
\bibitem[Fulton \& Petigura(2017)]{Fulton2017} 
Fulton, B., \& Petigura, E.\ 2017, doi:10.5281/zenodo.580821.
\bibitem[Fulton, Petigura, Blunt, \& Sinukoff(2018)]{Fulton2018} 
Fulton, B.~J., Petigura, E.~A., Blunt, S., \& Sinukoff, E.\ 2018, \pasp, 130, 044504.
\bibitem[Gaia Collaboration et al.(2016)]{Gaia2016}
Gaia Collaboration, Prusti, T., de Bruijne, J.~H.~J., et al.\ 2016, \aap, 595, A1.
\bibitem[Gaia Collaboration et al.(2021)]{Gaia2021}
Gaia Collaboration, Brown, A.~G.~A., Vallenari, A., et al.\ 2021, \aap, 649, A1.
\bibitem[Gelman et al.(2013)]{Gelman2013}
Gelman, A. et al.\ 2013, Bayesian Data Analysis, Third Edition (Chapman and Hall)
\bibitem[Gettel et al.(2012)]{Gettel2012} 
Gettel, S., Wolszczan, A., Niedzielski, A., et al.\ 2012, \apj, 745, 28.
\bibitem[Gray \& Corbally(1994)]{Gray1994} 
Gray, R.~O. \& Corbally, C.~J.\ 1994, \aj, 107, 742. 
\bibitem[Green et al.(2019)]{Green2019} 
Green, G.~M., Schlafly, E., Zucker, C., et al.\ 2019, \apj, 887, 93. 
\bibitem[Goldreich \& Ward(1973)]{Goldreich1973} 
Goldreich, P. \& Ward, W.~R.\ 1973, \apj, 183, 1051. 
\bibitem[Grunblatt et al.(2017)]{Grunblatt2017} 
Grunblatt, S.~K., Huber, D., Gaidos, E., et al.\ 2017, \aj, 154, 254.
\bibitem[Grunblatt, Howard, \& Haywood(2016)]{Grunblatt2016} 
Grunblatt, S.~K., Howard, A., \& Haywood, R.\ 2016, American Astronomical Society Meeting Abstracts \#227, 227, 122.01.
\bibitem[Haisch et al.(2001a)]{Haisch2001a} 
Haisch, K.~E., Lada, E.~A., \& Lada, C.~J.\ 2001, \apjl, 553, L153.
\bibitem[Haisch et al.(2001b)]{Haisch2001b} 
Haisch, K.~E., Lada, E.~A., \& Lada, C.~J.\ 2001, \aj, 121, 2065. 
\bibitem[Han et al.(2010)]{Han2010} 
Han, I., Lee, B.~C., Kim, K.~M., et al.\ 2010, \aap, 509, A24.
\bibitem[Hatzes et al.(2005)]{Hatzes2005} 
Hatzes, A.~P., Guenther, E.~W., Endl, M., et al.\ 2005, \aap, 437, 743.
\bibitem[Hatzes et al.(2018)]{Hatzes2018} 
Hatzes, A.~P., Endl, M., Cochran, W.~D., et al.\ 2018, \aj, 155, 120.
\bibitem[Hayashi et al.(1985)]{Hayashi1985} 
Hayashi, C., Nakazawa, K., \& Nakagawa, Y.\ 1985, Protostars and Planets II, 1100
\bibitem[Heeren et al.(2021)]{Heeren2021} 
Heeren, P., Reffert, S., Trifonov, T., et al.\ 2021, arXiv:2102.01999
\bibitem[Hekker et al.(2006)]{Hekker2006} 
Hekker, S., Reffert, S., Quirrenbach, A., et al.\ 2006, \aap, 454, 943.
\bibitem[Henry et al.(2000)]{Henry2000} 
Henry, G.~W., Fekel, F.~C., Henry, S.~M., et al.\ 2000, \apjs, 130, 201.
\bibitem[Huber et al.(2013)]{Huber2013} 
Huber, D., Carter, J.~A., Barbieri, M., et al.\ 2013, Science, 342, 331.
\bibitem[Huber et al.(2017)]{Huber2017} 
Huber, D., Zinn, J., Bojsen-Hansen, M., et al.\ 2017, The Astrophysical Journal, 844, 102.
\bibitem[Izumiura(1999)]{Izumiura1999} 
Izumiura, H.\ 1999, Observational Astrophysics in Asia and its Future, 77 
\bibitem[Izumiura(2005)]{Izumiura2005} 
Izumiura, H.\ 2005, Journal of Korean Astronomical Society, 38, 81 
\bibitem[Jofr{\'e} et al.(2015)]{Jofre2015} 
Jofr{\'e}, E., Petrucci, R., Saffe, C., et al.\ 2015, \aap, 574, A50.
\bibitem[Johnson et al.(2006)]{Johnson2006} 
Johnson, J.~A., Marcy, G.~W., Fischer, D.~A., et al.\ 2006, \apj, 652, 1724.
\bibitem[Johnson et al.(2007)]{Johnson2007} 
Johnson, J.~A., Fischer, D.~A., Marcy, G.~W., et al.\ 2007, \apj, 665, 785. 
\bibitem[Johnson et al.(2010)]{Johnson2010} 
Johnson, J.~A., Bowler, B.~P., Howard, A.~W., et al.\ 2010, \apjl, 721, L153.
\bibitem[Jones et al.(2006)]{Jones2006} 
Jones, H.~R.~A., Butler, R.~P., Tinney, C.~G., et al.\ 2006, \mnras, 369, 249.
\bibitem[Jones, Jenkins, Rojo, \& Melo(2011)]{Jones2011} 
Jones, M.~I., Jenkins, J.~S., Rojo, P., \& Melo, C.~H.~F.\ 2011, \aap, 536, A71.
\bibitem[Jones et al.(2014)]{Jones2014} 
Jones, M.~I., Jenkins, J.~S., Bluhm, P., Rojo, P., \& Melo, C.~H.~F.\ 2014, \aap, 566, A113.
\bibitem[Jones et al.(2016)]{Jones2016} 
Jones, M.~I., Jenkins, J.~S., Brahm, R., et al.\ 2016, \aap, 590, A38.
\bibitem[Kambe et al.(2013)]{Kambe2013} 
Kambe, E., Yoshida, M., Izumiura, H., et al.\ 2013, \pasj, 65, 15.
\bibitem[Kass \& Raftery (1996)]{Kass1996}
Kass, Robert E. \& Raftery, Adrian E.\ 1996, Journal of the American Statistical Association, 90, 430, 773.
\bibitem[Kolmogorov(1933)]{Kolmogorov1933}
Kolmogorov, A.\ 1933, Giornale dell’Istituto Italiano degli Attuari, 4: 83–91.
\bibitem[Kozai(1962)]{Kozai1962} 
Kozai, Y.\ 1962, \aj, 67, 591.
\bibitem[Kuiper(1951)]{Kuiper1951} Kuiper, G.~P.\ 1951, Proceedings of the National Academy of Science, 37, 1. doi:10.1073/pnas.37.1.1
\bibitem[Lagrange et al.(2009)]{Lagrange2009} 
Lagrange, A.-M., Desort, M., Galland, F., Udry, S., \& Mayor, M.\ 2009, \aap, 495, 335 
\bibitem[Lidov(1962)]{Lidov1962} 
Lidov, M.~L.\ 1962, \planss, 9, 719.
\bibitem[Lillo-Box et al.(2014)]{Lillo-Box2014} 
Lillo-Box, J., Barrado, D., Moya, A., et al.\ 2014, \aap, 562, A109.
\bibitem[Lillo-Box et al.(2016)]{Lillo-Box2016} 
Lillo-Box, J., Barrado, D., \& Correia, A.~C.~M.\ 2016, \aap, 589, A124.
\bibitem[Lindegren et al.(2021)]{Lindegren2021} 
Lindegren, L., Klioner, S.~A., Hern{\'a}ndez, J., et al.\ 2021, \aap, 649, A2. 
\bibitem[Liu et al.(2014)]{Liu2014} 
Liu, Y.~J., Tan, K.~F., Wang, L., et al.\ 2014, \apj, 785, 94.
\bibitem[Mayor \& Udry(2008)]{Mayor2008} 
Mayor, M., \& Udry, S.\ 2008, Physica Scripta Volume T, 130, 014010.
\bibitem[Mason et al.(2001)]{Mason2001} 
Mason, B.~D., Wycoff, G.~L., Hartkopf, W.~I., et al.\ 2001, \aj, 122, 3466. 
\bibitem[Medina et al.(2018)]{Medina2018} 
Medina, A.~A., Johnson, J.~A., Eastman, J.~D., \& Cargile, P.~A.\ 2018, \apj, 867, 32.
\bibitem[Mortier et al.(2015)]{Mortier2015} 
Mortier, A., Faria, J.~P., Correia, C.~M., et al.\ 2015, \aap, 573, A101.
\bibitem[Mortier \& Collier Cameron(2017)]{Mortier2017} 
Mortier, A., \& Collier Cameron, A.\ 2017, \aap, 601, A110.
\bibitem[Narang et al.(2018)]{Narang2018} 
Narang, M., Manoj, P., Furlan, E., et al.\ 2018, \aj, 156, 221.
\bibitem[Nelder \& Mead(1965)]{Nelder1965}
J. A. Nelder, R. Mead, A Simplex Method for Function Minimization, The Computer Journal, Volume 7, Issue 4, January 1965, Pages 308–313,
\bibitem[Nicholls et al.(2009)]{Nicholls2009} 
Nicholls, C.~P., Wood, P.~R., Cioni, M.-R.~L., et al.\ 2009, \mnras, 399, 2063. 
\bibitem[Niedzielski et al.(2009)]{Niedzielski2009} 
Niedzielski, A., Nowak, G., Adam{\'o}w, M., et al.\ 2009, \apj, 707, 768.
\bibitem[Niedzielski et al.(2015)]{Niedzielski2015} 
Niedzielski, A., Villaver, E., Wolszczan, A., et al.\ 2015, \aap, 573, A36.
\bibitem[Niedzielski et al.(2007)]{Niedzielski2007} 
Niedzielski, A., Konacki, M., Wolszczan, A., et al.\ 2007, \apj, 669, 1354.
\bibitem[Paxton et al.(2011)]{Paxton2011} 
Paxton, B., Bildsten, L., Dotter, A., et al.\ 2011, \apjs, 192, 3.
\bibitem[Paxton et al.(2013)]{Paxton2013} 
Paxton, B., Cantiello, M., Arras, P., et al.\ 2013, \apjs, 208, 4. 
\bibitem[Paxton et al.(2015)]{Paxton2015} 
Paxton, B., Marchant, P., Schwab, J., et al.\ 2015, \apjs, 220, 15. 
\bibitem[Paxton et al.(2018)]{Paxton2018} 
Paxton, B., Schwab, J., Bauer, E.~B., et al.\ 2018, \apjs, 234, 34. 
\bibitem[Pepe et al.(2002)]{Pepe2002} 
Pepe, F., Mayor, M., Galland, F., et al.\ 2002, \aap, 388, 632.
\bibitem[Tala Pinto et al.(2020)]{Pinto2020} 
Tala Pinto, M., Reffert, S., Quirrenbach, A., et al.\ 2020, \aap, 644, A1. 
\bibitem[Pollack et al.(1996)]{Pollack1996} 
Pollack, J.~B., Hubickyj, O., Bodenheimer, P., et al.\ 1996, Icarus, 124, 62.
\bibitem[Reffert et al.(2015)]{Reffert2015} 
Reffert, S., Bergmann, C., Quirrenbach, A., Trifonov, T., \& K{\"u}nstler, A.\ 2015, \aap, 574, A116.
\bibitem[Reichert et al.(2019)]{Reichert2019} 
Reichert, K., Reffert, S., Stock, S., et al.\ 2019, \aap, 625, A22. 
\bibitem[Ribas et al.(2015)]{Ribas2015} 
Ribas, {\'A}., Bouy, H., \& Mer{\'\i}n, B.\ 2015, \aap, 576, A52.
\bibitem[Saar et al.(1998)]{Saar1998} 
Saar, S.~H., Butler, R.~P., \& Marcy, G.~W.\ 1998, \apjl, 498, L153
\bibitem[Saio et al.(2015)]{Saio2015} 
Saio, H., Wood, P.~R., Takayama, M., et al.\ 2015, \mnras, 452, 3863.
\bibitem[Safronov (1969)]{Safronov1969}
Safronov, V. 1969, Evolution of the Protoplanetary Cloud and Formation of the Earth and Planets (Moscow: Nauka)
\bibitem[Santos et al.(2017)]{Santos2017} 
Santos, N.~C., Adibekyan, V., Figueira, P., et al.\ 2017, \aap, 603, A30
\bibitem[Sato et al.(2002)]{Sato2002} 
Sato, B., Kambe, E., Takeda, Y., Izumiura, H., \& Ando, H.\ 2002, \pasj, 54, 873.
\bibitem[Sato et al.(2005)]{Sato2005} 
Sato, B., Kambe, E., Takeda, Y., et al.\ 2005, \pasj, 57, 97.
\bibitem[Sato et al.(2012)]{Sato2012} 
Sato, B., Omiya, M., Harakawa, H., et al.\ 2012, \pasj, 64, 135.
\bibitem[Sato et al.(2013)]{Sato2013} 
Sato, B., Omiya, M., Harakawa, H., et al.\ 2013, \pasj, 65, 85.
\bibitem[Sato et al.(2015)]{Sato2015} 
Sato, B., Hirano, T., Omiya, M., et al.\ 2015, \apj, 802, 57.
\bibitem[Setiawan et al.(2003)]{Setiawan2003} 
Setiawan, J., Pasquini, L., da Silva, L., von der L{\"u}he, O., \& Hatzes, A.\ 2003, \aap, 397, 1151.
\bibitem[Schwarz (1978)]{Schwarz1978} 
Schwarz, G.\ 1978, Annals of Statistics, 6(2): 461-464.
\bibitem[Smirnov(1948)]{Smirnov1948}
Smirnov, N.\ 1948, Annals of Mathematical Statistics, 19 (2): 279–281.
\bibitem[Soszy{\'n}ski et al.(2021)]{Soszynski2021} 
Soszy{\'n}ski, I., Olechowska, A., Ratajczak, M., et al.\ 2021, \apjl, 911, L22.
\bibitem[Stassun, Collins, \& Gaudi(2017)]{Stassun2017} 
Stassun, K.~G., Collins, K.~A., \& Gaudi, B.~S.\ 2017, \aj, 153, 136.
\bibitem[Su{\'a}rez Mascare{\~n}o et al.(2020)]{Suarez2020} Su{\'a}rez Mascare{\~n}o, A., Faria, J.~P., Figueira, P., et al.\ 2020, \aap, 639, A77. doi:10.1051/0004-6361/202037745
\bibitem[Takarada et al.(2018)]{Takarada2018} 
Takarada, T., Sato, B., Omiya, M., et al.\ 2018, \pasj, 70, 59.
\bibitem[Takeda(1995)]{Takeda1995} 
Takeda, Y.\ 1995, \pasj, 47, 287
\bibitem[Takeda \& Tajitsu(2015)]{Takeda2015} 
Takeda, Y., \& Tajitsu, A.\ 2015, \mnras, 450, 397.
\bibitem[Takeda, Sato, \& Murata(2008)]{Takeda2008} 
Takeda, Y., Sato, B., \& Murata, D.\ 2008, \pasj, 60, 781.
\bibitem[Udry et al.(2019)]{Udry2019} 
Udry, S., Dumusque, X., Lovis, C., et al.\ 2019, \aap, 622, A37.
\bibitem[van Leeuwen(2007)]{vanLeeuwen2007} 
van Leeuwen, F.\ 2007, \aap, 474, 653.
\bibitem[Winn et al.(2009)]{Winn2009} 
Winn, J.~N., Johnson, J.~A., Albrecht, S., et al.\ 2009, The Astrophysical Journal, 703, L99.
\bibitem[Wittenmyer, Endl, Cochran, \& Levison(2007)]{Wittenmyer2007} 
Wittenmyer, R.~A., Endl, M., Cochran, W.~D., \& Levison, H.~F.\ 2007, \aj, 134, 1276.
\bibitem[Wittenmyer et al.(2017a)]{Wittenmyer2017a} 
Wittenmyer, R.~A., Jones, M.~I., Zhao, J., et al.\ 2017, \aj, 153, 51.
\bibitem[Wittenmyer et al.(2017b)]{Wittenmyer2017b} 
Wittenmyer, R.~A., Jones, M.~I., Horner, J., et al.\ 2017, \aj, 154, 274.
\bibitem[Wittenmyer et al.(2011)]{Wittenmyer2011} 
Wittenmyer, R.~A., Endl, M., Wang, L., et al.\ 2011, \apj, 743, 184.
\bibitem[Wood et al.(2004)]{Wood2004} 
Wood, P.~R., Olivier, E.~A., \& Kawaler, S.~D.\ 2004, \apj, 604, 800. 
\bibitem[Yu et al.(2018)]{Yu2018} 
Yu, J., Huber, D., Bedding, T.~R., et al.\ 2018, \mnras, 480, L48.
\bibitem[Zechmeister \& K{\"u}rster(2009)]{Zechmeister2009} 
Zechmeister, M., \& K{\"u}rster, M.\ 2009, \aap, 496, 577.
\end{thebibliography}
\end{document}